%% file: 0-main.tex
\documentclass[final,12pt]{article}
\pdfoutput=1
\usepackage{jheppub}
\usepackage{relsize}
\usepackage{units}

\usepackage{amssymb}

\usepackage{lineno}

\usepackage[section]{placeins}

\usepackage{color}
\usepackage{graphicx}
\usepackage{subfigure}
\usepackage{amsmath}

\input{maths}

\numberwithin{equation}{section}

\newcommand{\comment}[1]{}

\newcommand*\patchAmsMathEnvironmentForLineno[1]{%
  \expandafter\let\csname old#1\expandafter\endcsname\csname #1\endcsname
  \expandafter\let\csname oldend#1\expandafter\endcsname\csname end#1\endcsname
  \renewenvironment{#1}%
     {\linenomath\csname old#1\endcsname}%
     {\csname oldend#1\endcsname\endlinenomath}}%
\newcommand*\patchBothAmsMathEnvironmentsForLineno[1]{%
  \patchAmsMathEnvironmentForLineno{#1}%
  \patchAmsMathEnvironmentForLineno{#1*}}%
\AtBeginDocument{%
\patchBothAmsMathEnvironmentsForLineno{equation}%
\patchBothAmsMathEnvironmentsForLineno{align}%
\patchBothAmsMathEnvironmentsForLineno{flalign}%
\patchBothAmsMathEnvironmentsForLineno{alignat}%
\patchBothAmsMathEnvironmentsForLineno{gather}%
\patchBothAmsMathEnvironmentsForLineno{multline}%
}

\begin{document}

\title{Model independent determination of the CKM phase $\gamma$ using input from \Dzero-\DzeroBar\ mixing}


\author{Samuel Harnew}
\author{and Jonas Rademacker}
\affiliation{H H Wills Physics Laboratory, University of Bristol, UK}
\emailAdd{Sam.Harnew@bristol.ac.uk}
\emailAdd{Jonas.Rademacker@bristol.ac.uk}

\abstract{ We present a new, amplitude model-independent method to
  measure the \CP\ violation parameter $\gamma$ in \BmtoDK\ and
  related decays. Information on charm interference parameters,
  usually obtained from charm threshold data, is obtained from charm
  mixing.  By splitting the phase space of the \Dee\ meson decay into
  several bins, enough information can be gained to measure \gam\
  without input from the charm threshold. We demonstrate the
  feasibility of this approach with a simulation study of \BmtoDK\
  with \DtoKpluspipipi. We compare the performance of our novel
  approach to that of a previously proposed binned analysis which uses
  charm interference parameters obtained from threshold data. While
  both methods provide useful constraints, the combination of the two
  by far outperforms either of them applied on their own. Such an
  analysis would provide a highly competitive measurement of \gam. Our
  simulation studies indicate, subject to assumptions about data
  yields and the amplitude structure of \DzerotoKpipipiDCS, a
  statistical uncertainty on \gam\ of $\sim 12^{\circ}$ with existing
  data and $\sim 4^{\circ}$ for the LHCb-upgrade.}

\maketitle


\section{Introduction}
\label{sec:intro}
\input{1-introduction}

\section{Formalism}
\label{sec:theory} 
\input{2-theory}
\section{Amplitude models and binning}
\subsection{Amplitude model}
\label{sec:model}
\input{3-model}

\subsection{Model-informed  binning}
\label{sec:binning}
\input{4-binning}

\section{Simulation studies}
\label{sec:toyStudies}
\input{5-toy-studies}


\section{Conclusion}
\label{sec:conclusion}
\input{6-conclusions}

\acknowledgments
\label{sec:acknowledgements}
\input{7-acknowledgements}




\bibliographystyle{JHEP}
\bibliography{bibliography}






\end{document}

%% file: maths.tex
\newcommand{\CLScanBox}[1]{\parbox[t]{0.28\textwidth}{\includegraphics[width=0.33\textwidth]{#1}}}
\newcommand{\CLYaxis}[1]{\rotatebox{90}{\mbox{}\hspace{2.5em}#1}\hspace{-1em}\mbox{}}

\newcommand{\Eqnref}[1]{Equation~\ref{#1}}
\newcommand{\eqnref}[1]{Eq.~\ref{#1}}

\newcommand{\eqnsref}[2]{Eqs.~\ref{#1}~-~\ref{#2}}

\newcommand{\eqnsrefList}[1]{Eqs.~#1}

\newcommand{\Tabref}[1]{Table~\ref{#1}}
\newcommand{\tabref}[1]{Tab.~\ref{#1}}

\newcommand{\secref}[1]{Sec.~\ref{#1}}

\newcommand{\secs}{Sections}

\newcommand{\Figref}[1]{Figure~\ref{#1}}
\newcommand{\figref}[1]{Fig.~\ref{#1}}
\newcommand{\Figrefs}[2]{Figures~\ref{#1} and~\ref{#2}}

\newcommand{\Imag}{\ensuremath{\mathit{Im}}}
\newcommand{\Real}{\ensuremath{\mathit{Re}}} 

\newcommand{\un}[2]{\ensuremath{#1\,\mathrm{#2}}}

\newcommand{\prt}[1]{\ensuremath{\mathit{#1}}}

\newcommand{\degrees}{\ensuremath{\mbox{}^o}}

\newcommand{\zeroCharge}{\ensuremath{\mathrm{0}}}

\newcommand{\gam}{\ensuremath{\gamma}}
\newcommand{\delB}{\ensuremath{\delta_B}}
\newcommand{\GammaD}{\ensuremath{\Gamma_D}}

\newcommand{\half}{\ensuremath{\frac{1}{2}}}

\def\DbarLHCb    {\kern 0.2em\overline{\kern -0.2em \ensuremath{D}}{}}

\def\Dzb     {\prt{\DbarLHCb^{\zeroCharge}}}

\newcommand{\Dee}{\prt{D}}
\newcommand{\Dbar}{\prt{\DbarLHCb}}
\newcommand{\Dzero}{\prt{D^\zeroCharge}}      
\newcommand{\DzeroBar}{\Dzb}
\newcommand{\Do}{\Dzero}
\newcommand{\Dobar}{\DzeroBar}
\newcommand{\DDbar}{\prt{\Dee\Dbar}}

\newcommand{\Deven}{\prt{D_{+}}}
\newcommand{\Dodd}{\prt{D_{-}}}
\newcommand{\Devo}{\prt{D_{\pm}}}

\newcommand{\fD}{\prt{\mathit{f}}}
\newcommand{\fDbar}{\prt{\overline{\mathit{f}}}}

\newcommand{\B}{\prt{B}}
\newcommand{\Bpm}{\prt{\B^{\pm}}}
\newcommand{\Bmp}{\prt{\B^{\mp}}}
\newcommand{\Bp}{\prt{\B^{+}}}
\newcommand{\Bm}{\prt{\B^{-}}}

\newcommand{\Kp}{\prt{K^{+}}}
\newcommand{\Km}{\prt{K^{-}}}
\newcommand{\Kpm}{\prt{K^{\pm}}}
\newcommand{\Kmp}{\prt{K^{\mp}}}

\newcommand{\pip}{\prt{\pi^{+}}}
\newcommand{\pim}{\prt{\pi^{-}}}
\newcommand{\pipm}{\prt{\pi^{\pm}}}
\newcommand{\pimp}{\prt{\pi^{\mp}}}

\newcommand{\cleoc}{\mbox{{C}{L}{E}{O}{-}{c}}}

\newcommand{\besiii}{\mbox{{B}{E}{S}{~}{I}{I}{I}}}

\newcommand{\besIII}{\besiii}

\newcommand{\markIII}{\mbox{{M}{A}{R}{K}{~}{I}{I}{I}}}

\newcommand{\lhcb}{{L}{H}{C}{b}}

\newcommand{\belleII}{{B}{E}{L}{L}{E}{-}{I}{I}}

\newcommand{\CP}{\ensuremath{\mathit{CP}}}

\newcommand{\ham}{\ensuremath{\hat{H}}}

\newcommand{\pspoint}{\ensuremath{\mathbf{p}}}

\newcommand{\intVolume}{\ensuremath{{\Omega}}}
\newcommand{\intVolumeBar}{\ensuremath{\bar{\intVolume}}}
\newcommand{\intVolumeSmall}{\ensuremath{\mathsmaller{\intVolume}}}

\newcommand{\intOmega}{\ensuremath{\int\limits_{\intVolume}}}

\newcommand{\deltaP}{\ensuremath{\mathrm{d}^n p}}

\newcommand{\inlinedphidp}{\ensuremath{\textstyle{ |\frac{\partial^n \phi}{\partial(p_1 \ldots p_n)} | }}}
\newcommand{\dphidp}{\ensuremath{\textstyle{ \left|\frac{\partial^n \phi}{\partial(p_1 \ldots p_n)} \right| }}}

\newcommand{\AMag}{\ensuremath{\mathcal{A}}}
\newcommand{\BMag}{\ensuremath{\mathcal{B}}}

\newcommand{\SupAmp}{\ensuremath{\mathcal{S}}}
\newcommand{\FavAmp}{\ensuremath{\mathcal{F}}}
\newcommand{\SupAmpP}{\ensuremath{\mathcal{S_{+}}}}
\newcommand{\FavAmpP}{\ensuremath{\mathcal{F_{+}}}}
\newcommand{\SupAmpM}{\ensuremath{\mathcal{S_{-}}}}
\newcommand{\FavAmpM}{\ensuremath{\mathcal{F_{-}}}}

\newcommand{\AMagOm}{\ensuremath{\AMag_{\intVolume}}}
\newcommand{\BMagOm}{\ensuremath{\BMag_{\intVolume}}}

\newcommand{\Z}{\ensuremath{\mathcal{Z}^f}}
\newcommand{\ZOm}{\ensuremath{\Z_{\intVolume}}}

\newcommand{\magZ}{\ensuremath{|\Z|}}

\newcommand{\ZKpipipi}{\ensuremath{\mathcal{Z}^{K3\pi}}}
\newcommand{\ZOmKpipipi}{\ensuremath{Z_{\intVolume}^{K3\pi}}}
\newcommand{\ZKpipipiOm}{\ZOmKpipipi}

\newcommand{\magZKpipipi}{\ensuremath{|\ZKpipipi|}}
\newcommand{\magZOmKpipipi}{\ensuremath{|\ZOmKpipipi|}}

\newcommand{\RD}{\ensuremath{R_D^f}}
\newcommand{\delD}{\ensuremath{\delta_D^f}}

\newcommand{\ROm}{\ensuremath{R_{\Omega}^f}}
\newcommand{\delOm}{\ensuremath{\delta_{\Omega}^f}}

\newcommand{\RDKpipipi}{\ensuremath{R_D^{K3\pi}}}
\newcommand{\delDKpipipi}{\ensuremath{\delta_D^{K3\pi}}}

\newcommand{\rDp}{\ensuremath{r_{\pspoint}}}
\newcommand{\delDp}{\ensuremath{\delta_{\pspoint}}}

\newcommand{\ImZKpipipi}{\ensuremath{\Imag \ZKpipipi}}
\newcommand{\ReZKpipipi}{\ensuremath{\Real \ZKpipipi}}

\newcommand{\rB}{\ensuremath{r_{B}}}
\newcommand{\rD}{\ensuremath{r_{\mathit{Df}}}}
\newcommand{\rDOm}{\ensuremath{r_{\mathit{D, \intVolume}}}}

\newcommand{\gamt}{\ensuremath{\Gamma_D t}}

\newcommand{\f}{\ensuremath{f}}
\newcommand{\fbar}{\ensuremath{\bar{f}}}

\newcommand{\fp}{\ensuremath{f_{\pspoint}}}

\newcommand{\KPlusMinuspipipi}{\prt{\Kpm\pimp\pipm\pimp}}
\newcommand{\KplusMinuspipipi}{\KPlusMinuspipipi} 

\newcommand{\DtoKpipipi}{\prt{\Dee \to \Km\pip\pim\pip}}
\newcommand{\DtoKPlusMinuspipipi}{\prt{\Dee \to \KPlusMinuspipipi}}

\newcommand{\DtoKpluspipipi}{\prt{\prt{\Dee \to \Kp\pim\pip\pim}}}
\newcommand{\DzeroToKpipipiCF}{\prt{\prt{\Dzero  \to \Km\pip\pim\pip}}}
\newcommand{\DzerotoKpipipiDCS}{\prt{\prt{\Dzero \to \Kp\pim\pip\pim}}}

\newcommand{\BmptoDK}{\prt{\Bmp \to \Dee \Kmp}}
\newcommand{\BtoDK}{\BmptoDK}

\newcommand{\BmtoDK}{\prt{\Bm \to \Dee \Km}}
\newcommand{\BptoDK}{\prt{\Bp \to \Dee \Kp}}

\newcommand{\BtoDKpipipiK}{\prt{\Bpm \to \Dee(K3\pi) \Kpm}}

\newcommand{\Dztof}{\prt{\Dzero(t) \to \f}}
\newcommand{\Dzbartof}{\prt{\DzeroBar(t) \to \f}}

\newcommand{\rateDztof}{\ensuremath{\Gamma( \Dztof )_{\intVolumeSmall}}}
\newcommand{\rateDzbartof}{\ensuremath{\Gamma ( \Dzbartof)_{\intVolumeSmall}}}

\newcommand{\bra}[1]{\ensuremath{\langle #1 |}}
\newcommand{\ket}[1]{\ensuremath{| #1 \rangle}}

\newcommand{\magninline}[1]{\ensuremath{|#1|}}

%% file: 1-introduction.tex
The measurement of \gam\ from \BmtoDK, \prt{\Dee \to f}~\cite{GLW1,
GLW2, ADS, DalitzGamma1, DalitzGamma2, Rademacker:2006zx} (where
\prt{f} represents a multibody final state accessible to both \Do\ and
\Dobar) depends on the correct description of the interference between
the \prt{\Do \to f} and \prt{\Dobar \to f} decay
amplitudes.\footnote{Charged conjugate modes are implied throughout
unless stated otherwise. The symbol \Dee\ is used to represent any
superposition of \Do\ and \Dobar.}
This can be obtained from an amplitude model of the
\Dee\ decay. However, this model dependence can lead to significant
systematic uncertainties. Alternative model-independent methods use
experimental input~\cite{Atwood:coherenceFactor,
ModelIndepGammaTheory} to remove this source of systematic
uncertainty. This input can be summarised in the complex interference
parameter $\Z = \RD e^{-i \delD}$, where \RD\ and \delD\ are the
coherence factor and average strong phase-difference introduced in
~\cite{Atwood:coherenceFactor}. \Z\ can be measured exploiting
quantum-correlated \DDbar\ pairs available at experiments operating at
the charm threshold, like \cleoc\ or
\besiii~\cite{Atwood:coherenceFactor, ModelIndepGammaTheory,
Lowery:2009id, Libby:2010nu, Briere:2009aa, CLEO:DeltaKpi,
Insler:2012pm, Libby:2014rea}.

We found previously that input from charm mixing, when combined with
constraints from threshold data, can substantially reduce the
uncertainty on \Z~\cite{selfcite}.  In this letter we present a new
method for an amplitude model-independent measurement of \gam\ based
on charm input from mixing that, by dividing the \Dee\ decay's phase
space into multiple bins, extracts sufficient information to perform a
model independent measurement of \gam\ without input from charm
threshold results.  We verify the feasibility of this method using
simulated data. We also study the performance of a binned analysis
with charm input from the charm threshold, rather than mixing, as proposed
in~\cite{Atwood:coherenceFactor}. While both methods provide
interesting constraints on \gam\ and related parameters, we find that
a combined approach far outperforms each method individually. Applied
to \BmtoDK, \DtoKpipipi, a substantially better precision on \gam\ and
related parameters can be achieved than with previously considered
methods for this decay mode, potentially making this one of the most
precise individual measurements.

This letter is organised as follows: based on the formalism described
in~\cite{selfcite} we show in \secref{sec:theory} that, when the \Dee\
decay's phase space is divided into multiple bins, it is possible to
extract \gam\ from a simultaneous analysis of \BtoDK\ and \Dee-mixing
without input from charm threshold data. In \secref{sec:binning} we
discuss how to divide the five-dimensional phase space of \DtoKpipipi\
into bins in a way that optimises the sensitivity to \gam. In
\secref{sec:toyStudies} we present the results of a simulation study
for the decay mode \BmtoDK, \DtoKpipipi, using sample sizes
corresponding to our estimates of plausible current and future \lhcb\
event yields. We estimate the precision on \gam\ and related
parameters for various data taking scenarios and approaches, with and
without input from the charm threshold. The key results of the
simulation study are summarised in \secref{sec:ResultSummary}
(\tabref{tab:allResults}). In \secref{sec:conclusion}, we
conclude.

%% file: 2-theory.tex
\subsection{Phase-space integrated amplitudes and interference
  parameter}
The measurement of \gam\ from \BmtoDK~\cite{GLW1, GLW2, ADS,
  DalitzGamma1, DalitzGamma2, Rademacker:2006zx,
  Atwood:coherenceFactor, ModelIndepGammaTheory} and the method for
extracting \Z\ from mixing introduced in~\cite{selfcite} both exploit
the interference of \Do\ and \Dobar\ decay amplitudes to the same
final state \fp, $\bra{\fp} \ham \ket{\Dzero}$ and $\bra{\fp} \ham
\ket{\DzeroBar}$. The subscript $\pspoint = (p_1, \ldots, p_n)$
identifies a point in $n$ dimensional phase space, with $n=3N_f-7$ for
a final state $f$ with a particle content of $N_f$
pseudoscalars. \ham\ is the interaction Hamiltonian relevant for the
decay. It is useful to define the magnitude of the ratio of these
amplitudes, \rDp, and their phase difference \delDp, at phase-space
point \pspoint, through
\begin{equation}
\label{eq:defineStrongPhaseDiff}
 \rDp e^{ i\delDp} = 
\frac{\bra{\fp} \ham \ket{\Dzero}}{\bra{\fp} \ham \ket{\DzeroBar}} .
\end{equation}
The decay rates integrated over regions or bins of phase space, which
we label with \intVolume, can be expressed in terms of the real, positive quantities
\begin{align}
\AMagOm 
& \equiv \sqrt{\intOmega |\bra{\fp} \ham \ket{\Dzero}|^{2} \dphidp \deltaP},
&
\BMagOm 
&\equiv \sqrt{\intOmega |\bra{\fp} \ham \ket{\DzeroBar}|^{2} \dphidp \deltaP},
\label{eqn:Damps}
\end{align}
and the complex parameter
\begin{align}
\ZOm \equiv \frac{1}{\AMagOm\BMagOm}
 \intOmega \bra{\fp} \ham \ket{\Dzero} \bra{\fp} \ham
 \ket{\DzeroBar}^{*} \dphidp \deltaP.
\label{eqn:normalisedZ}
\end{align}
In these expressions, \inlinedphidp\ represents the density of states
at phase space point \pspoint. The complex interference parameter
\ZOm\ has a magnitude between $0$ and $1$. It encodes the relevant
interference effects in phase-space region \intVolume. As the
integrand in the definition of \ZOm\ is proportional to $e^{i\delDp}$,
\magninline{\ZOm} is maximal if \delDp\ is constant over the
integration region, while highly fluctuating \delDp\ tends to result
in small \magninline{\ZOm}. The complex interference parameter \ZOm\
can also be expressed in terms of the coherence factor
$R^f_{\intVolume}$ and average strong phase difference
$\delta^f_{\intVolume}$ introduced in~\cite{Atwood:coherenceFactor},
or in terms of the $c_{\intVolume}$ and $s_{\intVolume}$ parameters
introduced in~\cite{ModelIndepGammaTheory}:
\begin{equation}
\label{eq:ZandRDandCiSi}
\ZOm = R^f_{\intVolume} e^{-i\delta_{\intVolume}^f} 
     = c_{\intVolume} + i\,s_{\intVolume}.
\end{equation}
\Eqnref{eq:ZandRDandCiSi} implies a normalisation of $c_{\intVolume}$
and $s_{\intVolume}$ that differs from that in the original
paper~\cite{ModelIndepGammaTheory}, but corresponds to the one used in
most subsequent publications~\cite{Libby:2010nu,Briere:2009aa,
  Bondar:CharmMixingCP, Belle_uses_CLEO_2011, LHCb2012DalitzGamma}.

\subsection{D mixing, time-dependent decay rates}

For simplicity, we assume \CP\ conservation in the neutral \Dee\
system, which has been shown to be a valid assumption to a frustrating
degree of accuracy~\cite{HFAG2013, LHCbMixingAndCPV}. The general case
is described for example in~\cite{selfcite}. We use the following
convention for the definition of the \CP\ even and odd \Dee\
eigenstates, \Deven\ and \Dodd:
\begin{equation}
\label{eq:DEvenOdd}
 \ket{\Devo} = \ket{\Dzero} \pm  \ket{\DzeroBar}
\end{equation}
which have masses $M_{\pm}$ and widths $\Gamma_{\pm}$. We also define
the mean lifetime $\Gamma_D$ and the usual dimensionless mixing
parameters $x$ and $y$:
\begin{align}
\Gamma_{D} & \equiv \half \left(\Gamma_- + \Gamma_+\right),
  &
x & \equiv \frac{M_- - M_+}{\Gamma} , 
  &  
y & \equiv \frac{\Gamma_- - \Gamma_+}{2\Gamma}.
\label{eq:xy}
\end{align}
\comment{double check signs!}  The mixing parameters $x$ and $y$ are
both small, approximately half a
percent~\cite{CDF:Mixing2008,Belle:Mixing2007,BaBar:Mixing2007,BaBar:Mixing2008,BaBar:Mixing2009,LHCb:Mixing,LHCbMixingAndCPV,HFAG2013}. The
above definitions imply $\CP\ \ket{\Do} = + \ket{\Dobar}$. An
alternative choice would be $\CP\ \ket{\Do} = - \ket{\Dobar}$,
resulting in a phase-shift of \ZOm, defined in
\eqnref{eqn:normalisedZ}, by $\pi$~\cite{selfcite}.

Although the method presented here is in principle applicable to any
\Dee\ decay to a final state accessible to both \Do\ and \Dobar, we
will restrict ourselves from here on to the case where $\bra{\fp} \ham
\ket{\Dzero}$ is doubly Cabibbo suppressed (DCS) and $\bra{\fp} \ham
\ket{\DzeroBar}$ is Cabibbo favoured (CF), as is the case for $f=
K^+\pi^-\pi^+\pi^-$. Such decays have the advantage that for the
suppressed, ``wrong sign'' (WS) decay, the mixing-induced amplitude
$A(\prt{\Do \to \Dobar \to f})$ and the direct amplitude $A(\prt{\Do
  \to f})$ are of comparable magnitude, leading to large interference
effects, and high sensitivity to \Z. On the other hand, the ``right
sign'' (RS) decay \prt{D^0 \to \fbar} is completely dominated by the
CF amplitude, with negligible interference effects, and thus provides
an excellent normalisation mode.
%
For this case
\begin{equation}
\rDOm \equiv \frac{\AMagOm^{DCS}}{\BMagOm^{CF}} \ll 1.
\end{equation}
where $\AMagOm$, $\BMagOm$ are defined in \eqnref{eqn:Damps}; the
superscripts are added for clarity.
The time dependent rates for a \Dee\ meson that was a \Do\ or a
\Dobar\ at time $t_0=0$, to decay to a final state $f$ within the
phase-space volume \intVolume\ at proper time $t$ are given, up to
third order in the small parameters $x$, $y$ and
$\rDOm$, by
\begin{equation}
\rateDztof \simeq \left[
  \AMagOm^{2} 
 +
 \AMagOm \BMagOm \left( y \Real{ \ZOm } + x  \Imag{ \ZOm } \right)
 \gamt 
 + 
 \BMagOm^{2} \frac{x^2 + y^2}{4} (\gamt)^{2} 
\right]   e^{-\gamt},
 \label{eqn:WSrate}
\end{equation}
for the WS rate, and
\begin{align}
\rateDzbartof &\simeq   \BMagOm^{2}  e^{-\gamt}
 \label{eqn:RSrate} 
\end{align}
for the RS rate, with corresponding expressions for the \CP\ conjugate
modes.
Many detector effects cancel in the ratio of WS to RS decays, given by
\begin{align}
\frac{\rateDztof}{\rateDzbartof} = 
 \rDOm^2 
 + \rDOm \left( y \Real{ \ZOm } + x  \Imag{ \ZOm } \right) (\gamt)
 + \frac{x^2 + y^2}{4} (\gamt)^{2}.
 \label{eqn:Ratiorate}
\end{align}
\subsection{\prt{B^{\mp} \to DK^{\mp}}, \gam, and \Z}
The decay \prt{B^{-} \to \Dee K^{-}}, and related decays, provide a
particularly clean way of measuring the CKM phase \gam. The details of
the analysis depend considerably on the final state \fD\ of the
subsequent \Dee\ decay, which must be accessible to both \Do\
and \Dobar~\cite{GLW1, GLW2, ADS, DalitzGamma1, DalitzGamma2,
  Rademacker:2006zx}. The sensitivity to \gam\ arises from the interference
of the decay amplitudes with the intermediary states \prt{\Do K^-} and
\prt{\Dobar K^-}, which we express as:
\begin{align}
 \FavAmpP \equiv \bra{ \DzeroBar \Kp } \ham \ket{ \Bp } ,
& \ \ \ \ \  \SupAmpP \equiv \bra{ \Dzero \Kp } \ham \ket{ \Bp } ,
\notag\\
 \FavAmpM \equiv \bra{ \Dzero \Km } \ham \ket{ \Bm } ,
& \ \ \ \ \  \SupAmpM \equiv \bra{ \DzeroBar \Km } \ham \ket{ \Bm } .
 \label{eqn:Bamps}
\end{align}
where $\mathcal{F}$ denotes colour and CKM favoured amplitudes,
while $\mathcal{S}$ denotes colour and CKM suppressed amplitudes.
The ratios of the suppressed to favoured amplitudes are given by
\begin{align}
\rB e^{i \left( \delB - \gamma \right)}
&= \frac{\SupAmpM}{\FavAmpM}
&
\rB e^{i \left( \delB + \gamma \right)}
&= \frac{\SupAmpP}{\FavAmpP}
\end{align}
where $\rB$ is the magnitude of those ratios, while $\delB$ and $\mp
\gamma$ are their strong and weak phase differences respectively.

Because \rB\ is small ($\sim
0.1$~\cite{LHCb2013GammaCombination,LHCb-CONF-2013-006}), the
interference effects and thus the sensitivity to \gam\ in \prt{\Bm \to
  \Dee K^-, \Dee \to f}, are enhanced if a final state is chosen such
that \prt{\Do \to f} is doubly Cabibbo suppressed, while \prt{\Dobar
  \to f} is Cabibbo favoured~\cite{ADS}, at the cost of an overall low
decay rate. The time and phase space integrated decay rate for these
suppressed \Bmp\ decays is given by
\begin{align}
\Gamma \left ( \BmtoDK , \Dee \to \f \right )_{\intVolume}
  \simeq 
  &  \FavAmp^{2} \AMagOm^{2} +  \SupAmp^{2} \BMagOm^{2} + 
\FavAmp \SupAmp \AMagOm \BMagOm 
     \left|\ZOm\right|  \cos (\delB -\delOm - \gam )
 \label{eqn:BsupRateM}
\\
\Gamma \left ( \BptoDK , \Dee \to \fbar \right )_{\intVolumeBar}
 \simeq 
  &  \FavAmp^{2} \AMagOm^{2} +  \SupAmp^{2} \BMagOm^{2} + 
\FavAmp \SupAmp \AMagOm \BMagOm 
     \left|\ZOm\right|  \cos (\delB -\delOm + \gam )
 \label{eqn:BsupRateP}
\end{align}
 The corresponding favoured decay \prt{\Bm \to \Dee K^-,
  \Dee \to \fDbar} is completely dominated by the favoured decay
amplitude with negligible interference effects and negligible
sensitivity to \gam, and has a much larger branching fraction. It
therefore provides an ideal normalisation or control mode. Its time
and phase-space integrated rate is given by:
\begin{align}
\Gamma  ( \BmtoDK , \Dee \to \fbar )_{\intVolumeBar} \simeq 
\Gamma  ( \BptoDK , \Dee \to \f )_{\intVolume} \simeq 
\FavAmp^{2} \BMagOm^{2}
 \label{eqn:BfavRate}
\end{align}
The ratios of the favoured and suppressed rates are given by
\begin{align}
\frac{\Gamma \left ( \BmtoDK, \Dee \to \f \right )_{\intVolume}} {\Gamma \left ( \BmtoDK, \Dee \to \fbar \right )_{ \intVolumeBar}} 
& = \rDOm^{2} + \rB^{2} + \rDOm \rB \left| \ZOm \right| \cos(\delB -\delOm - \gam )  \label{eqn:BdecayRatioM}
\\
\frac{\Gamma \left ( \BptoDK, \Dee \to \fbar \right )_{\intVolumeBar}} {\Gamma \left ( \BptoDK, \Dee \to \f \right )_{ \intVolume}} 
& = \rDOm^{2} + \rB^{2} + \rDOm \rB \left| \ZOm \right| \cos (\delB
-\delOm + \gam ) .  \label{eqn:BdecayRatioP}
\end{align}
These can also be expressed in terms of the Cartesian coordinates
\begin{align}
x_{\pm} &\equiv \Real \left( r_B e^{i\left(\delB \pm \gamma\right)} \right)
&
y_{\pm} &\equiv \Imag \left(r_B e^{i\left(\delB \pm \gamma\right)}\right)
\end{align}
using the relations
\begin{align}
\rB \left| \ZOm \right| \cos(\delB -\delOm \pm \gam )
&=
 x_{\pm}\Real{ \ZOm } + y_{\pm}\Imag{ \ZOm } 
\;\;\;\mbox{and}
&
\rB^2 &= x_{\pm}^2 + y_{\pm}^2 .
\end{align}
Effects due to \Do-\Dobar\ mixing have been ignored in the expressions
for the \BmptoDK, \prt{\Dee \to \f (\fbar)} decay rates, which is justified given
the expected statistical precision. These effects can be included if
required~\cite{Rama:2013voa}.

\subsection{Parameter counting using ratios}
\label{sec:ParaCountingRatios}
Taking $\Gamma_D$, $x$, and $y$ from external inputs,
\eqnsrefList{\ref{eqn:Ratiorate}, \ref{eqn:BdecayRatioM},
  \ref{eqn:BdecayRatioP}} depend on three unknown parameters for each
pair of \CP-conjugate phase space bins (\intVolume, \intVolumeBar):
\rDOm, $\Real{ \ZOm }$ and $\Imag{ \ZOm }$; and three that are the
same in all bins: \gam, \delB\ and \rB. The time-dependent fit to the
tagged charm decay rates (\eqnref{eqn:Ratiorate}) provides two
constraints on these parameters for each bin (the constant and the
coefficient of the linear term). The \BtoDK\ decay rate ratios
(\eqnsrefList{\ref{eqn:BdecayRatioM}, \ref{eqn:BdecayRatioP}}) provide
another two constraints. For $N$ bin pairs, there are therefore $4N$
constraints and $3N + 3$ unknown parameters. To extract all unknown
parameters from the data therefore requires $4N \ge 3N+3
\Leftrightarrow N \ge 3$. If instead we wish to measure $x_{\pm},
y_{\pm}$, we need $N \ge 4$.
%
\subsection{Parameter counting using rates}
\label{sec:ParaCountingRates}
Taking again $\Gamma_D$, $x$, and $y$ from external inputs,
\eqnsrefList{\ref{eqn:WSrate}, \ref{eqn:RSrate}, \ref{eqn:BsupRateM},
  \ref{eqn:BsupRateP}, \ref{eqn:BfavRate}} depend on four unknown
parameters for each pair of \CP-conjugate phase space bins:
$\AMagOm^{2}$, $\BMagOm^{2}$, $\Real{\ZOm}$, and $\Imag{\ZOm}$; and
four that are the same in all bins: \gam, \delB, $r_B = \SupAmp/\FavAmp$,
$\FavAmp^{2}$. \eqnsref{eqn:WSrate}{eqn:RSrate} provide three constraints for
each bin, and \eqnsref{eqn:BsupRateM}{eqn:BfavRate} another
three. Hence, to extract all of these parameters, we require $6N \ge
4N + 4 \Leftrightarrow N \ge 2$. A fit to extract $x_{\pm},
y_{\pm}$ requires $N \ge 3$.
\subsection{Multiple solutions}
\label{sec:ambiguities}
\begin{figure}
\begin{center}
\includegraphics[width=0.5\textwidth]{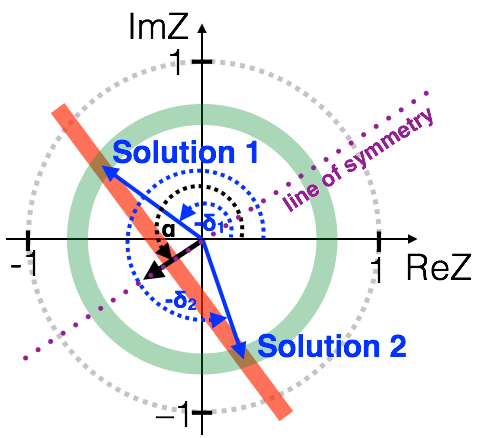}
\end{center}
\caption{The combined constraints on $\ZOm = \ROm e^{-i\delOm}$ from
  charm mixing (red line with slope -y/x) and \BtoDK\ (green solid
  circle) lead to two possible solutions, whose sum (short black
  arrow) is always perpendicular to the charm
  constraint. (In the figure, the subscript $\Omega$ and superscript
  $f$ are omitted for clarity.) The grey broken circular line
  indicates the boundary of the physically allowed region.\label{fig:multipleSolutions}}
\end{figure}
As described in \cite{selfcite}, the charm mixing input constrains
each $\ZOm = \ROm e^{-i\delOm}$ to a line of slope $-y/x$ in the
$\Real{\ZOm} - \Imag{\ZOm}$ plane. The input from the \BtoDK\ adds
information on the magnitude of \ZOm, leaving two possible solutions
for each \ZOm, which have the same magnitude but different phases:
$-\delta_{\Omega\;1}^f$ and $-\delta_{\Omega\;2}^f$, as illustrated in
\figref{fig:multipleSolutions}. These solutions are symmetric with
respect to a line of symmetry that is perpendicular to the constraint
from charm mixing. Their sum is always along this line of symmetry and
has the phase $ \alpha = - \half( \delta_{\Omega\;1}^f +
\delta_{\Omega\;2}^f)$. Because $\alpha$ depends only on the
charm mixing parameters (with $\tan\alpha = x/y$) it is the same for
all phase-space bins. It is easy to show that, as a consequence of
this relationship, the system of equations remains invariant under the
following operation:
\begin{equation}
\left( \left\{\delOm\right\}, \delB, \gamma\right)
\to
\left(\left\{-2\alpha - \delOm\right\},  -2\alpha - \delB, - \gamma\right)
.
\end{equation}
There is also the more obvious invariance under the
simultaneous shift by $\pi$ of $\delB$ and $\gamma$:
\begin{equation}
\left(\left\{\delOm\right\}, \delB, \gamma\right)
\to
\left(\left\{\delOm\right\},  \delB + \pi, \gamma +\pi \right),
\end{equation}
leading to an overall four-fold ambiguity in \gam\ and \delB. In
\secref{sec:ExternalConstraints} we show how external input from the
charm threshold~\cite{Libby:2014rea,Lowery:2009id} can be used to
reduce this to a 2-fold ambiguity.

%% file: 3-model.tex
Up to this point, the discussion has not been specific to any
particular final state of the \Dee\ decay. For the remainder of this
letter, we will require a specific amplitude model to test the binning
method (\secref{sec:binning}) and perform simulation studies
(\secref{sec:toyStudies}). We will concentrate on the case where the
\Dee\ meson decays to \KplusMinuspipipi. Our amplitude model for the
CF \prt{\DzeroBar \to K^+\pi^-\pi^+\pi^-} decay is based on that found
by the \markIII\ experiment~\cite{MarkIII_K3piModel}. There is
currently no model available for the DCS decay \prt{\Dzero \to
K^+\pi^-\pi^+\pi^-}. Any experiment in a position to use the method
described here would have sufficient DCS decays to obtain such a
model. For the purpose of this study, we have created a series of
plausible DCS models by randomly varying the magnitudes and phases of
the amplitude components of \markIII's CF model. Amongst these we
select a representative sample of 100 DCS models that give, together
with the \markIII\ model for the CF decay, global complex coherence
parameters \ZKpipipi\ distributed approximately according to the
\cleoc\ measurement~\cite{Libby:2014rea}. Most studies are based on
our default model, which we chose based on its \ZKpipipi\ value of
$0.26 + i0.24 = 0.36e^{i(42\pi / 180 )}$, which matches the central
value measured in~\cite{Libby:2014rea}.

%% file: 4-binning.tex
The model-independent method for measuring \gam\ described in
\secref{sec:theory} relies on dividing the $\Dzero \to \f$ phase
space, which is five dimensional for \DtoKpipipi, into several
bins. In principle, any binning will work, for example the rectangular
five dimensional binning used in~\cite{LHCb2013:Miranda}. However, to
optimise the sensitivity of our approach, we follow the ideas for a
model-informed binning described
in~\cite{Bondar:2007ir,Bondar:CharmMixingCP}. Because \ZOm\ is a
factor in all \gam\ sensitive terms, the sensitivity to \gam\
increases with larger values of $|\ZOm|$ in each bin. A strategy that
ensures large $|\ZOm|$ is to split phase space into bins of similar
phase difference $\delDp$. We use an amplitude model to assign a value
of \delDp\ to each event. The optimised binning is then achieved by
splitting the one-dimensional \delDp\ distribution into continuous
intervals, each of which constitutes one bin (which could in principle
be discontinuous in 5-dimensional phase space). We choose the size of
the intervals such that there is a similar number of suppressed
\BtoDK\ events in each bin. A wrong model would result in a
sub-optimal binning, resulting in smaller, but still
model-independently measured, $|\ZOm|$ in each bin. While this would
reduce the sensitivity, which would be evident from the statistical
uncertainty estimated from the fit, it would not introduce a
model-dependent bias.
\newcommand{\ZScatterPlot}[1]{\includegraphics[width=0.45\linewidth]{#1}}
 \begin{figure}[t]
 \centering
 \ZScatterPlot{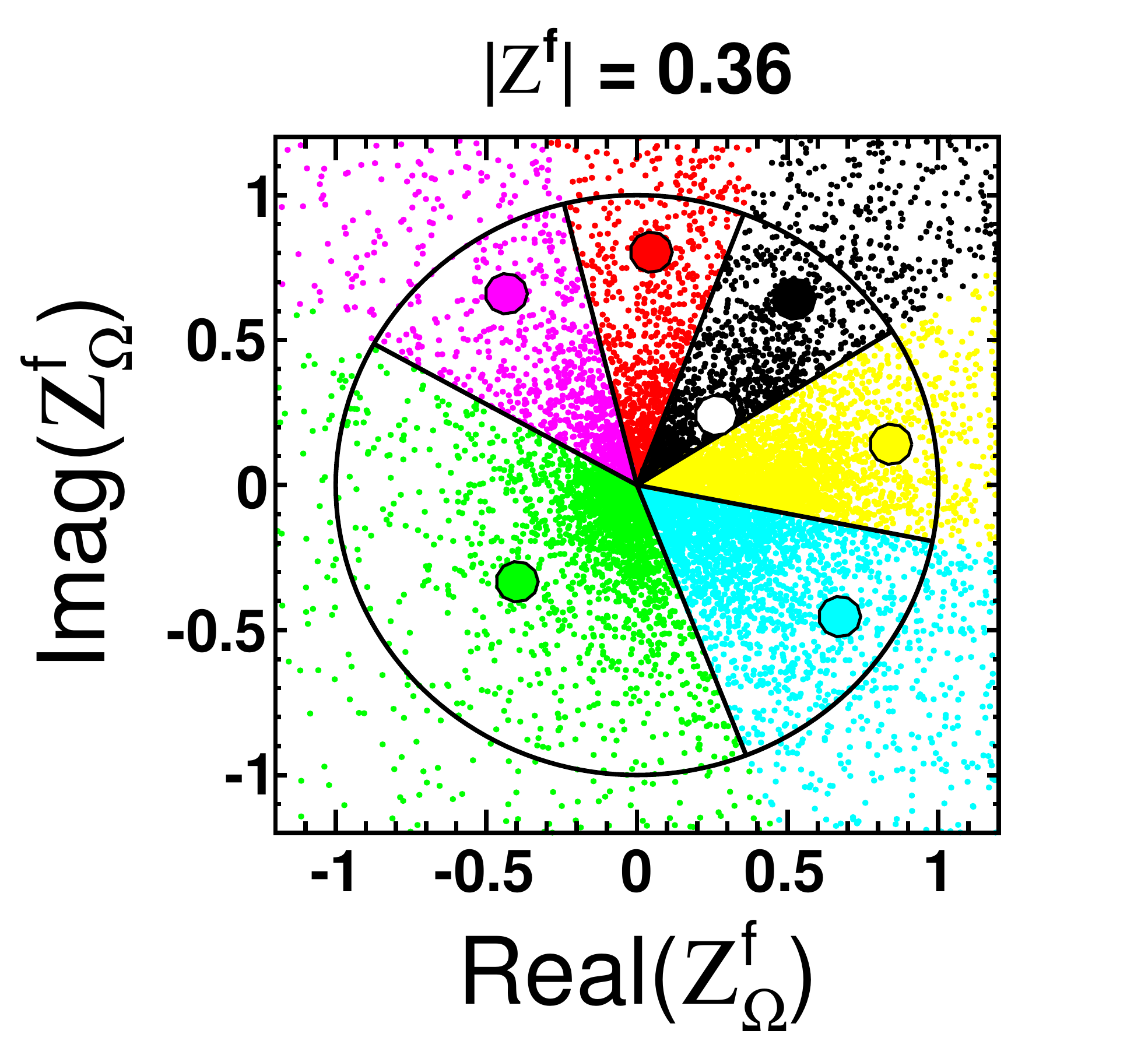}
 \ZScatterPlot{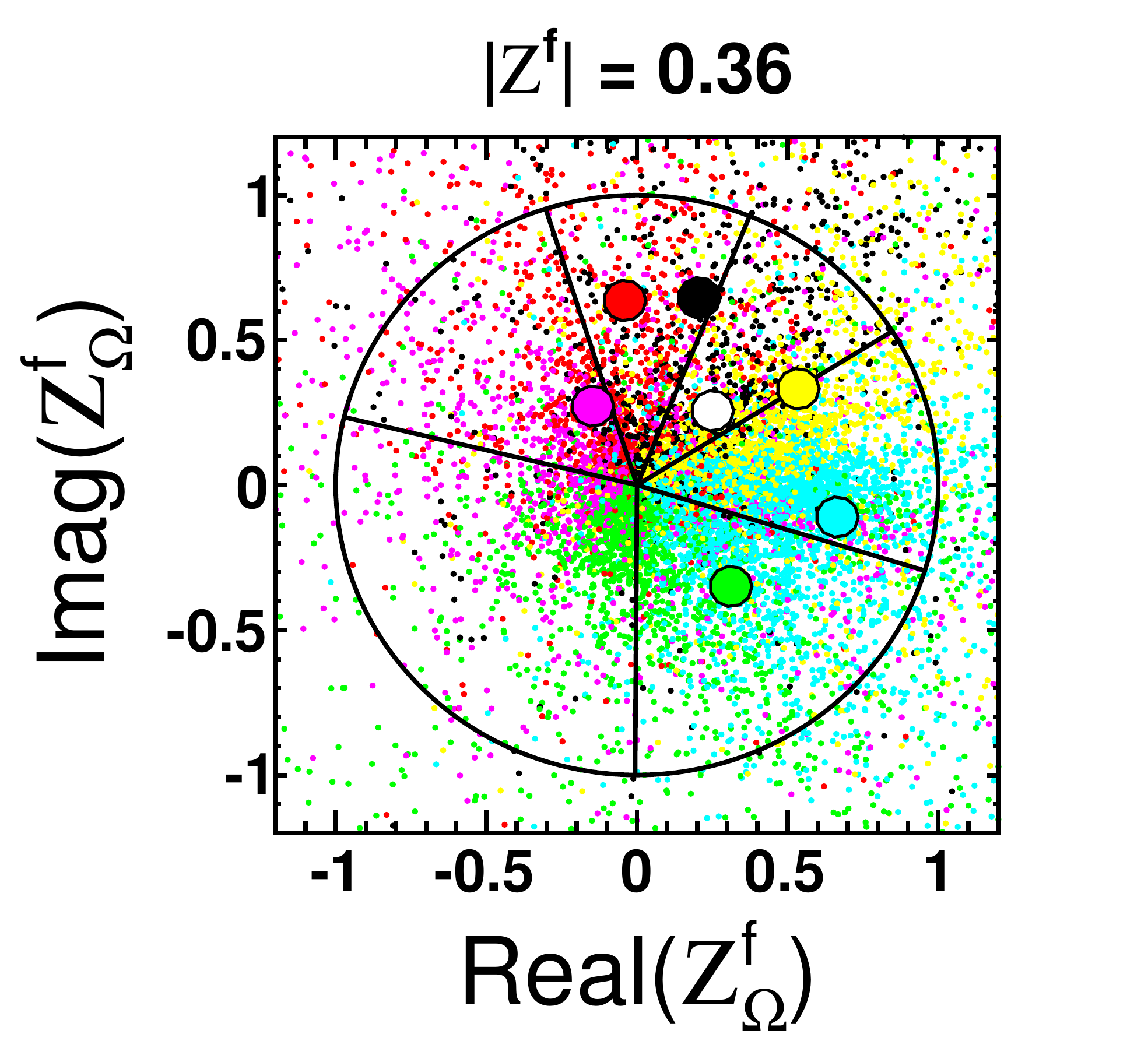}
   \caption{ The plot shows simulated events (small dots), complex
     coherence parameters \ZKpipipiOm\ (colour-filled circles) for each bin,
     and the global coherence parameters \ZKpipipi\ (white-filled circle),
     represented in the \ReZKpipipi-\ImZKpipipi\ plane, with bin
     assignments based on a perfect and an imperfect amplitude model,
     as described in the text.\comment{Should we change f to K3pi in the figure?}
     \label{fig:binningGoodAndBad}}
\end{figure}
\Figref{fig:binningGoodAndBad} shows the binned \ZKpipipiOm\ obtained
from the default model, on the left hand side for a binning based on a
perfect model and on the right for a binning based on an imperfect
model. The perfect model is identical to the one used for the event
generation. The imperfect model is obtained from the perfect one by
multiplying each amplitude component's magnitude by a random factor
between $0.8$ and $1.2$ (corresponding to a fit fraction variation of
0.64 -- 1.44), and by adding to each component a random phase between
$-0.3$ and $+0.3$ radians. \Figref{fig:binningGoodAndBad} shows
simulated events represented in the \ReZKpipipi--\ImZKpipipi\
plane. The events are generated according to the phase space density
of states.  The position of the small dots represents the true value
of $\frac{1}{\AMagOm\BMagOm} \bra{\fp} \ham \ket{\Dzero} \bra{\fp}
\ham \ket{\DzeroBar}^{*}$, while the colour-coding represents the bin
they have been assigned to. For the left hand plot, this assignment is
done with the perfect model, for the right hand plot with an imperfect
model. The circular ``pie chart'' represents the bins in $\delDp$
based on the model used for the binning. The \ZKpipipiOm\ values
extracted are the average over the true values of
$\frac{1}{\AMagOm\BMagOm} \bra{\fp} \ham \ket{\Dzero} \bra{\fp} \ham
\ket{\DzeroBar}^{*}$ for the events in the bin they have been assigned
to (which includes events beyond the plot boundaries). The
model-independent method proposed above does of course not require the
knowledge of $\frac{1}{\AMagOm\BMagOm} \bra{\fp} \ham \ket{\Dzero}
\bra{\fp} \ham \ket{\DzeroBar}^{*}$ to measure \ZKpipipiOm, this
information is only used for this illustration. The
\ZKpipipiOm\ values are shown as colour-filled circles. The global
complex coherence parameter \ZKpipipi\ is shown as a white-filled
circle. While the imperfect model leads to smaller $|\ZKpipipiOm|$,
they are still on average larger than the global $|\ZKpipipi|$.

To quantify this observation, we repeated the study with the full set of
100 representative models and different numbers of bins.
\begin{figure}[ht]
\centering
\includegraphics[width=0.5\linewidth]{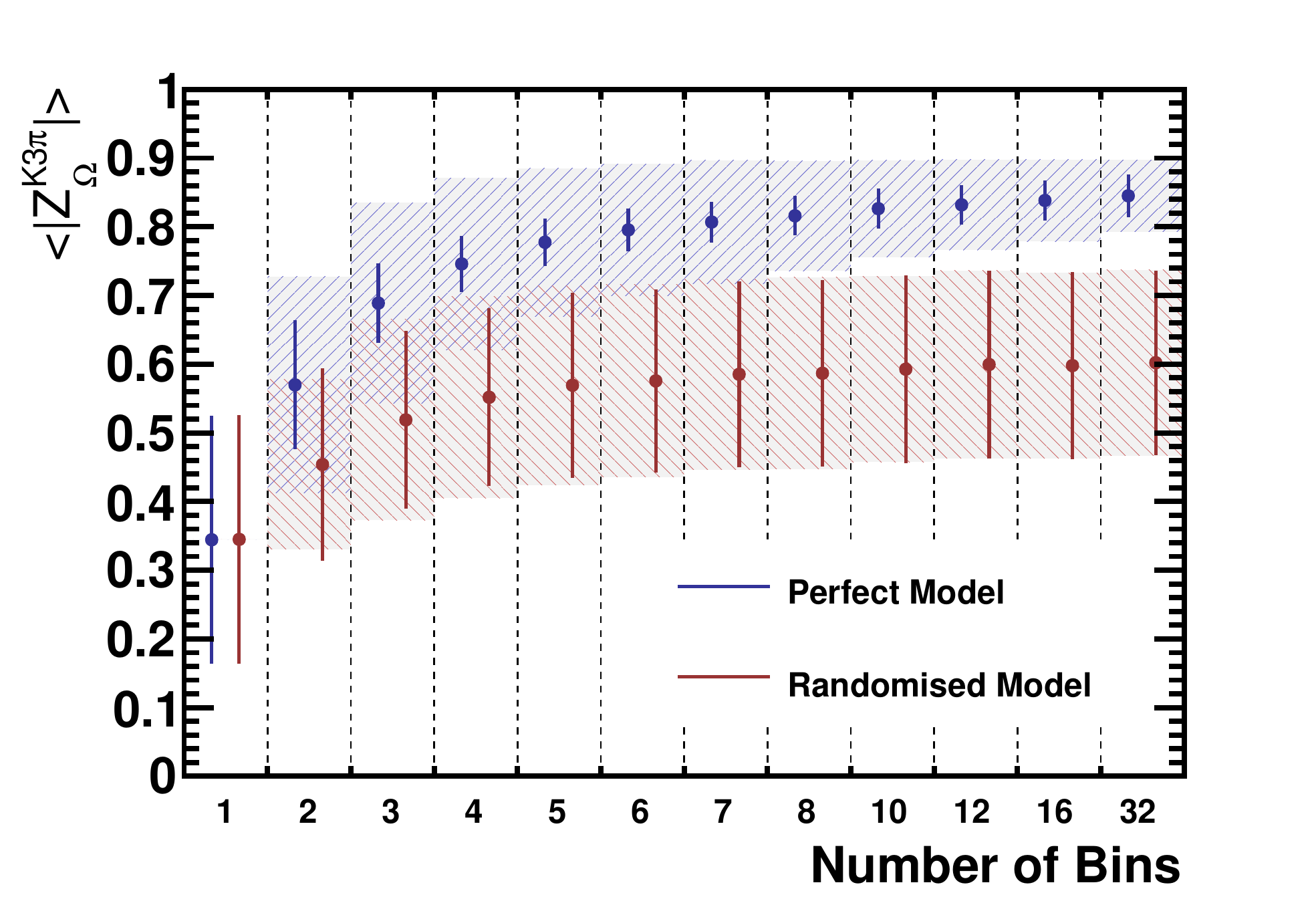}
        \caption{The average coherence factor with different number of
          bins in strong phase difference, for the set of 100
          representative models, with perfect binning (blue, on top)
          and imperfect binning described in the text (red, lower).
          The error bars represent the standard deviation of the
          mean \ZKpipipiOm\ of each model, i.e. they represent the ``between
          model scatter''. The shaded
          areas represent the average of the standard deviation of \ZKpipipiOm\
          within the models (i.e. the ``within model scatter'').
        \label{fig:CoherenceVersusNumberOfBins}}
\end{figure}
The results are summarised in \figref{fig:CoherenceVersusNumberOfBins}
which shows the average \magZOmKpipipi\ as a function of the number of bins
for the case where the binning is based on a perfect model, and for
the case where the model used for binning is randomised as described
above. The study shows that even a rather ``bad'' model provides
typical binned coherence factors that are substantially larger than
the global coherence factor.

%% file: 5-toy-studies.tex
In order to demonstrate the validity of our method, and to evaluate
its sensitivity, we perform fits to simulated data.

\subsection{Simulated data samples}
The data are generated according to the CF amplitude model based on the
\markIII\ analysis of \DzeroToKpipipiCF~\cite{MarkIII_K3piModel} .
For the DCS amplitude describing \DzerotoKpipipiDCS\ we choose from
the large number of models we generated (see \secref{sec:binning}) the
one that, when combined with the CF model, reproduces best the
measured value of \ZKpipipi~\cite{Libby:2014rea} as our default
model. We also consider other DCS models to evaluate the stability of
our results.

\begin{table}
\centering
\begin{tabular}{l |r |r r r}
        \ & \multicolumn{2}{c}{  \prt{\Bpm \to D(K3\pi) K^{\pm}}} & \vline & 
        \multicolumn{1}{l}{\prt{D^{*\pm} \to }} \\
        \ & suppressed    & favoured &  \vline & 
        \multicolumn{1}{l}{\mbox{}\hspace{0.2em}\prt{D(K3\pi)\pi^{\pm}}}\\
         \hline
LHCb run~I (\un{3}{fb^{-1}} @ \un{7-8}{TeV}) & 120 & 10k & \vline &  8M\hspace{1em} \\
LHCb run~II (\un{8}{fb^{-1}} @ \un{13}{TeV}) & 800 & 60k & \vline &  50M\hspace{1em}  \\
LHCb upgrade (\un{50}{fb^{-1}} @ \un{13}{TeV}) & 9000 & 700k & \vline &  600M\hspace{1em} \\
\end{tabular}
\caption{Event yields assumed in the
  simulation studies, based on reported event yields for
  \un{1}{fb^{-1}} at LHCb~\cite{LHCb2013ADSObservation,
    LHCb2013:Miranda}. The event yields are inclusive, for example,
    LHCb run~II yields includes those from LHCb run~I.
    The fraction of WS events in \prt{D^{*\pm} \to
    D(K3\pi)\pi^{\pm}}
  depends on the input variables; typically it is $0.38\%$.
  \label{tab:eventYields}}
\end{table}
We study three scenarios with different event yields, based on
plausible extrapolations of the yields reported for \un{1}{fb^{-1}} at
LHCb~\cite{LHCb2013ADSObservation, LHCb2013:Miranda}: ``LHCb run~I'',
where we extrapolate event yields to LHCb's already recorded
\un{3}{fb^{-1}}; ``LHCb run II'', plausible event yields at the end
of the next LHC data taking period with approximately twice the
collision energy; and ``LHCb upgrade'', estimated event yields for the
LHCb upgrade. We take into account the increase in the heavy flavour
cross section at higher collision energies, and the expected
improvement in trigger efficiency at the LHCb
upgrade~\cite{LHCbUpgradeCDR}. The sample sizes we use in our
simulation studies, are given in \tabref{tab:eventYields}. These
extrapolations have of course large uncertainties.

We take into account the time-dependent detection efficiency that is
typical for hadronic heavy flavour decays at LHCb, where the trigger
is based on detecting displaced vertices, disfavouring small decay
times. We use the same efficiency function as in~\cite{selfcite}. We
ignore all other detector effects and backgrounds, given the
clean data samples at LHCb even for the suppressed \BtoDKpipipiK\ 
modes~\cite{LHCb2013ADSObservation}, this is a reasonable
simplification for the purpose of these feasibility studies. Simulated
data are generated with the following parameter values:
$\gamma = 69.7\degrees$
$\delta_B = 112.0\degrees$,
$r_B = 0.0919$, and
$r_D^2 = \frac{1}{300}$.

\subsection{Fit method and parametrisation}
\label{sec:method}

Our default approach is to perform a simultaneous $\chi^2$ fit to the
decay rates \eqnsrefList{\ref{eqn:WSrate}, \ref{eqn:RSrate},
  \ref{eqn:BsupRateM}, \ref{eqn:BsupRateM} and \ref{eqn:BfavRate}} in terms of the
fit parameters $\rDOm$, $\Real{\ZOm}$, 
$\Imag{\ZOm}$, \BMagOm, \FavAmp, \gam, \delB\ and $\rB$. As a cross check, we also performed
binned likelihood fits and found that they lead to equivalent results,
but take longer to converge.

As long as all phase space bins are well populated, we find that the
fit results are not crucially dependent on the number of bins. In our
default scenario we divide phase space into 4~bins for Run~I,
6~bins for Run~II and 8~bins for the upgrade.

We allow the charm mixing parameters
$x$ and $y$ to vary in the fit, but constrain their value with a
two-dimensional Gaussian constraint to their world-average
using, for the LHCb Run~I scenario~\cite{HFAG_CHARM2013}:
\begin{align}
x & = 0.526 \pm 0.161 \% & y &= 0.668 \pm 0.088 \% & \rho_{xy} = 0.188,
\end{align}
where $\rho_{xy}$ is the correlation coefficient between $x$ and
$y$. We expect substantial improvements on this measurement from
\lhcb, its upgrade, and \belleII\ in the future. Lacking detailed
forecasts, for the purpose of this study, we assume that the
uncertainties on $x$ and $y$ scale with the inverse square-root of
LHCb event yields of the relevant data taking scenario, while the
correlation coefficient remains constant. We fix the well-measured
average \Dee\ lifetime to $\tau_D=1/\GammaD =
\un{410.1}{fs}$~\cite{PDG2014}.


While the default approach is to fit the
decay rates, in an experimental measurement it may be favourable
to fit the decay rate ratios \eqnsrefList{\ref{eqn:Ratiorate},
  \ref{eqn:BdecayRatioM} and \ref{eqn:BdecayRatioP}}. 
In this case we loose sensitivity to the parameters $\BMagOm$ and $\FavAmp$.
Using both fit methods on the same simulated dataset, we find that both
approaches give the same results on the parameters they share.
In \secref{sec:ExternalConstraints} we will demonstrate how fitting
the rates, as opposed to the ratios, allows us to add additional constraints
to the fit.

\subsection{Algorithms}
In order to cope with the various local $\chi^2$ minima that are
present in addition to the four global minima, we use a two-stage
fitting process. The first step is a fit with the GENEVA~\cite{geneva}
package which is specifically designed to deal with multiple
minima. We use GENEVA's parameter estimates as input to
MINUIT~\cite{MINUIT} and perform a second fit to refine the parameter
estimate. To further reduce the risk of converging on false minima, we
repeat this process 75 times with many randomly chosen starting values
for all fit parameters. Finally, we choose the fit result that gives
the smallest $\chi^2$ as our central value. In order to avoid
unphysical values of \ZOm, which also can lead to further secondary
minima, we add for each volume \intVolume\ a term that increases the
$\chi^2$ if \ZOm\ leaves the physical region:
\begin{equation}
\chi^2_{\mathrm{constr\;\ZOm}} = 
\left\{
\begin{array}{cl}
\left((|\ZOm| - 1)/0.5\right)^2 & \mbox{if } |\ZOm| > 1 \\
0 & \mbox{else}
\end{array}
\right\}
\end{equation}
\subsection{Confidence regions in \gam, \delB, \rB\ and $x_{\pm}$, $y_{\pm}$}
\label{sec:confidenceScans}
\begin{figure}
\begin{tabular}{c ccc}
     & \delB\ vs \gam & \rB\ vs \gam & $x_{\pm}$ vs $y_{\pm}$ \\
\CLYaxis{LHCb run~1} & 
\CLScanBox{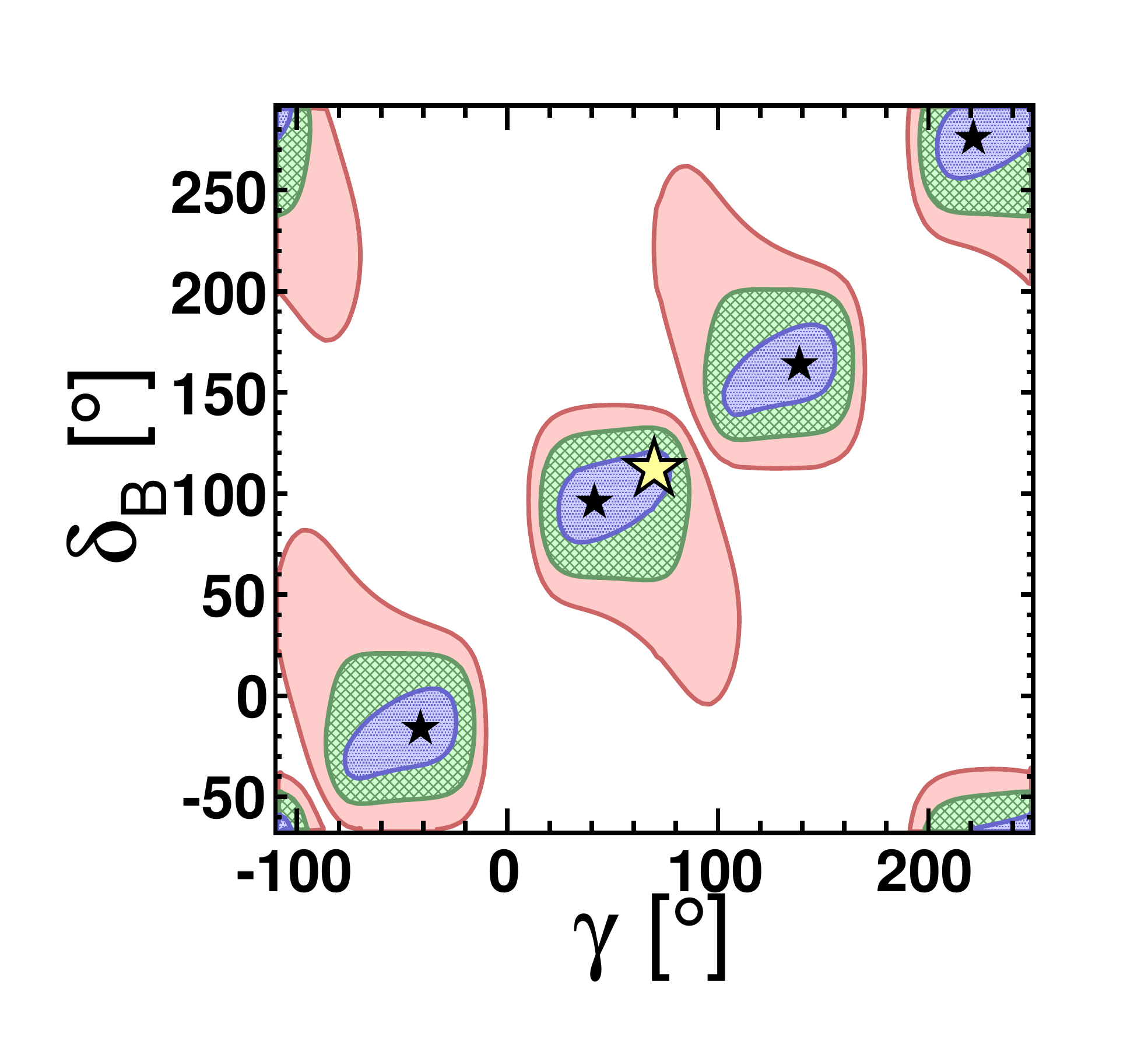} &
\CLScanBox{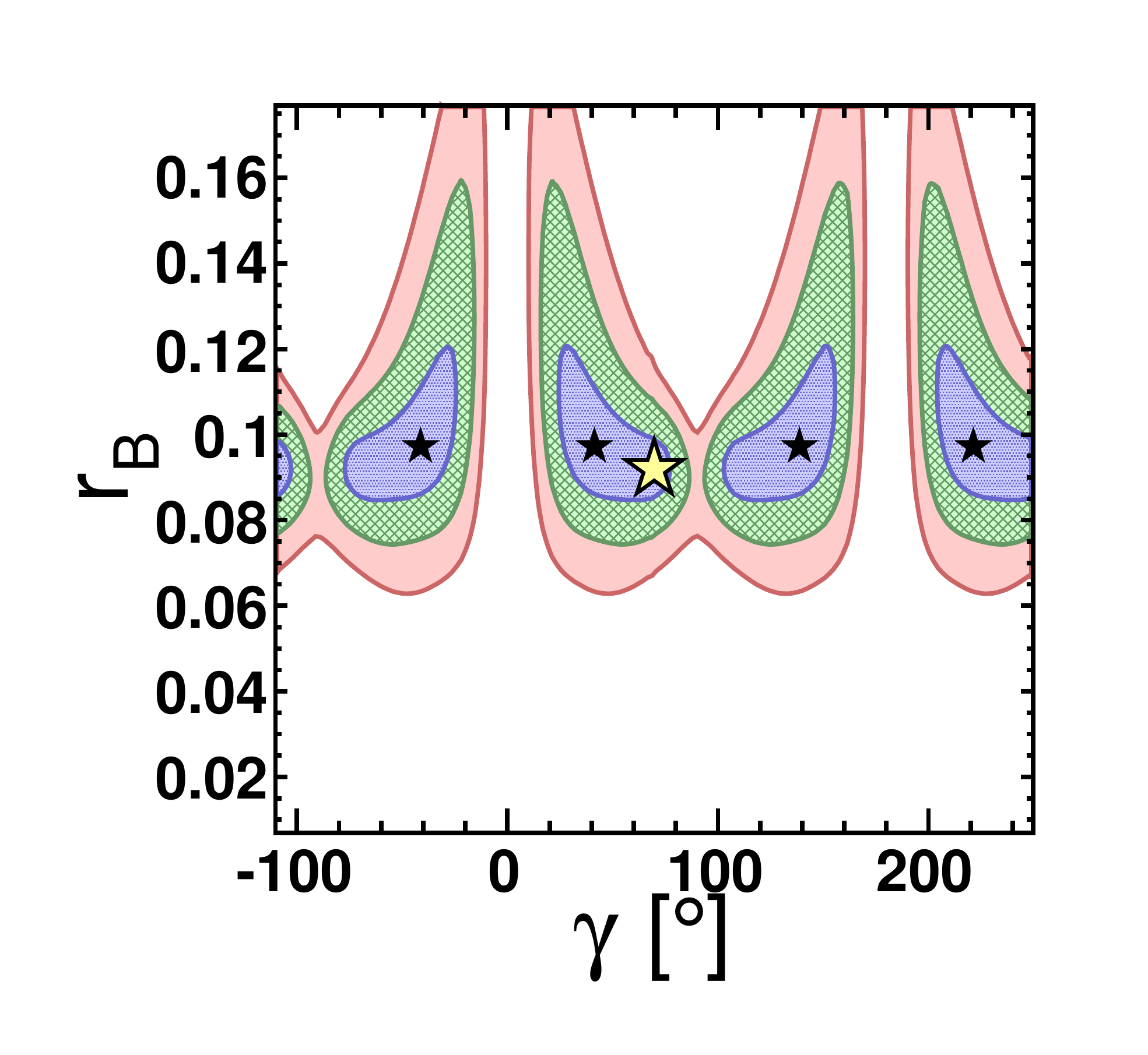}    &
\CLScanBox{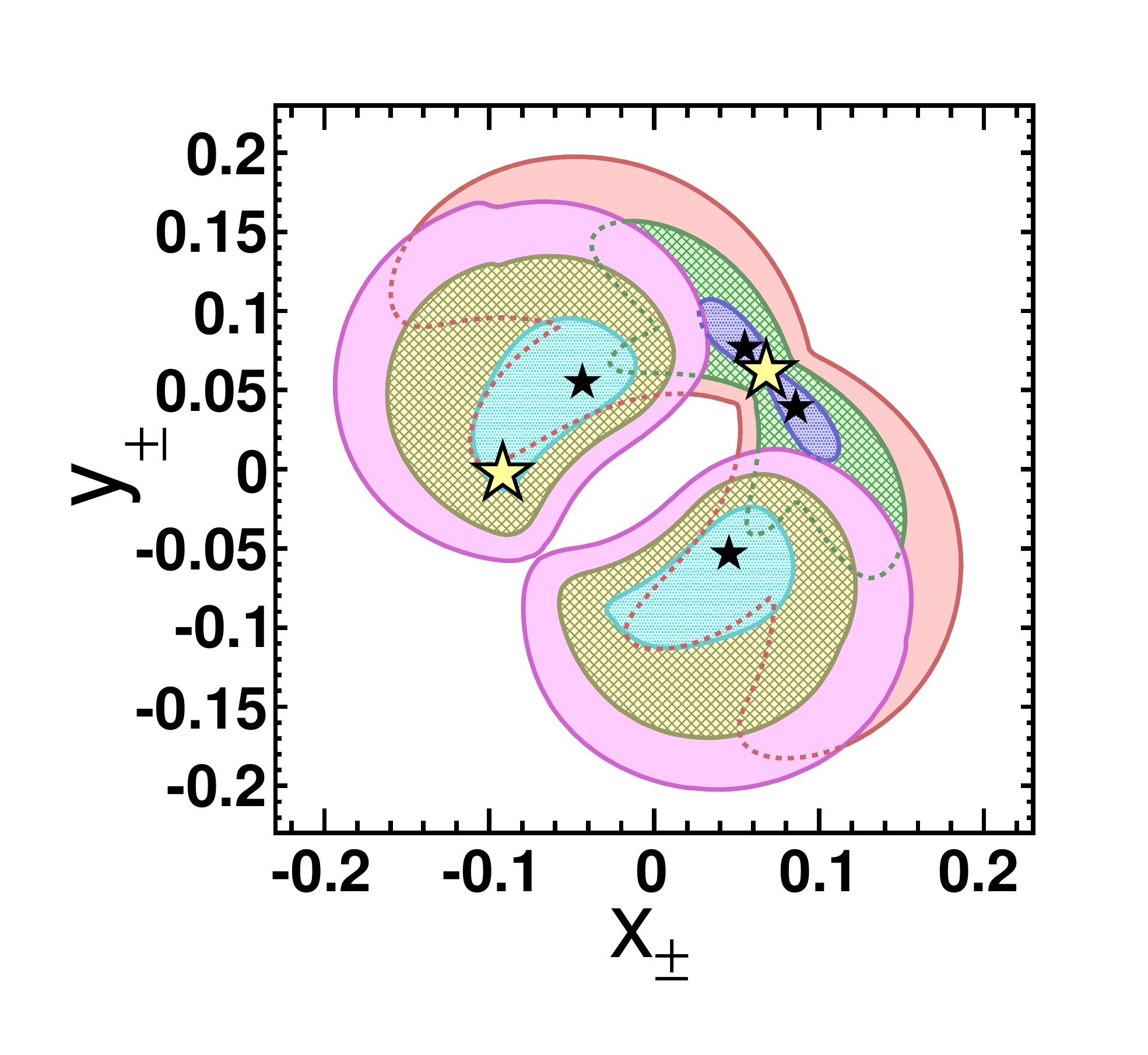}\\
\CLYaxis{LHCb run~II} &
\CLScanBox{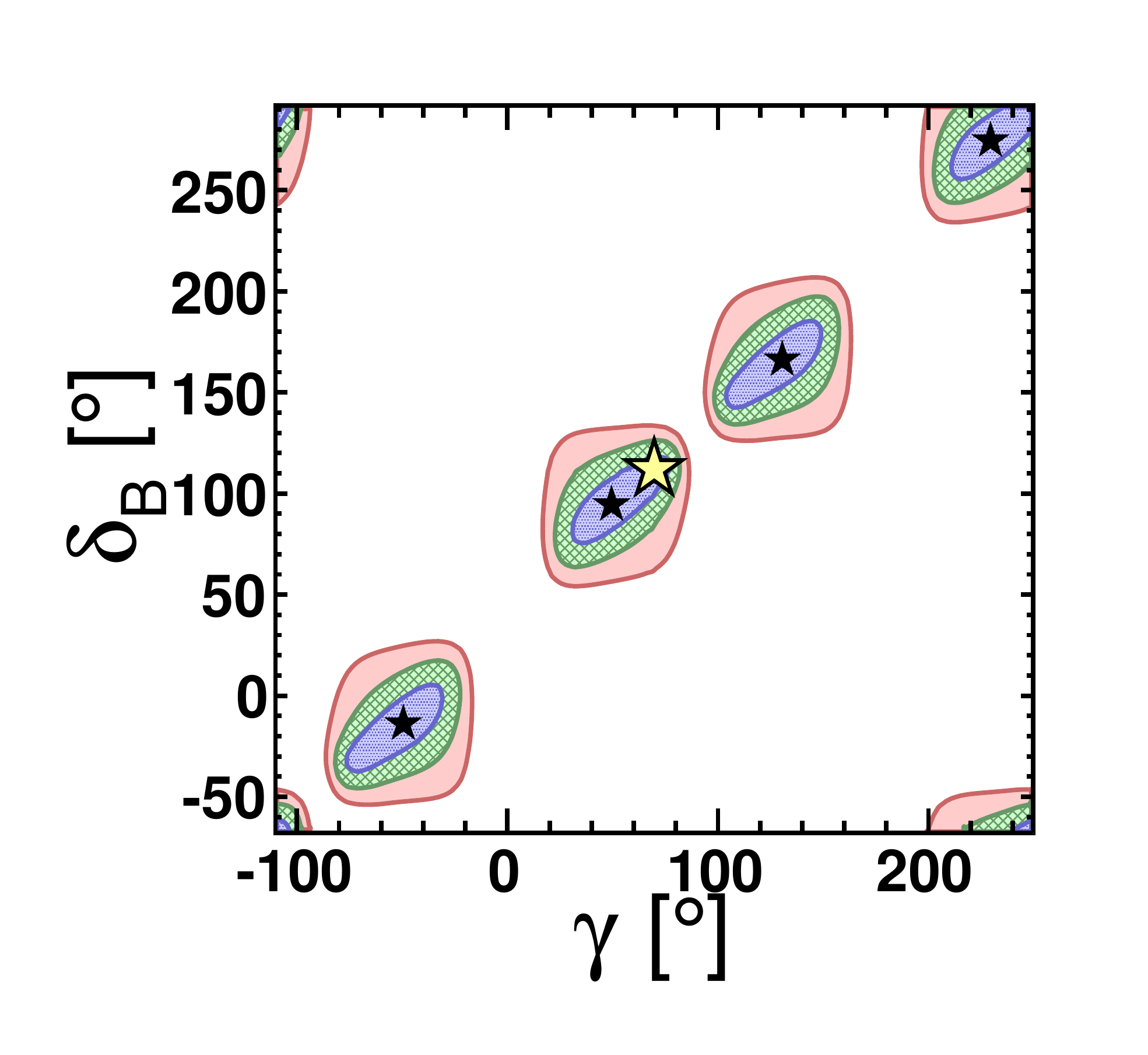} &
\CLScanBox{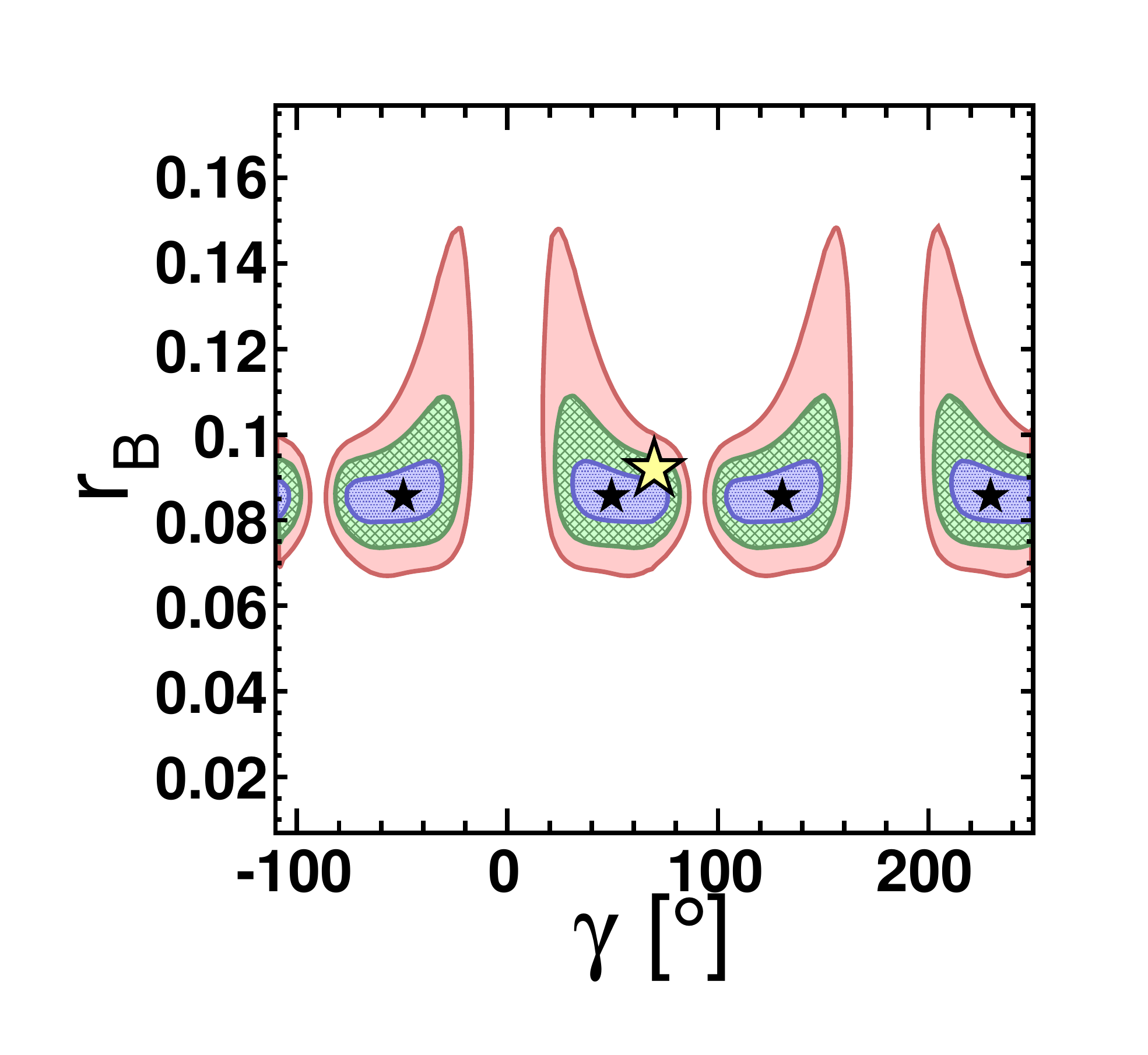}    &
\CLScanBox{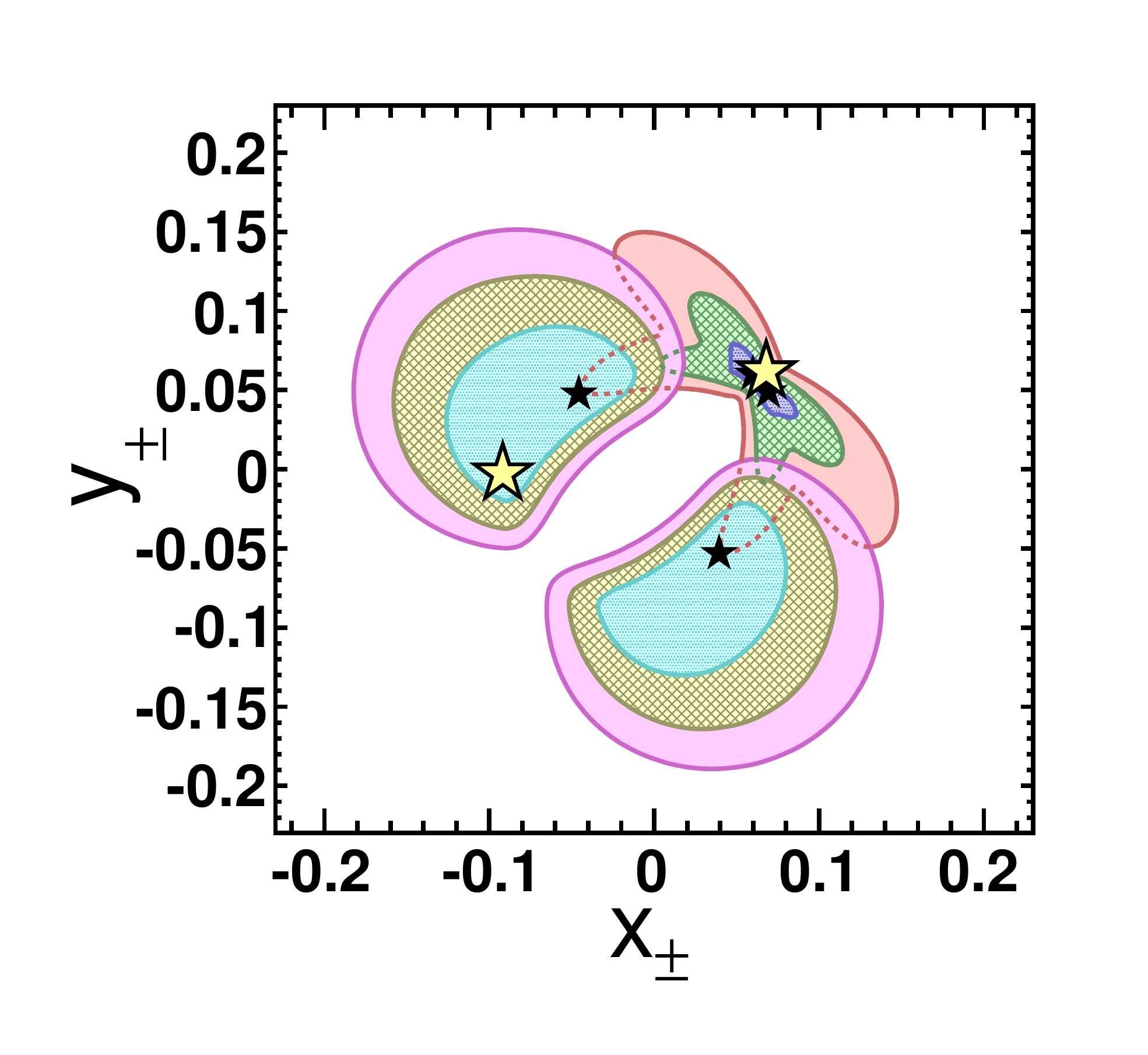}\\
\CLYaxis{LHCb upgrade} &
\CLScanBox{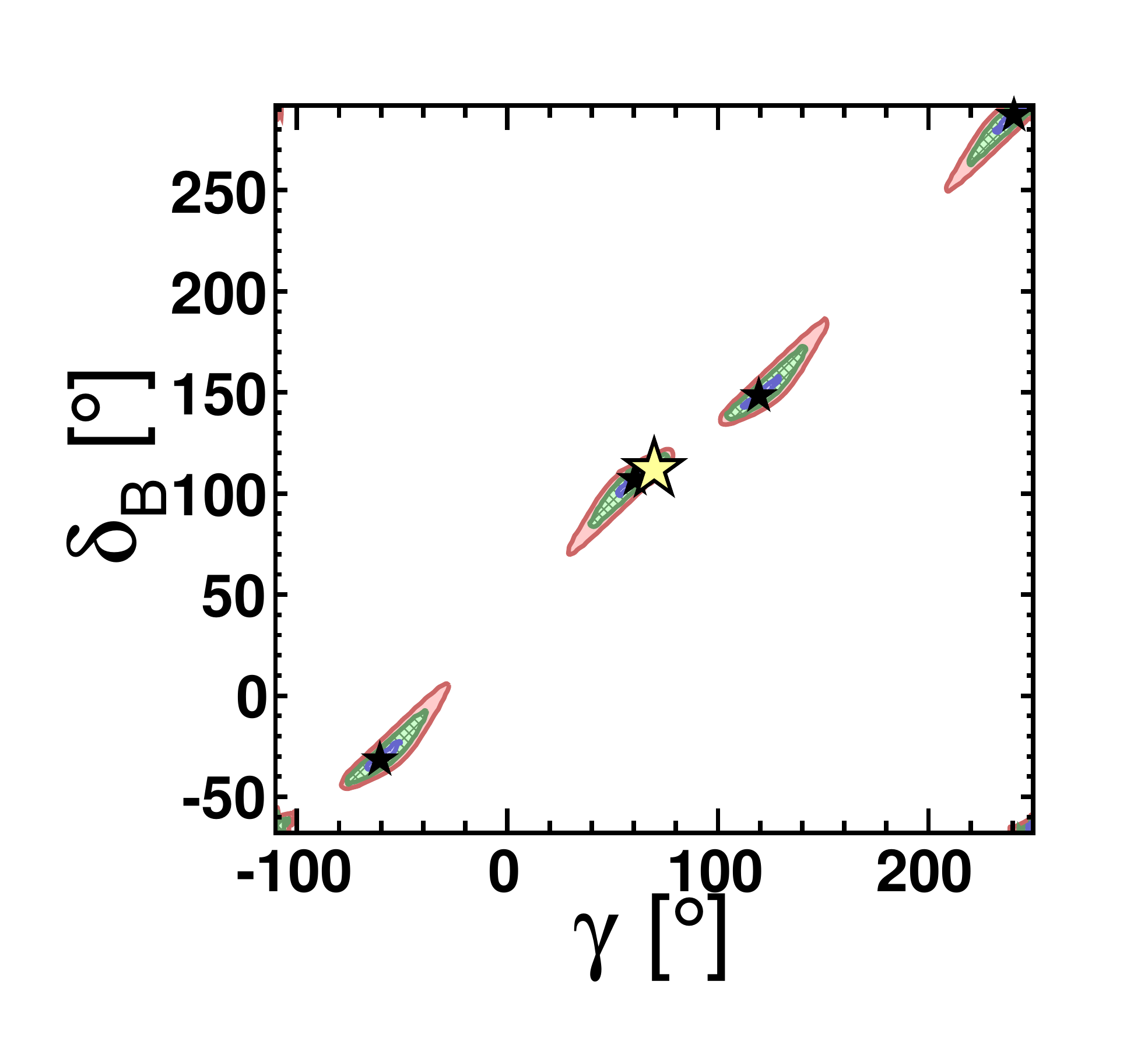} &
\CLScanBox{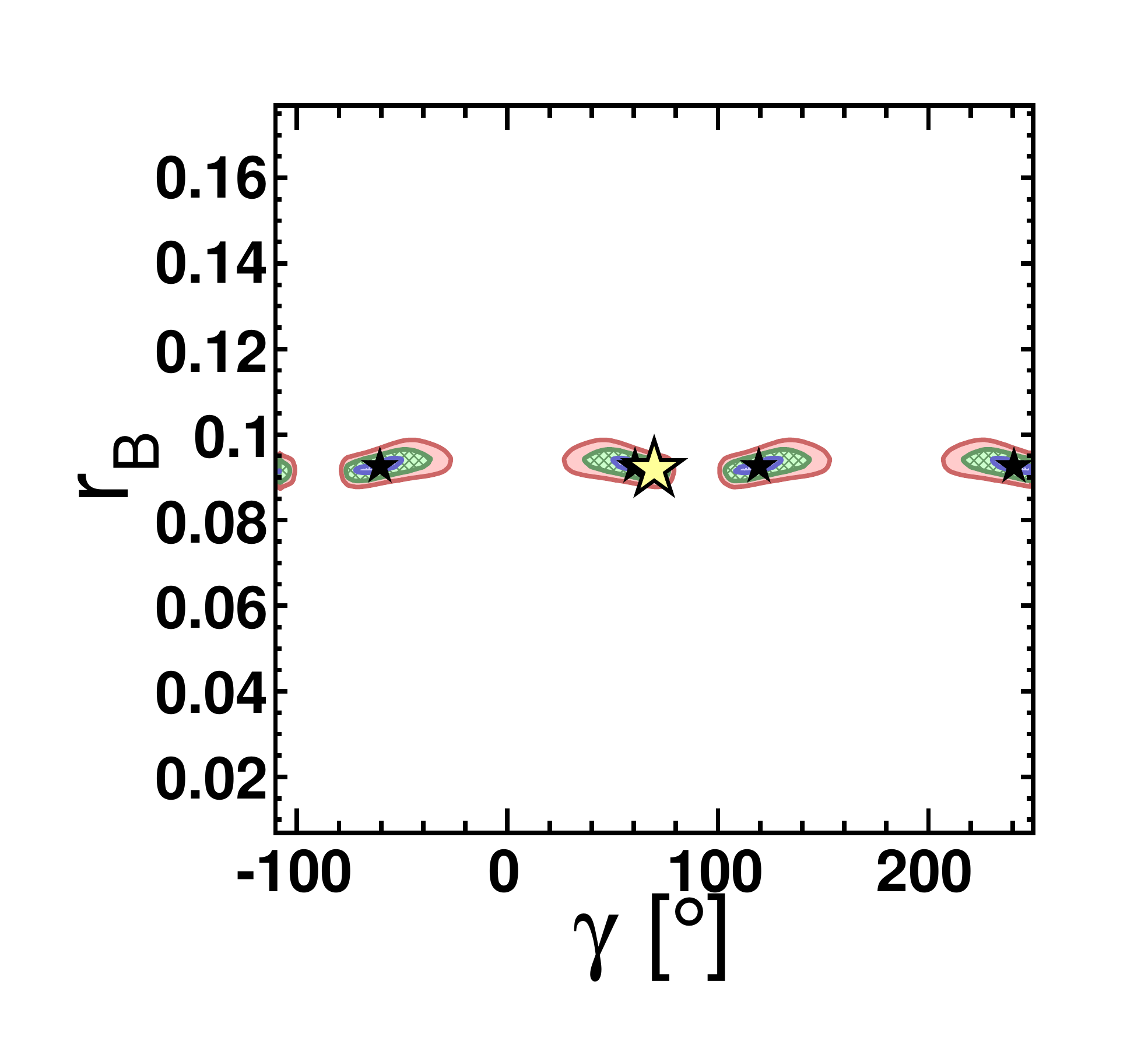}    &
\CLScanBox{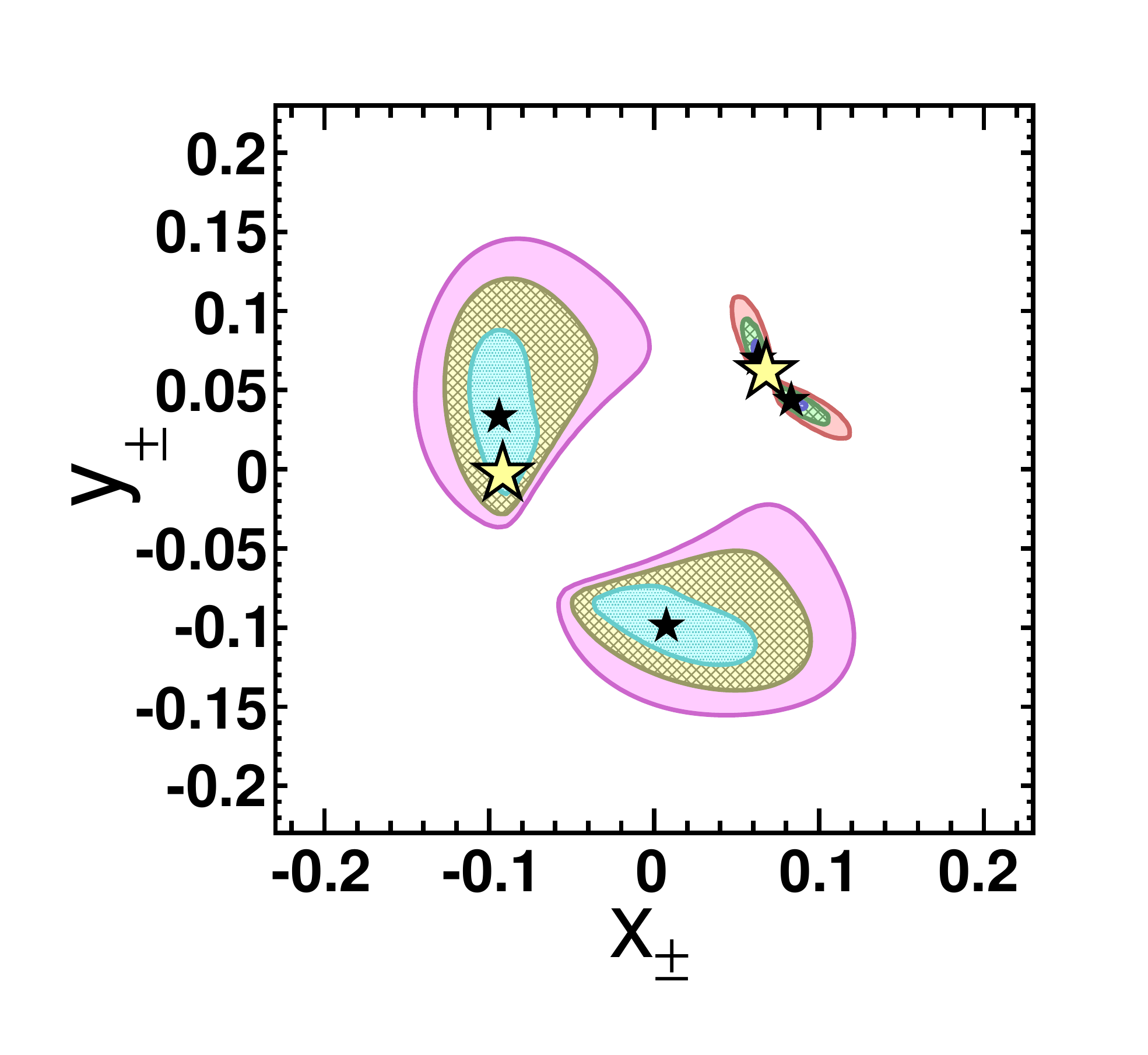}\\
\end{tabular}
\caption{Confidence-level scans for $\gamma$, $\delta$ and $\rB$ in
  the first two columns, and $x_{\pm}, y_{\pm}$ in the third column,
  for simulated events according to the different scenarios given in
  \tabref{tab:eventYields}. The $2-D$ plots show $\sqrt{\Delta \chi^2}
  = 1, 2, 3$ contours. The yellow star indicates the input value and
  the black stars the (multiple) $\chi^2$ minima. When secondary local
  minima are present, as in
  \Figrefs{fig:wrongModel}{fig:addLocalConstraints}, we indicate their
  positions with black crosses. The plots in the last column show
  contours for $x_{+}, y_{+}$ (with minima in the second and fourth
  quadrant) and $x_{-}, y_{-}$ (with two minima in the first
  quadrant).\label{fig:CL_scans}}
\end{figure}
We construct confidence regions in the parameters of interest based on
the $\chi^2$ difference, $\Delta \chi^2$, of the fit where the
relevant parameters are fixed to the values to be probed, relative to
the $\chi^2$ of the best fit result when all parameters float. With
$\sigma \equiv \sqrt{\Delta \chi^2}$, the probability or confidence
level, CL, that the true value of the fit parameter is amongst those
with a smaller $\chi^2$ is approximately
\begin{equation}
\mathrm{CL} = 1-p =
\frac{1}{\sqrt{2\pi}}\int\limits_{-\sigma}^{+\sigma} e^{-\half y^2}dy
\label{eq:PGaussian}
\end{equation}
justifying the interpretation of $\sigma$ in terms of Gaussian
confidence levels. \Eqnref{eq:PGaussian} also defines the $p$-value,
used in \secref{sec:ScansAndNumbers}. We tested the applicability of
\eqnref{eq:PGaussian} to our fit in extensive simulation studies. We
observe good coverage for the default amplitude model and the vast
majority of other amplitude models, for all three data taking
scenarios. Amongst the large number of amplitude models we consider,
there are however some where we find significant deviations from exact
coverage (mostly over-coverage), suggesting that these studies ought
to be repeated once an amplitude model has been obtained from data.

\Figref{fig:CL_scans} shows 2-dimensional scans in terms of $1, 2,
3\sigma$ confidence regions for $\gamma$ vs $\delta$, $\gamma$ vs
$\rB$, and $y_{\pm}$ vs $x_{\pm}$ for each of the three data taking
scenarios. 
The results show that the precision on $x_-, y_-$ (or $\delta -
\gamma$) is much better than that on $x_+, y_+$ (or $\delta +
\gamma$). We found this behaviour in many of the \Dee\ amplitude
models we studied (see \figref{fig:CL_otherModels}), and that it
appears to depend predominantly on the values for \delB, and \gam.

\subsubsection{Using the wrong model}
To study the impact of an imperfect binning, we repeated the
sensitivity study using the imperfect binning discussed in
\secref{sec:binning}, and applied it to our default Run~II scenario.
\begin{figure}
\begin{tabular}{c ccc}
\CLScanBox{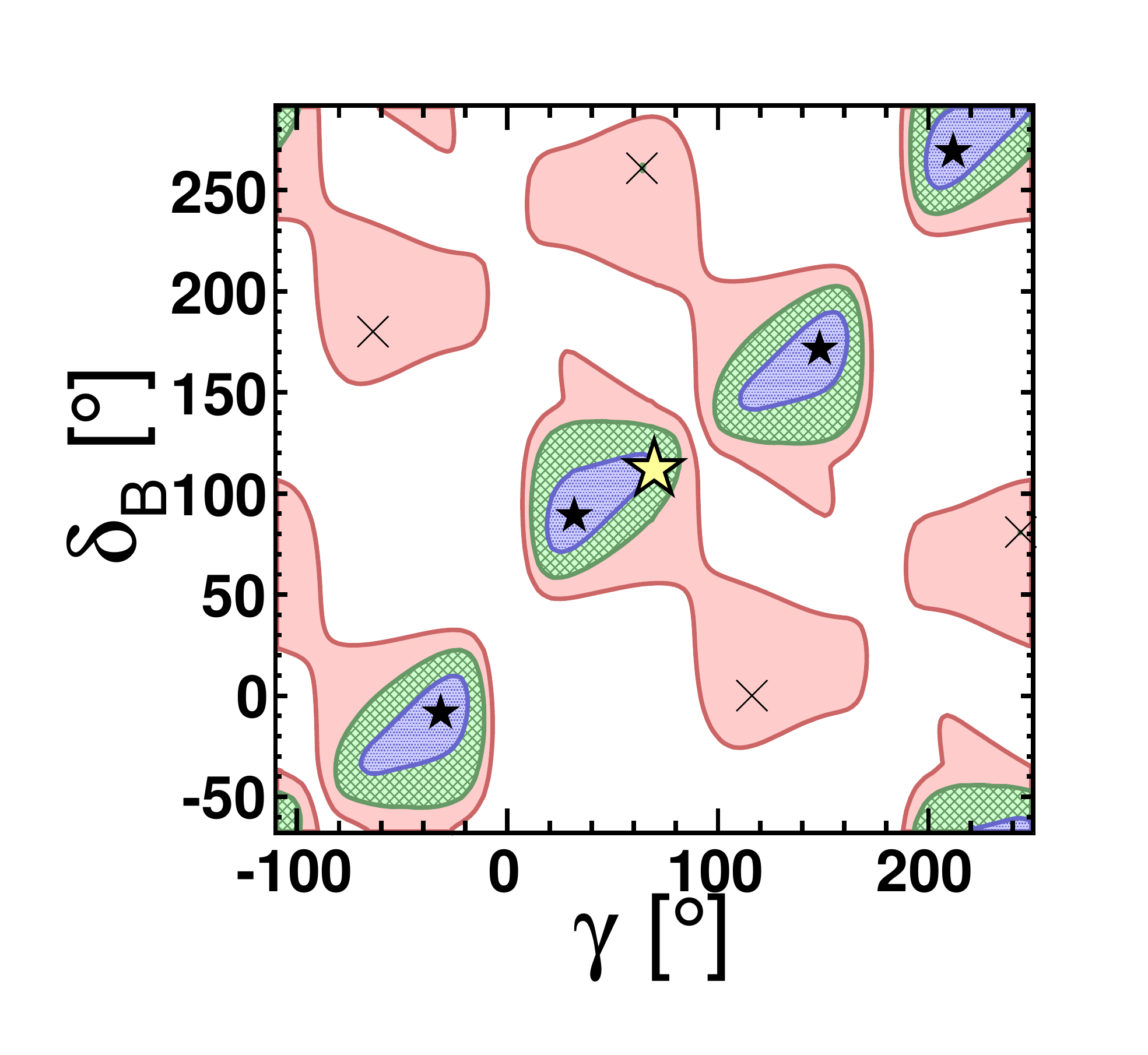} &
\CLScanBox{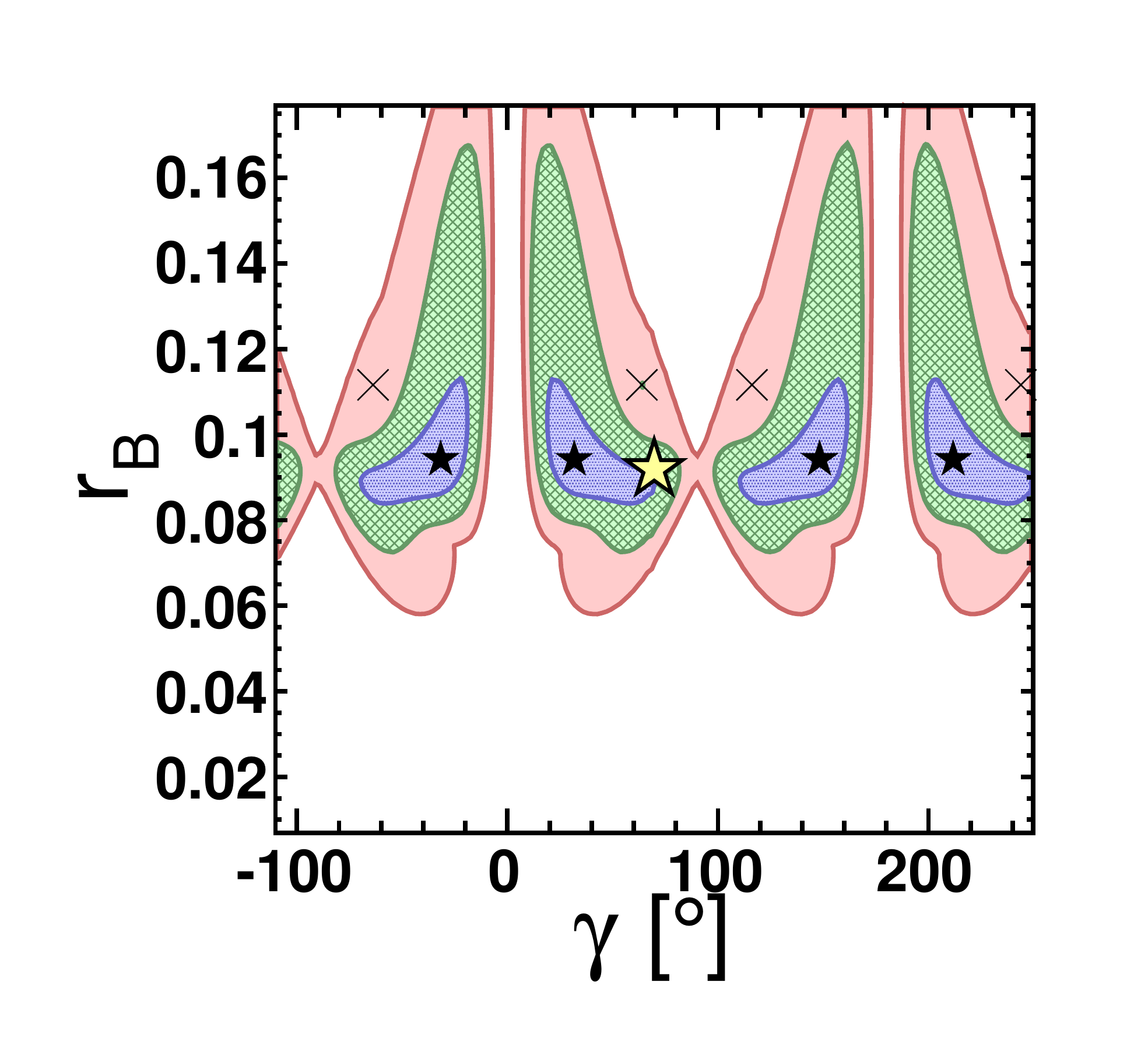} &
\CLScanBox{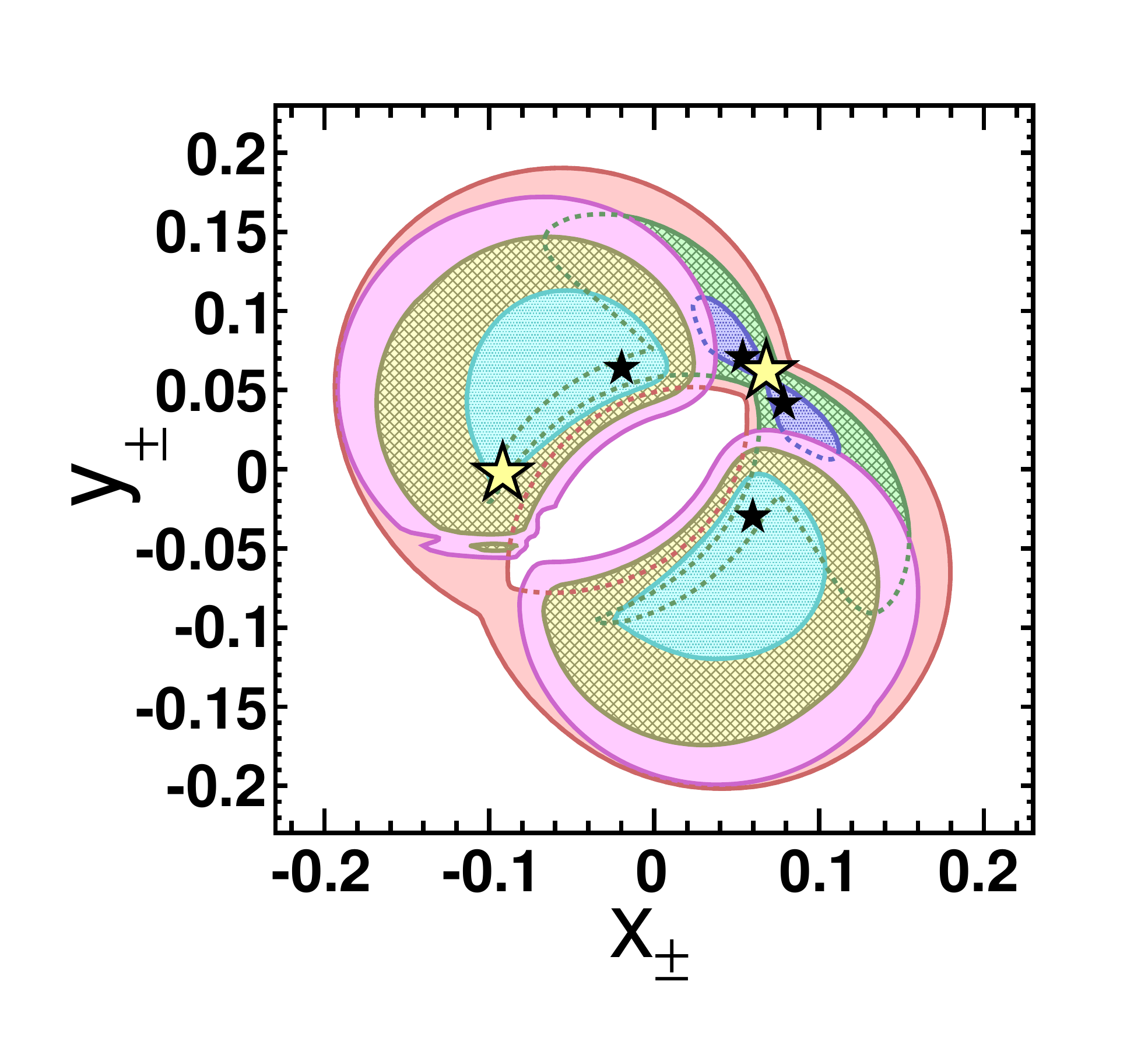} \\
\end{tabular}
  \caption{CL scans for simulated data generated
    with the default model, but binned based on the randomised model
    described in \secref{sec:binning} (same format as in
    \figref{fig:CL_scans}).\label{fig:wrongModel}}
\end{figure}
Comparing the results, shown in \figref{fig:wrongModel}, to those in
\figref{fig:CL_scans} shows that the imperfect binning results in a
visible reduction in sensitivity especially at the $3\sigma$ level,
but it does not lead to a catastrophic deterioration of the fit, which
retains a similar precision at the $1\sigma$ level.

\subsection{Studies with other models}
\begin{figure}
\begin{tabular}{c ccc}
\CLYaxis{Models} & 
\CLScanBox{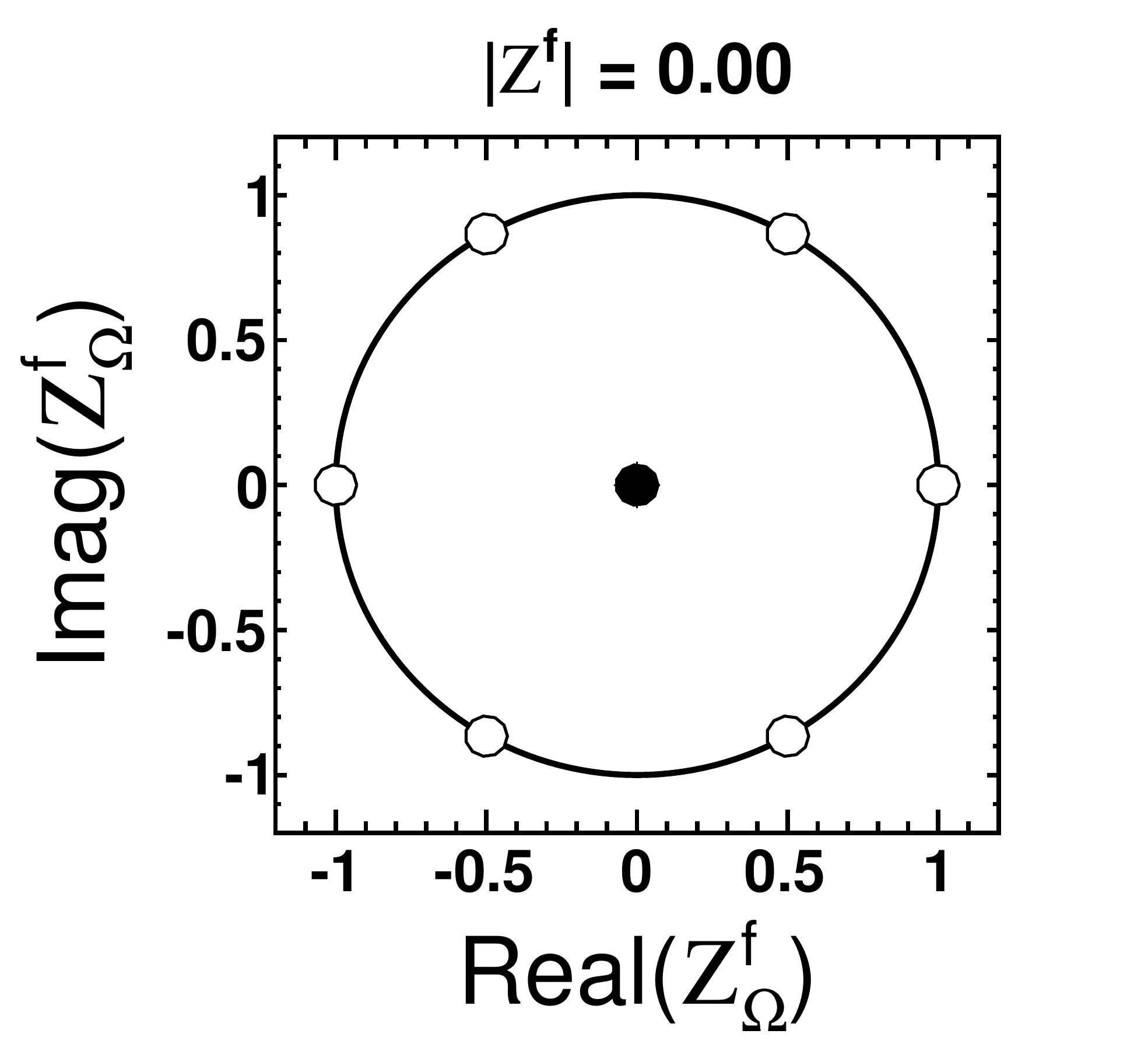} &
\CLScanBox{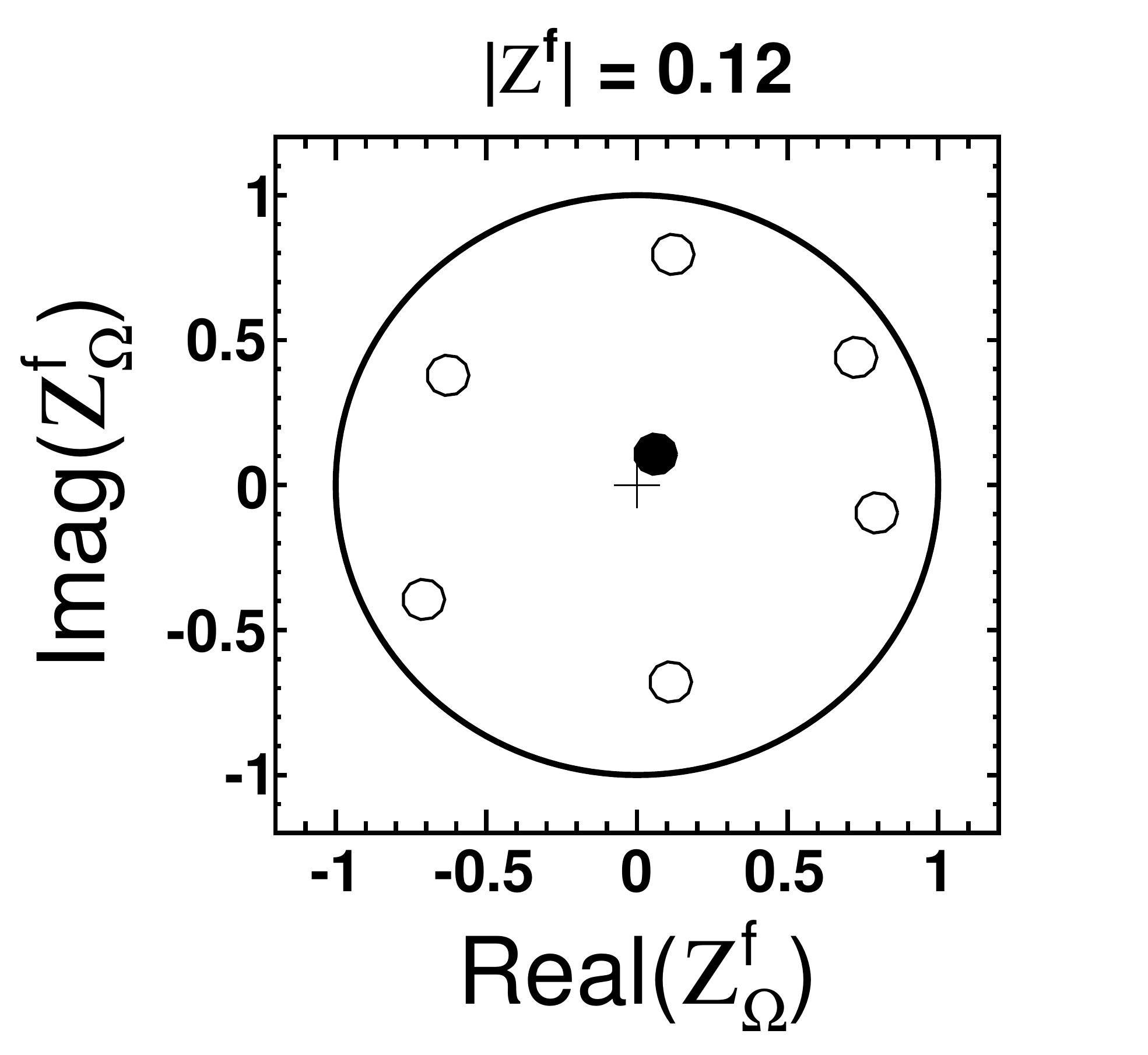} &
\CLScanBox{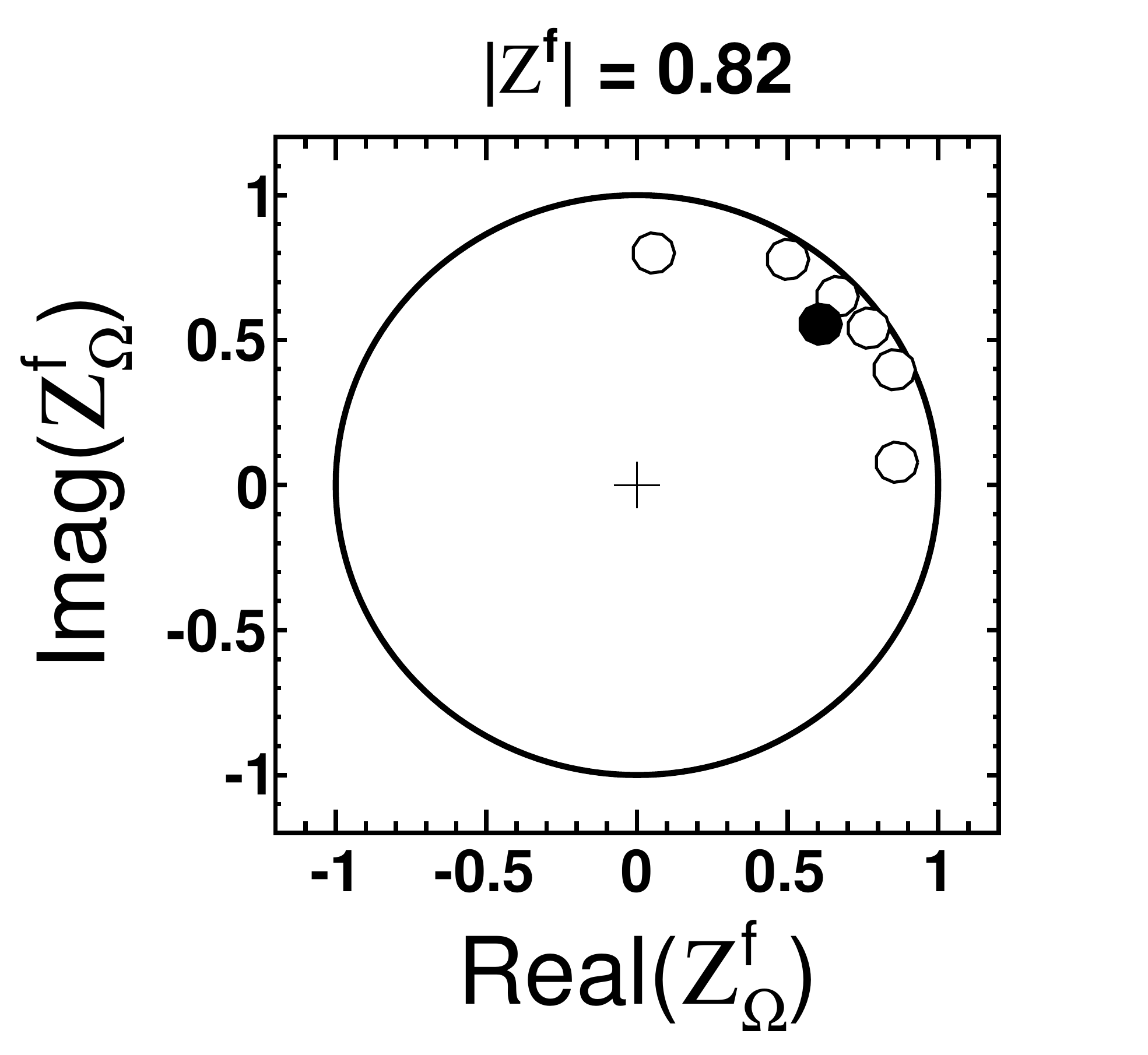} \\
\CLYaxis{LHCb run~II} &
\CLScanBox{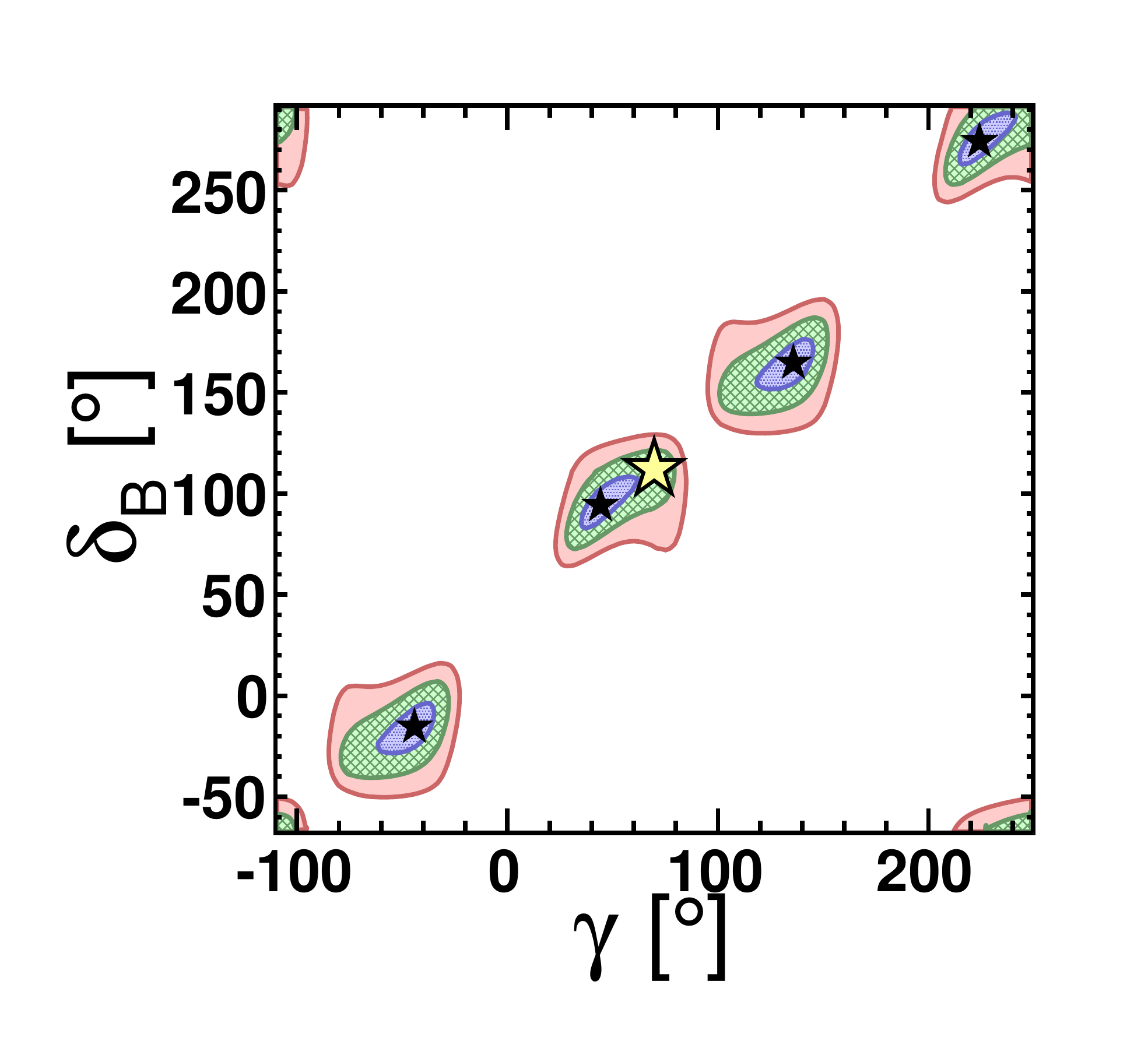} &
\CLScanBox{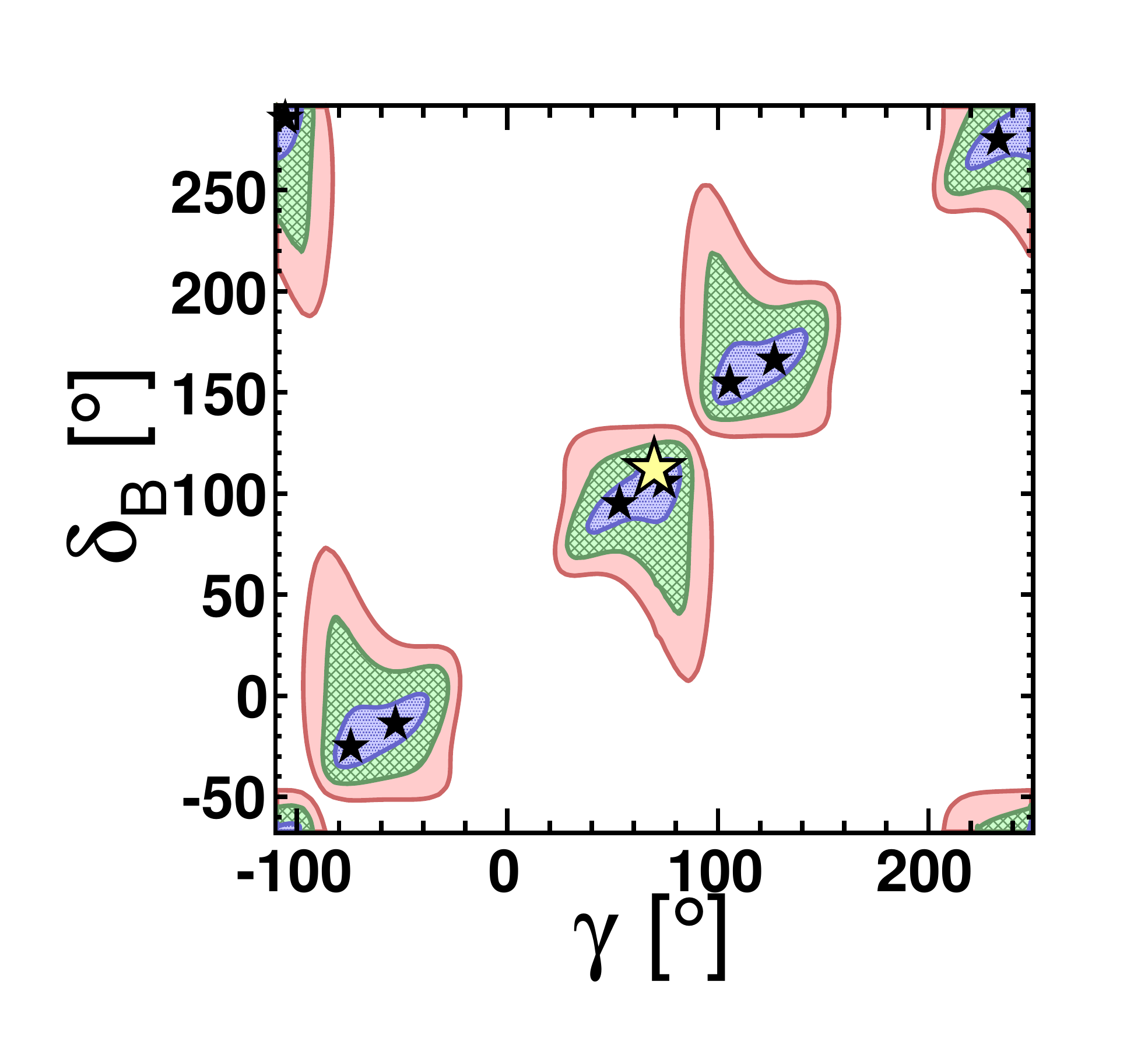} &
\CLScanBox{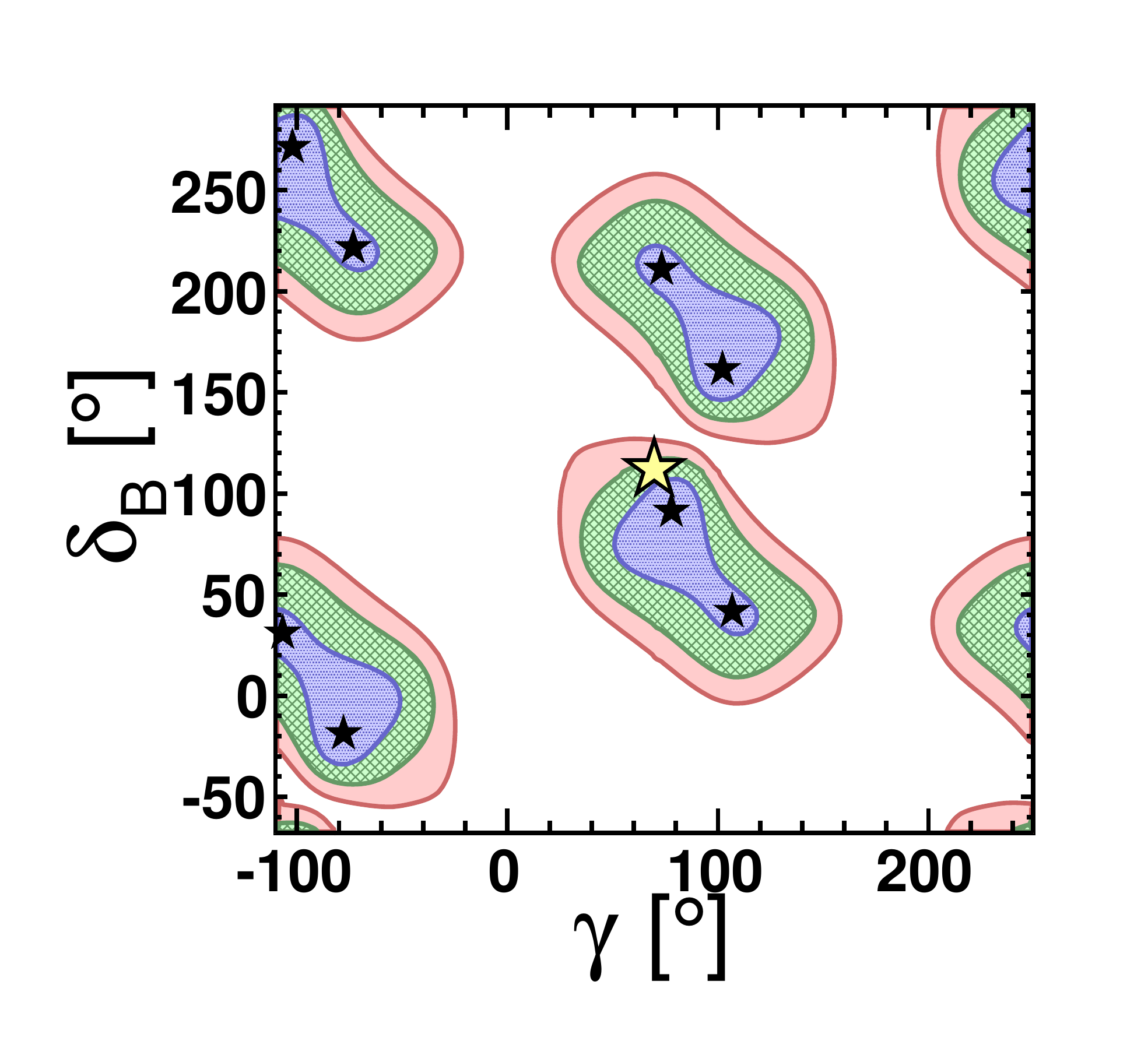} \\
\end{tabular}
\caption{CL scans for three alternative models, for the LHCb run~II data
  taking scenario.  The top row shows the \ZKpipipiOm\ values and the central
  value of \ZKpipipi\ for the each model. The second row show the CL
  scans in the $\gam-\delB$ plane, for the LHCb run~II
  scenario.\label{fig:CL_otherModels}}
\end{figure}
To study the dependence of our results on the particular amplitude
model for the DCS \DzerotoKpipipiDCS\ decay, we repeated the studies
with a variety of amplitude models. 
CL scans in the \gam-\delB\ plane for three examples, for the LHCb
run~II data taking scenario, are shown in
\figref{fig:CL_otherModels}. 
The first column shows an artificial ``ideal'' model, set up to have
bins with evenly distributed \delOm, and $|\ZOm| = 1$, $\BMagOm=1$,
$\AMagOm=\rD$ for all \intVolume; this also implies $\magZ=0$.  The
second and third column show models taken from the set of randomly
generated models; one where \magZKpipipi\ is smaller than \cleoc's
central value, and another where it is larger. The results illustrate
a general tendency we observe, which is that the precision improves
for models with a fairly even spread of \delDp, while clustering of
\delDp, a feature typical for models with large \magZKpipipi, leads to
reduced sensitivity.

\subsection{Additional input  from the charm threshold}
\label{sec:ExternalConstraints}
We consider two ways of incorporating additional information from the
charm threshold. One is to incorporate constraints on the global
coherence factor \Z. Such constraints are already available for
\DtoKpipipi\ and a few other decay modes, based on \cleoc\
data~\cite{Libby:2014rea, Lowery:2009id, Insler:2012pm}, and could
significantly improve with input from \besIII, who have collected
$3.5$ times as much integrated luminosity at the charm
threshold. These constraints can be added either to a phase-space
integrated analysis of \Dee\ mixing and \BtoDK\ as proposed
in~\cite{selfcite} or to the binned analysis introduced
here. Alternatively, charm threshold data can be analysed in the same
phase space bins as \BtoDK\ and charm mixing. This, as we will show
below, will add additional information that substantially improves the
measurement. Below we discuss each method in turn.

\subsubsection{Phase-space integrated analysis with input from the
  charm threshold}
\begin{figure}
\centering


 


\besIII\ + \Dee\ mixing, phase-space integrated analysis

\raggedright
  
\begin{tabular}{ccc}
\CLScanBox{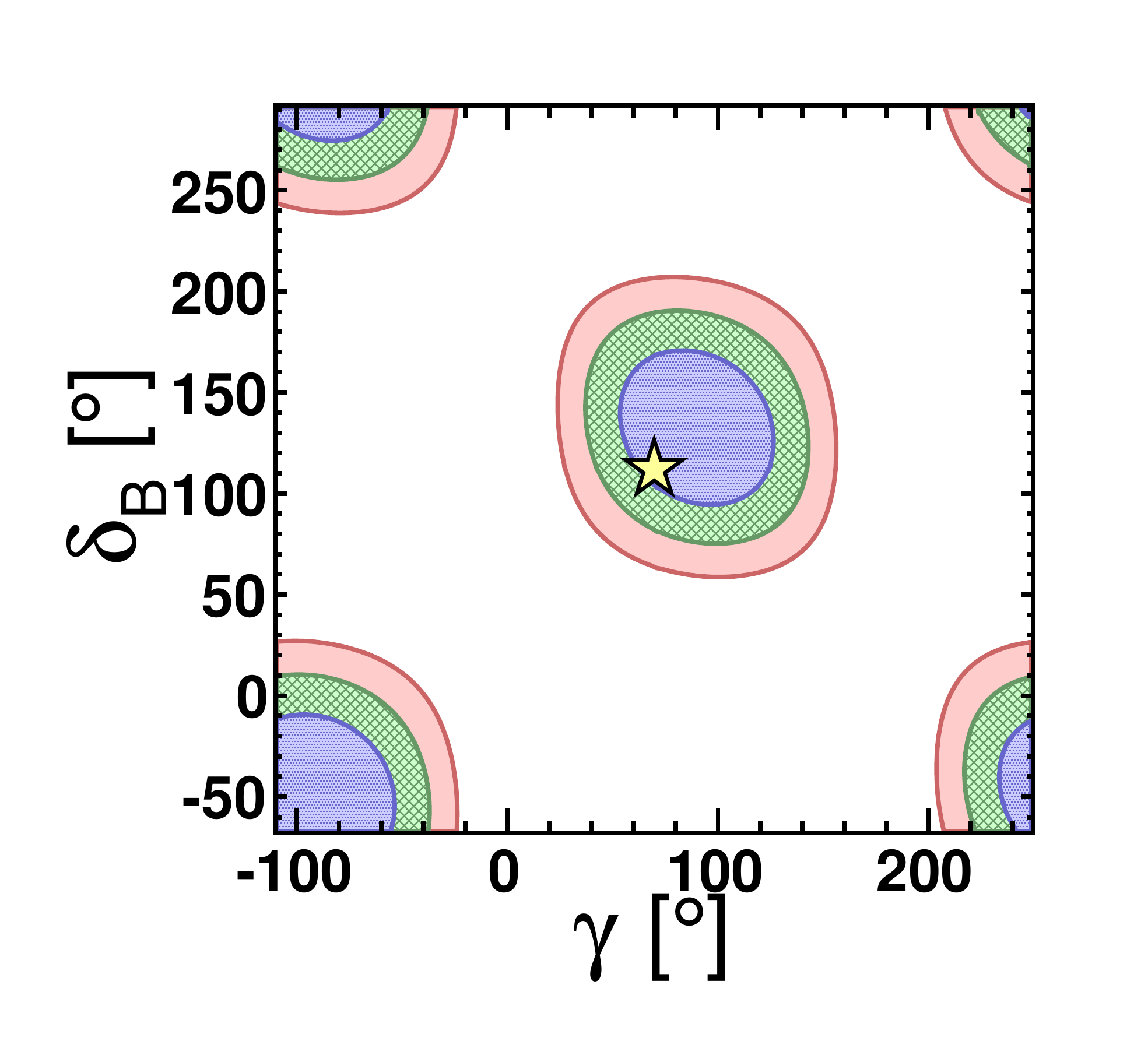} &
\CLScanBox{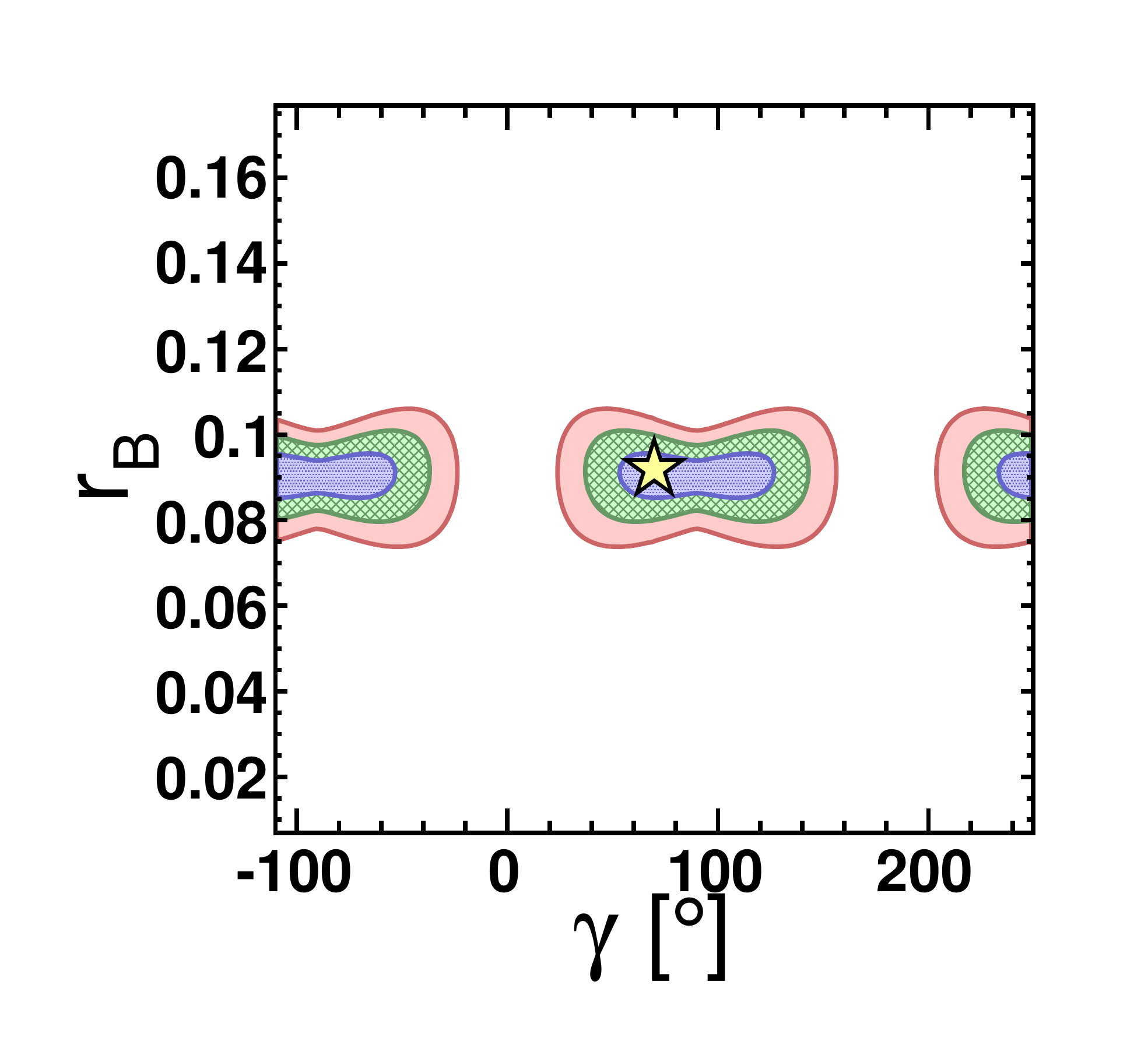} &
\CLScanBox{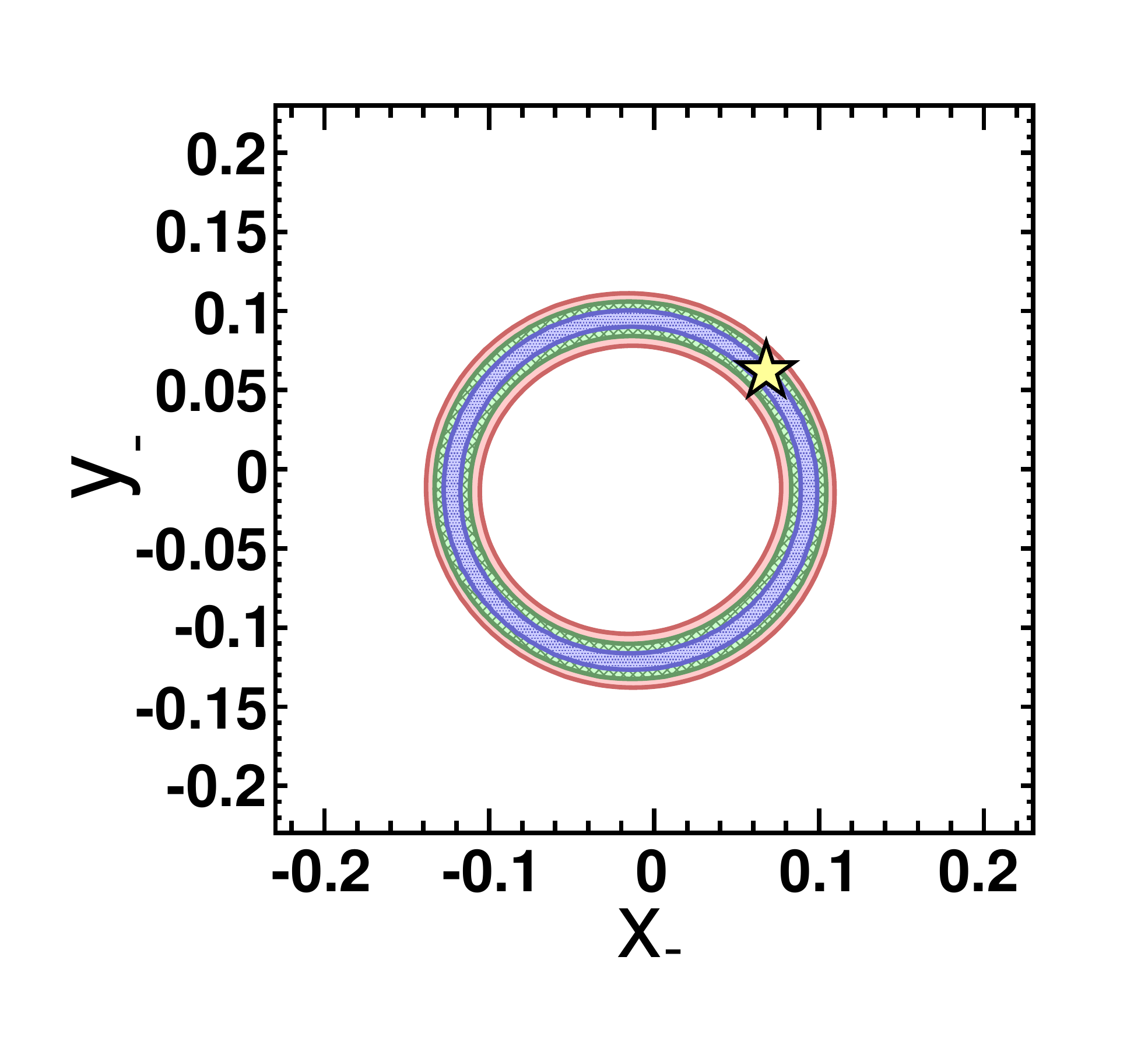} 
\end{tabular}
\caption{Constraints on \rB, \delB, \gam, $x_{-}$, $y_{-}$ obtained
  using the phase-space integrated approach proposed
  in~\cite{Atwood:coherenceFactor}, 
  with additional constraints from mixing~\cite{selfcite}. In
  contrast to all other results shown in this letter, neither \Dee\
  mixing nor \BtoDK\ data are separated into multiple phase space
  bins. The study uses global constraints on \ZKpipipi\ extrapolated
  to \besIII\ statistics~\cite{Libby:2014rea}, and the LHCb run~II
  data scenario.\label{fig:OneBinScans}}
\end{figure}
In contrast to all other results presented in this letter, for this
analysis, neither the charm mixing data, nor the \BtoDK\ data are
divided into multiple phase space bins. We incorporate constraints on
\ZKpipipi\ obtained from charm threshold data
following~\cite{Atwood:coherenceFactor}, and perform fits to simulated
data with and without input from a phase-space integrated \Dee\ mixing
analysis as proposed in~\cite{selfcite}. \Figref{fig:OneBinScans}
shows confidence regions obtained for such a phase-space integrated
analysis based on the LHCb run~II scenario, with input from the charm
threshold extrapolated to \besIII\ statistics~\cite{Libby:2014rea},
including input from charm mixing. While with this method, there is
insufficient information to obtain point-estimates, $68\%$ confidence
regions can still be interpreted in terms of uncertainties on \gam,
\delB\ and \rB, as described in
\secref{sec:ScansAndNumbers}. Averaging over $50$ simulated
experiments, we find $\sigma(\gamma) = 56^{\circ}$ ($64^{\circ}$),
$\sigma(\delB) = 53^{\circ}$ ($66^{\circ}$) and $\sigma(\rB) =0.92
\cdot 10^{-2}$ ($4.1 \cdot 10^{-2}$) with (without) input from \Dee\
mixing. While the constraints on \gam\ and \delB\ are rather weak, the
precision on \rB\ is excellent. As \cite{Libby:2014rea} have shown,
input from such an analysis would play an important role in a global
fit to measure \gam.


\subsubsection{Global constraints from the charm threshold,
  with a binned \BtoDK\ and \Dee\ mixing analysis}
\label{sec:globalThresholdBinnedMixing}
Performing the fit on the absolute decay rates (see
\secs~\ref{sec:ParaCountingRates} and~\ref{sec:method}) rather than
the fractions, it is possible to incorporate constraints on the total
coherence factor \Z\ from the charm threshold while still performing
the binned analysis of \BtoDK\ and charm mixing data as described
above, using the relation
\begin{align}
\sum\limits_{\mathrm{all\;\intVolume_i}} \AMag_{\Omega_i} \BMag_{\Omega_i} \Z_{\Omega_i}
= \AMag \BMag \Z .
\end{align}
In the above expressions, $\AMag, \BMag, \Z$ are the equivalent
quantities to $\AMagOm, \BMagOm, \ZOm$ for a volume that encompasses
the entire phase space.
\begin{figure}
\begin{tabular}{ccc}
\besIII\ (global) & D mixing (binned) & \besIII\ (global) \\
(w/o D mixing)    & alone    & with binned D mixing \\
\CLScanBox{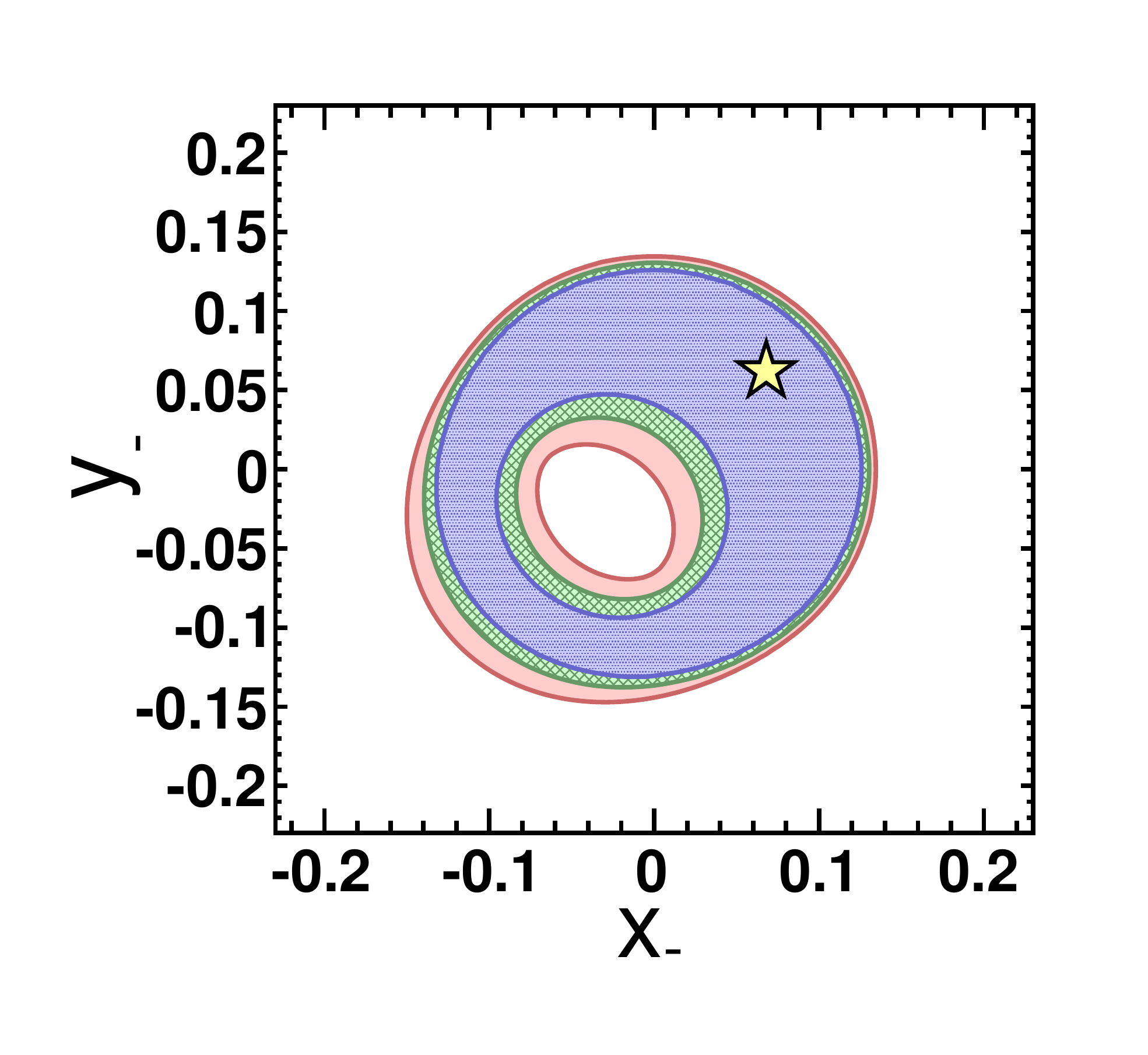} &
\CLScanBox{fig/2Dscan_xpm_ypm_run2_perfectModel_6bins_model49} &
\CLScanBox{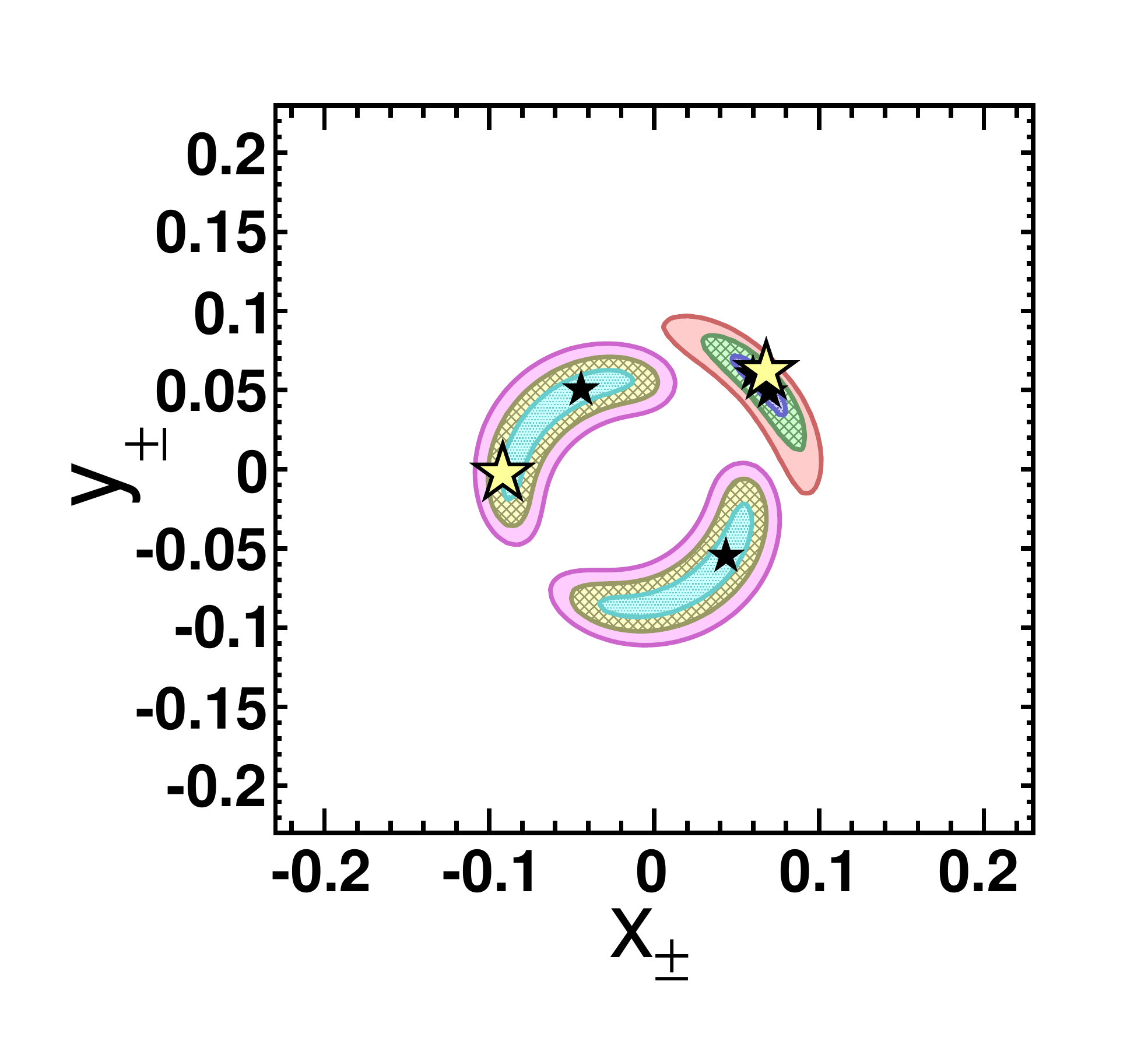}
\end{tabular}
\caption{Constraints on $x_{\pm}$ and $y_{\pm}$, obtained by combining
  simulated \BtoDK\ data (LHCb Run~II statistics) with different
  constraints from charm. Left: future (\besIII) charm threshold
  constraints on \ZKpipipi\ (only the effect on $x_-, y_-$ is shown, results for
  $x_+, y_+$ are similar). Centre: D mixing constraints. Right:
  Both. (Same format as in \figref{fig:CL_scans}.)\label{fig:addGlobalConstraints}}
\end{figure}
\Figref{fig:addGlobalConstraints} illustrates the significant benefit
of such additional constraints, numerical results can be found in
\tabref{tab:allResults}. The predicted \besIII\ uncertainties on
\ZKpipipi\ are taken from~\cite{Libby:2014rea}.

\subsubsection{Binned constraints from the charm threshold}
\label{sec:binnedInput}
\begin{figure}
\begin{tabular}{ccc}
\multicolumn{2}{c}{\besIII\ (binned)}    & \besIII\ binned \\
\multicolumn{2}{c}{(w/o D mixing input)} & and binned D mixing \\
\CLScanBox{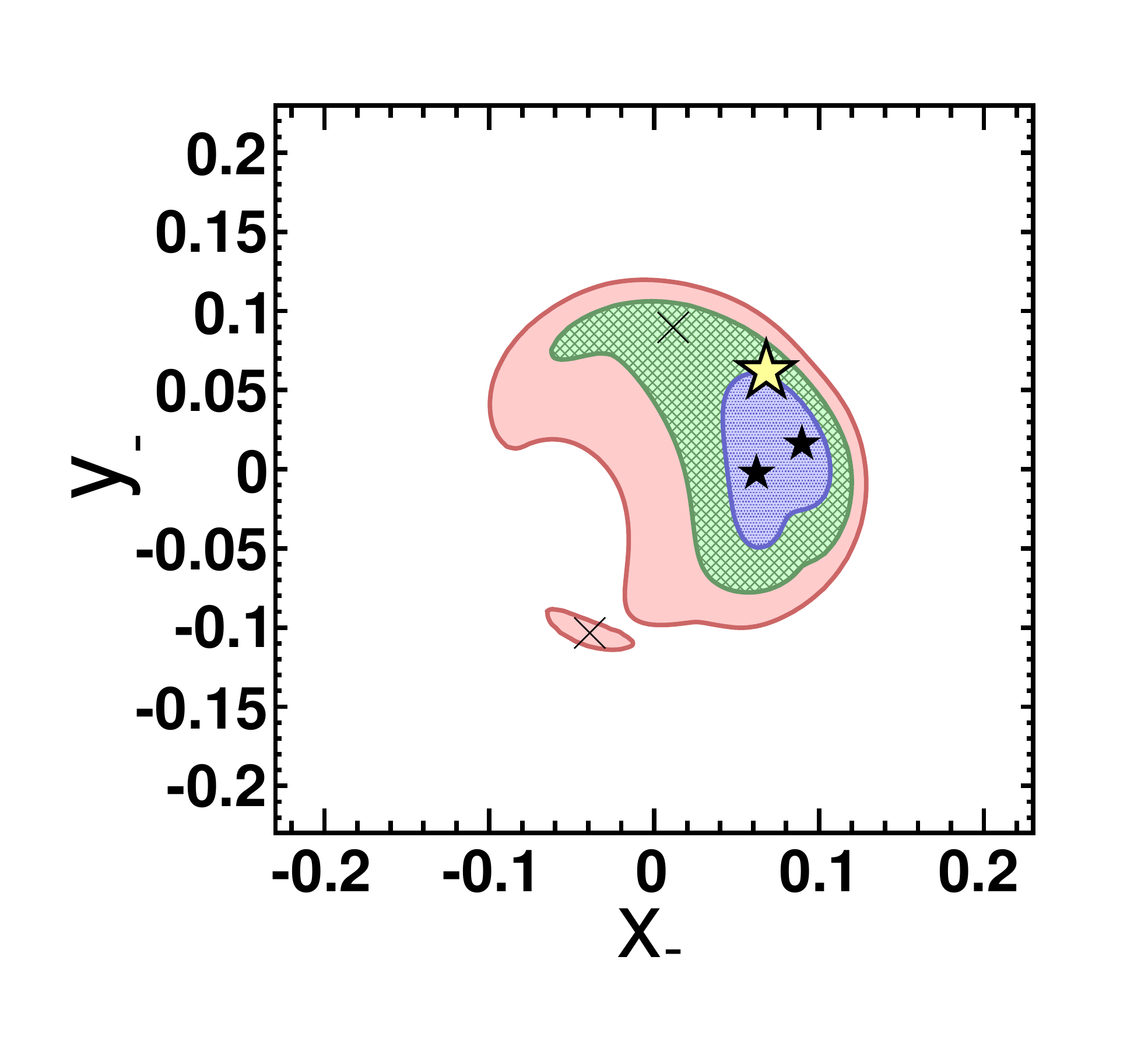} &
\CLScanBox{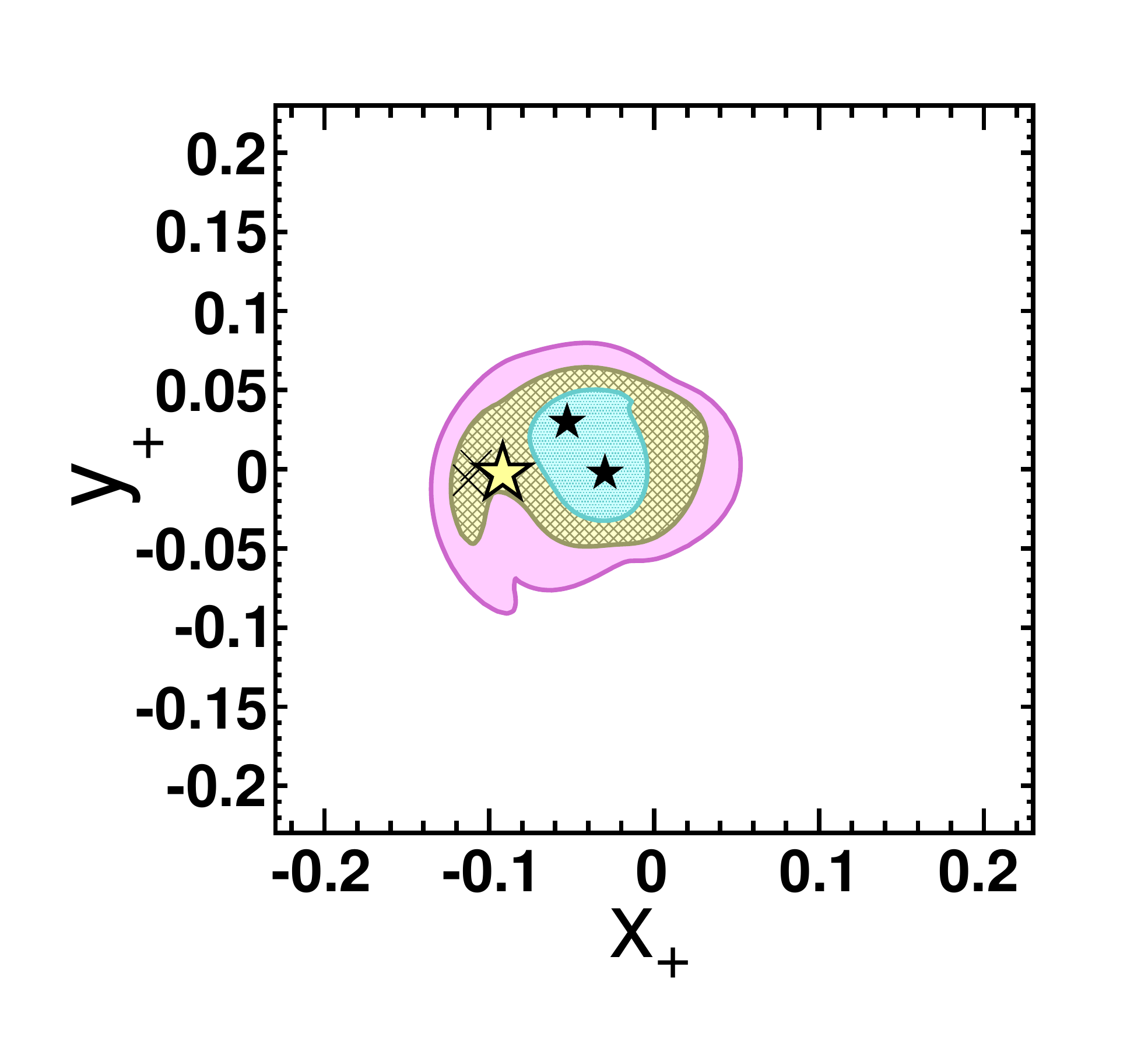} &
\CLScanBox{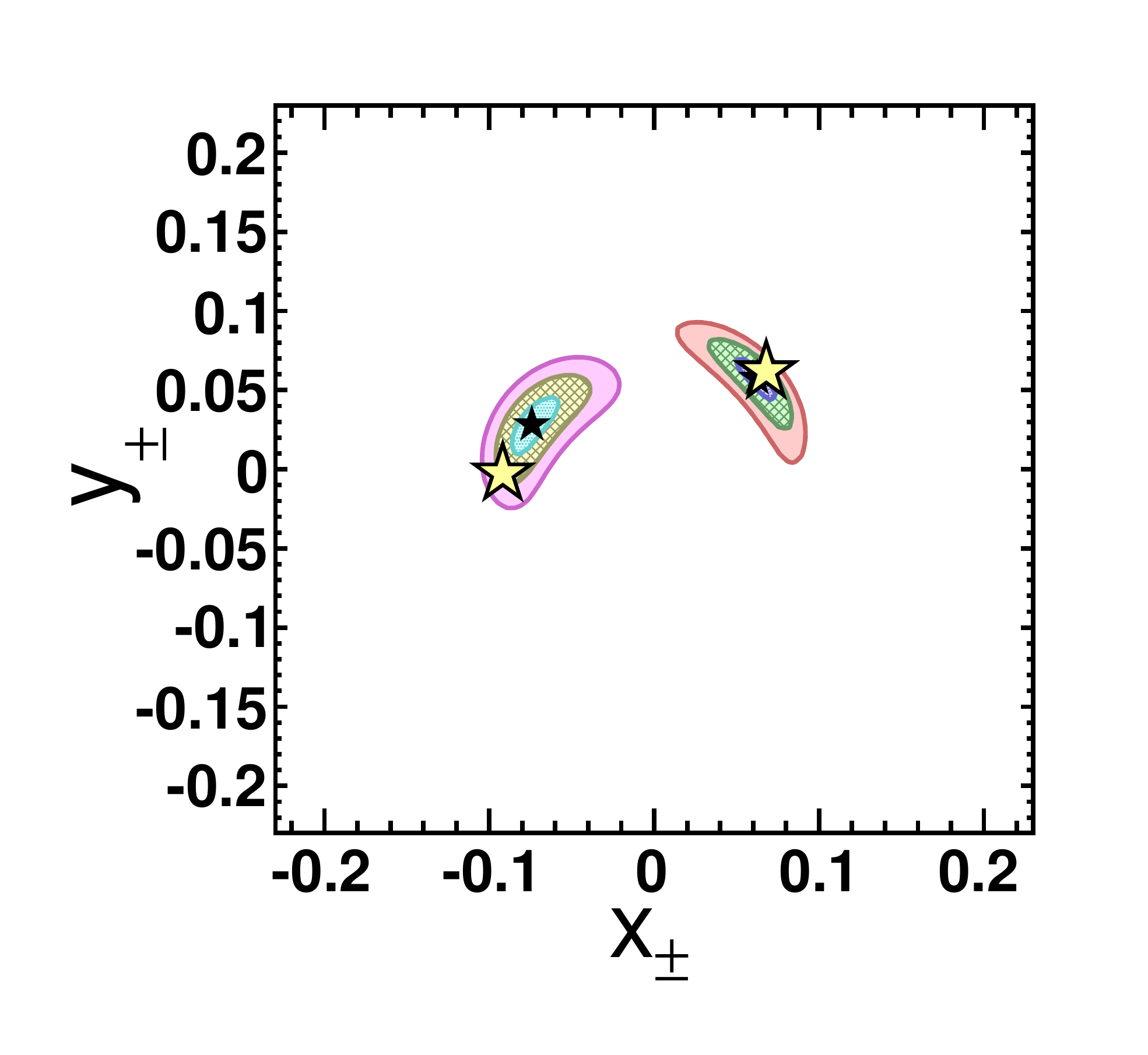}
\end{tabular}
\caption{Constraints on $x_{\pm}$ and $y_{\pm}$, obtained by combining
  simulated \BtoDK\ data (LHCb Run~II statistics) with different
  constraints from charm. Two plots on the left: future (\besIII)
  charm threshold constraints on binned \ZKpipipiOm. Right: that,
  combined with D mixing. (Same format as in
  \figref{fig:CL_scans}.)\label{fig:addLocalConstraints}}
\end{figure}
In this section we compare the performance of a binned analysis
relying on charm threshold data for the charm interference parameter,
as proposed in~\cite{Atwood:coherenceFactor}, with the novel method
proposed in this letter, and with a combined approach using binned
threshold and charm mixing data.
We analyse the charm threshold data in the same phase-space bins as
\BtoDK\ and charm mixing. This provides a constraint from threshold
data on each individual \ZKpipipiOm, rather than only their weighted
sum as in \secref{sec:globalThresholdBinnedMixing}. To estimate the
uncertainties on \ZKpipipiOm\ from such an analysis, we take the
results on \ZKpipipi\ from~\cite{Libby:2014rea}, and assume that
uncertainties scale with the inverse square-root of the number of
signal events used for the measurement. Given the fairly large
uncertainty on \ZKpipipi\ from \cleoc\ data, we assume that these data
can be divided into at most three bins while still providing
meaningful constraints on \ZKpipipiOm\ in each bin. With \besIII\
statistics, we expect it will be possible to match the binnings
defined in \secref{sec:method}, with up to eight
bins. \Figref{fig:addLocalConstraints} illustrates in the
$x_{\pm}-y_{\pm}$ plane the dramatic effect that the combination of
mixing constraints and binned \ZKpipipiOm\ constraints from a future
analysis of \besIII\ threshold data could have. Not only are the
uncertainties on $x_{\pm}, y_{\pm}$ much reduced compared to either
constraint being applied individually (see \tabref{tab:allResults} for
numerical results), but the \besIII\ input also removes the previously
existing ambiguities in $x_{\pm}$ and
$y_{\pm}$. \Figref{fig:gammaScans}, described below, confirms this
observation for 1-dimensional parameters scans of $x_{\pm}$ and \gam.

\subsection{1-D scans and quantified uncertainties}
\label{sec:ScansAndNumbers}
\begin{figure}
\begin{tabular}{ccc}
D mixing (binned)       &  BES~III binned &  D mixing (binned) with  \\
 alone           &    alone        &  BES~III binned \\
\CLScanBox{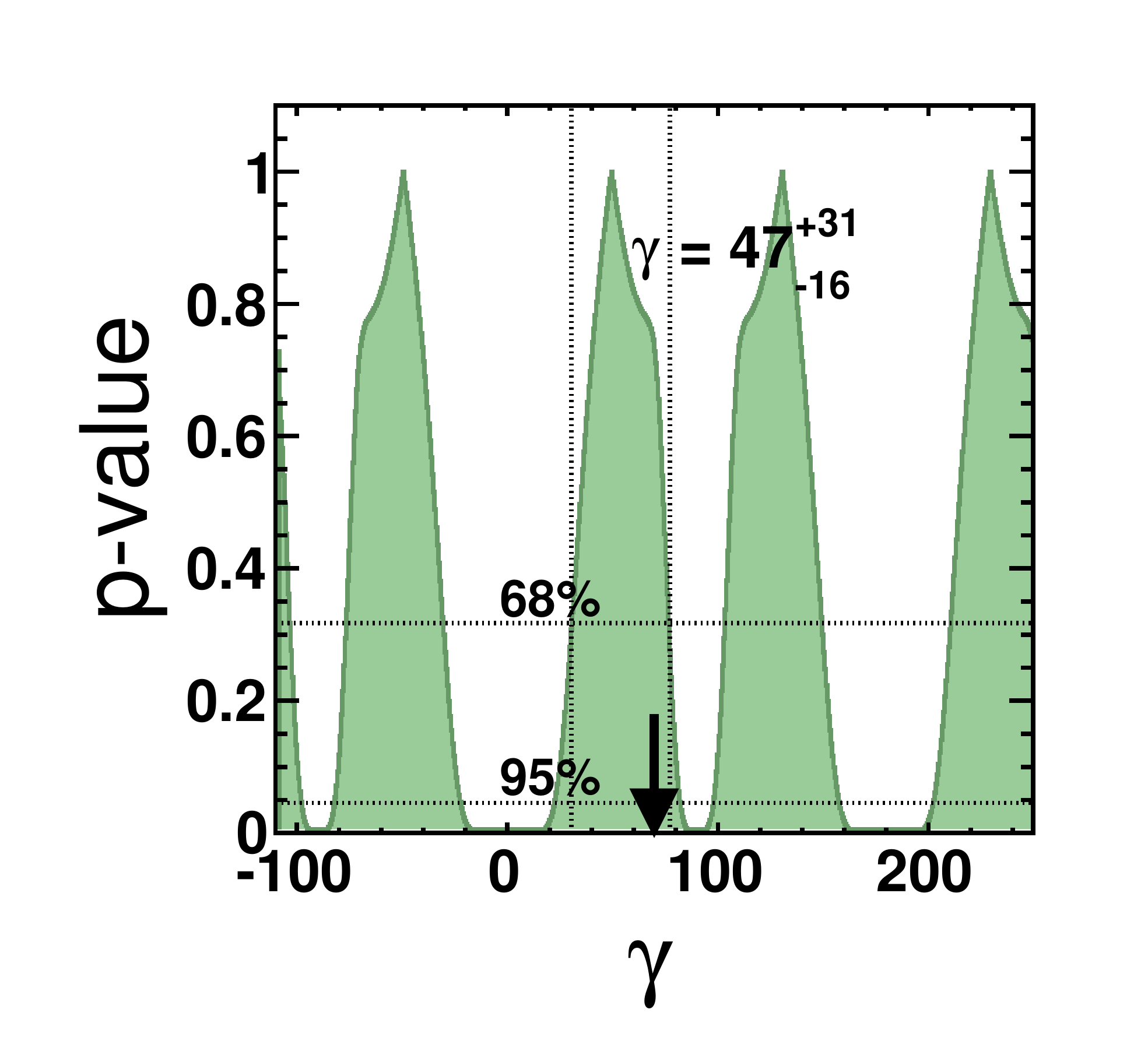} &
\CLScanBox{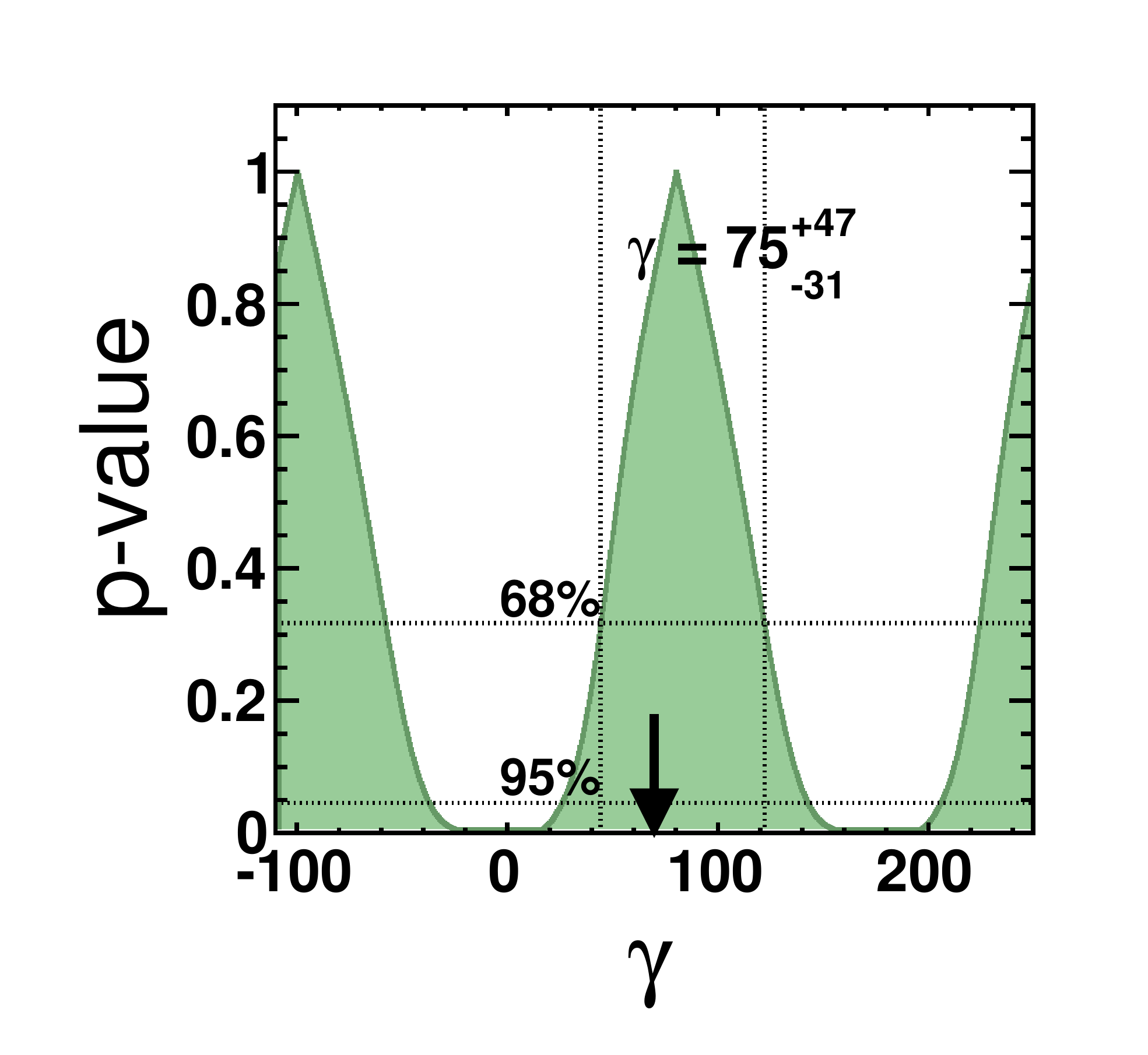} &
\CLScanBox{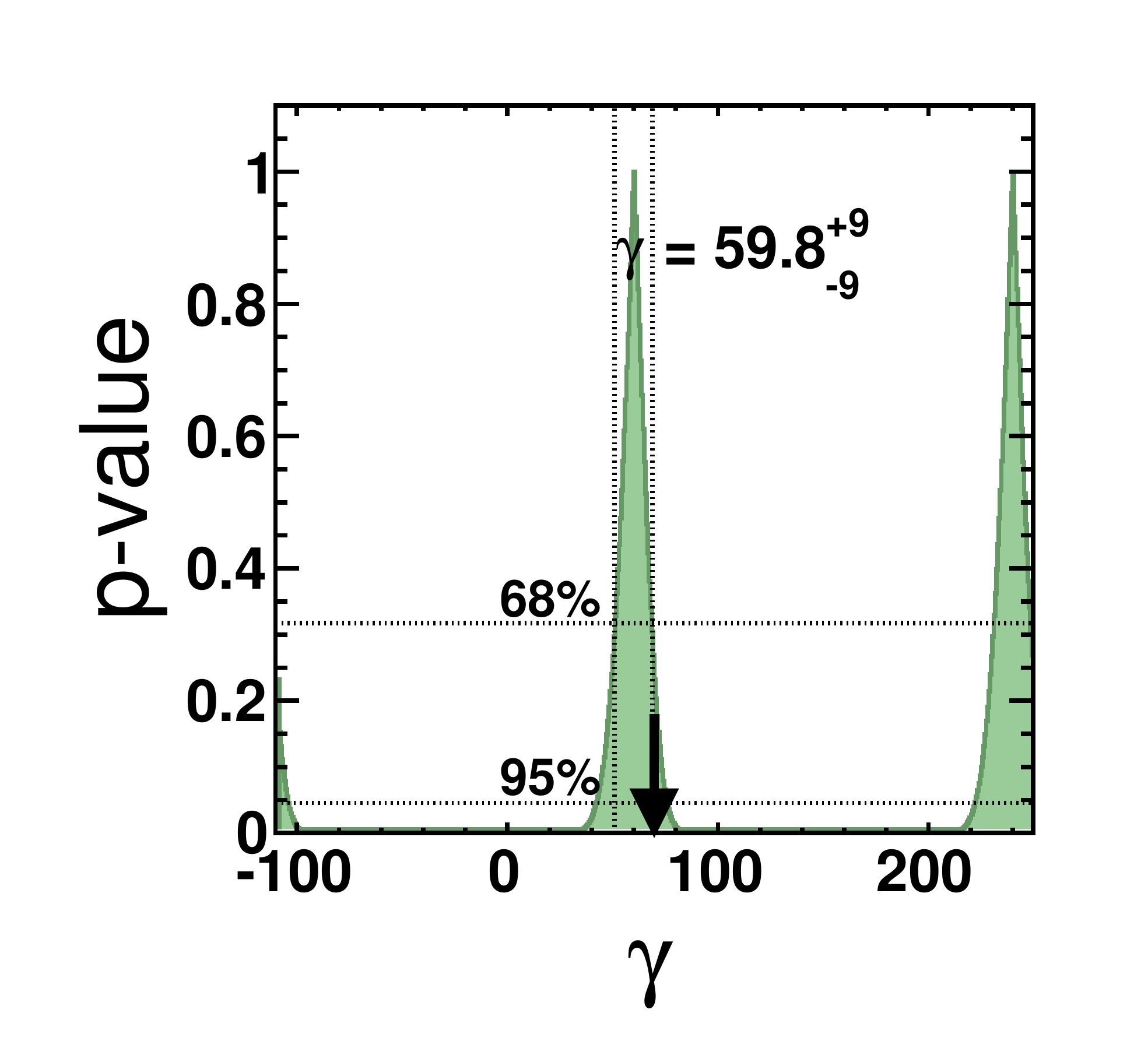}
\\ 
\CLScanBox{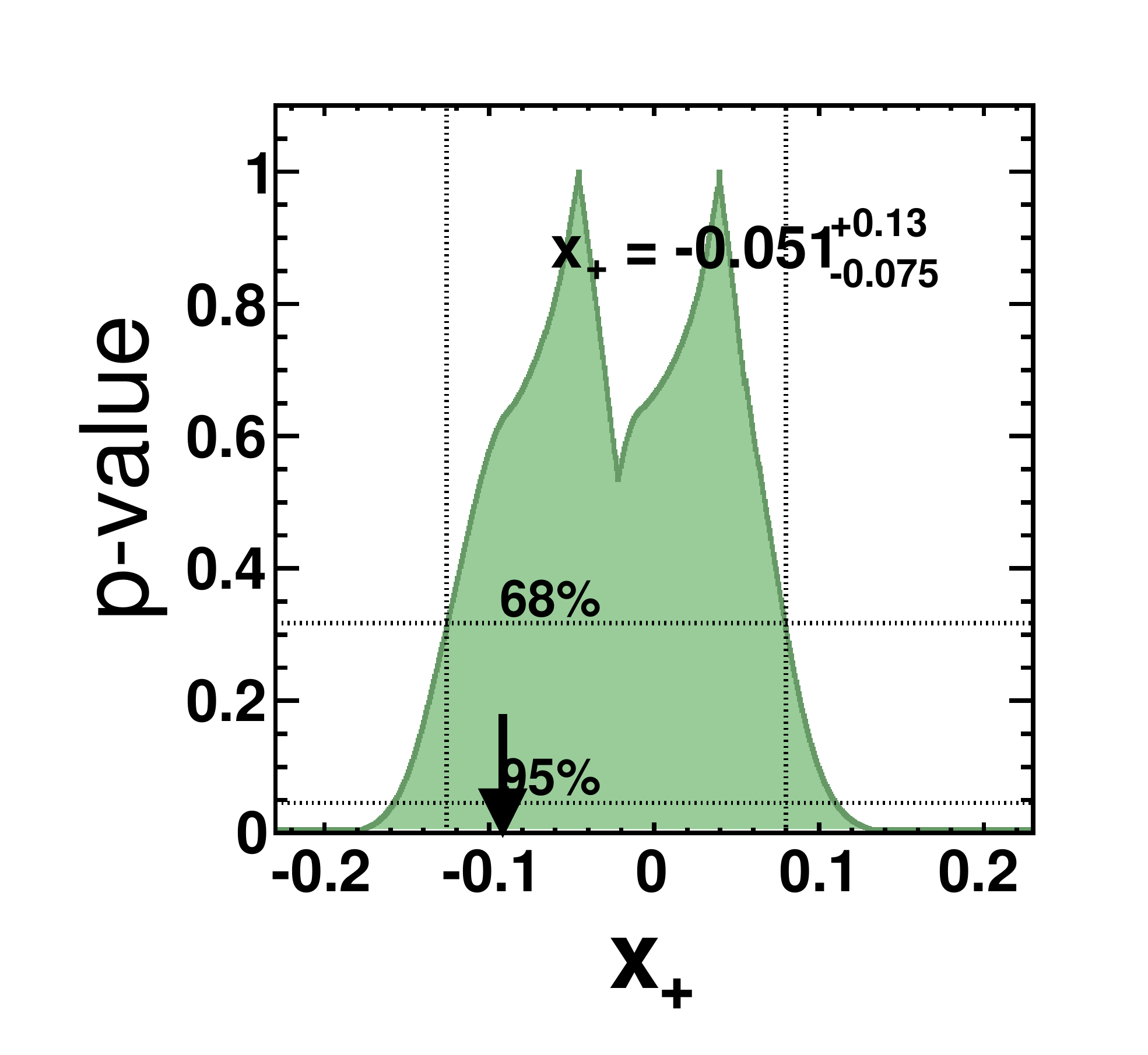} &
\CLScanBox{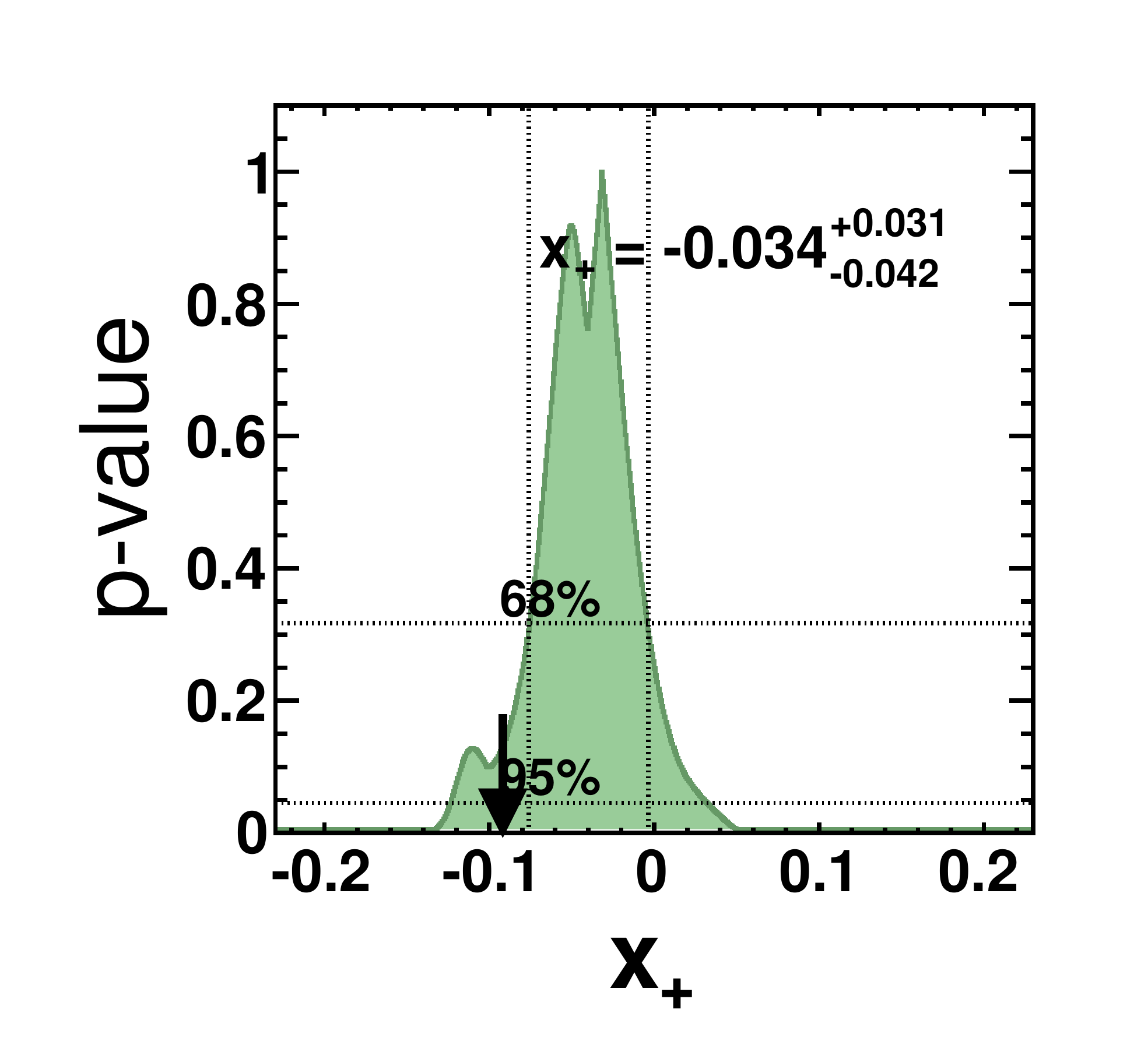} &
\CLScanBox{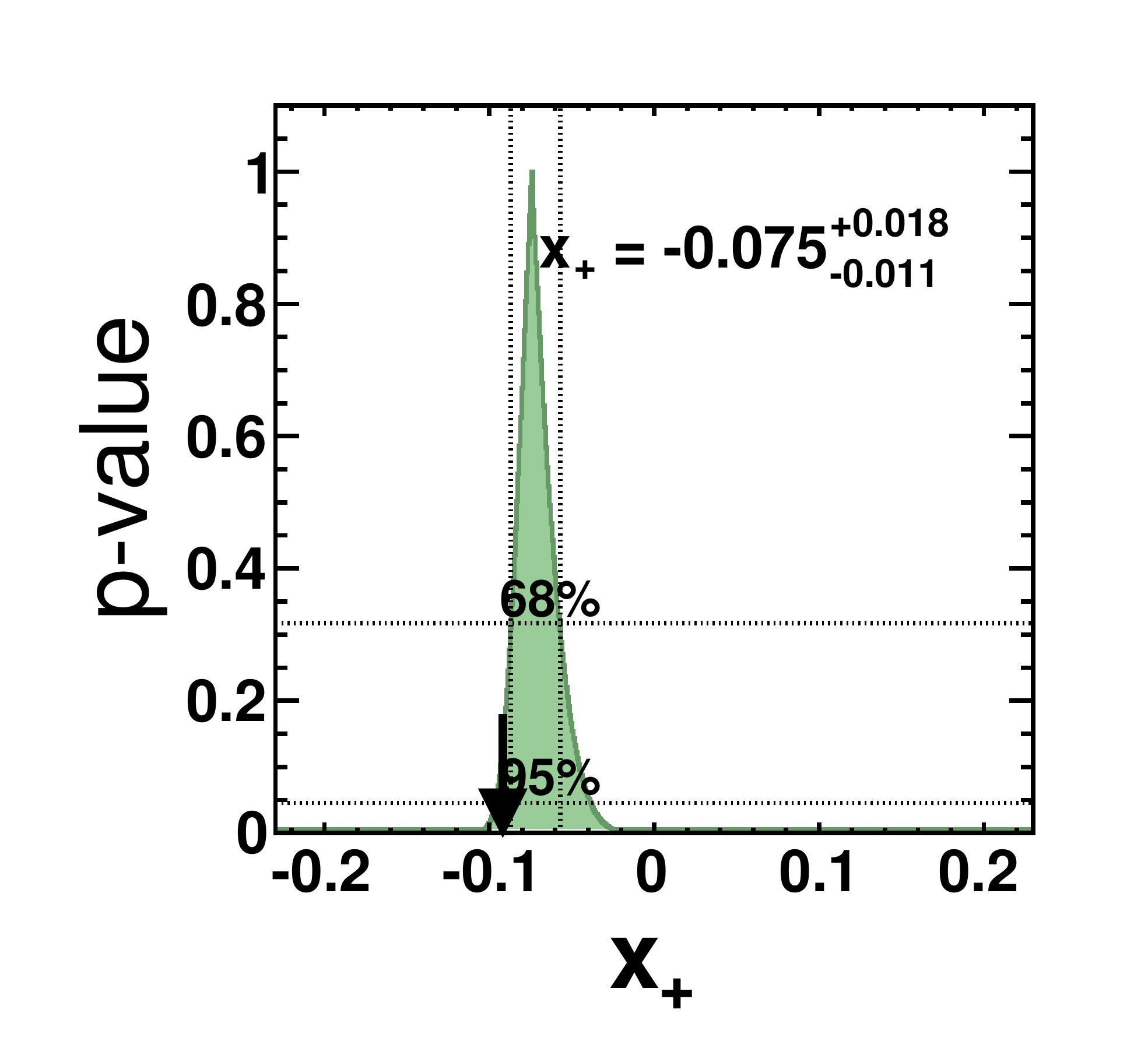}
\\ 
\CLScanBox{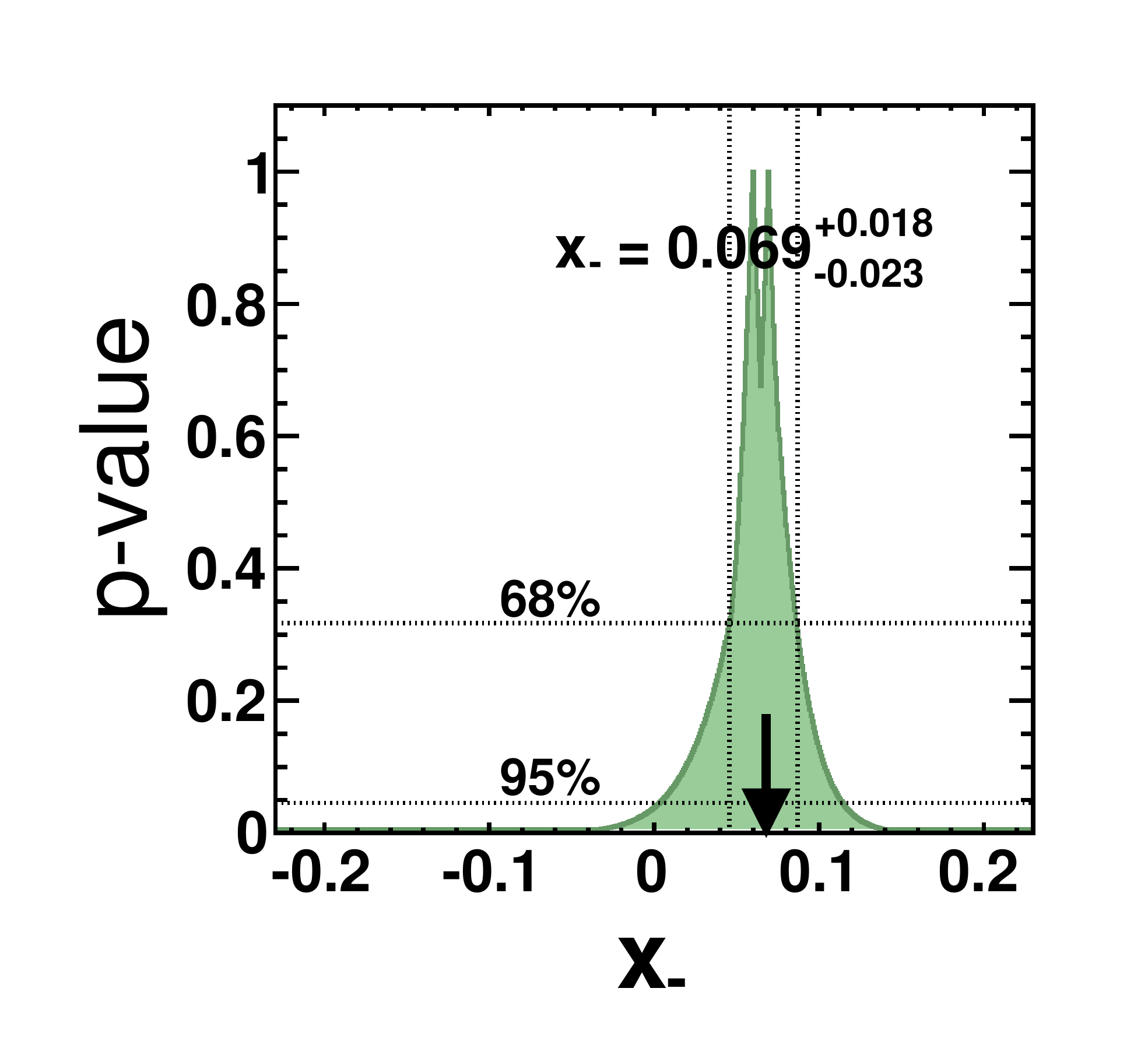} &
\CLScanBox{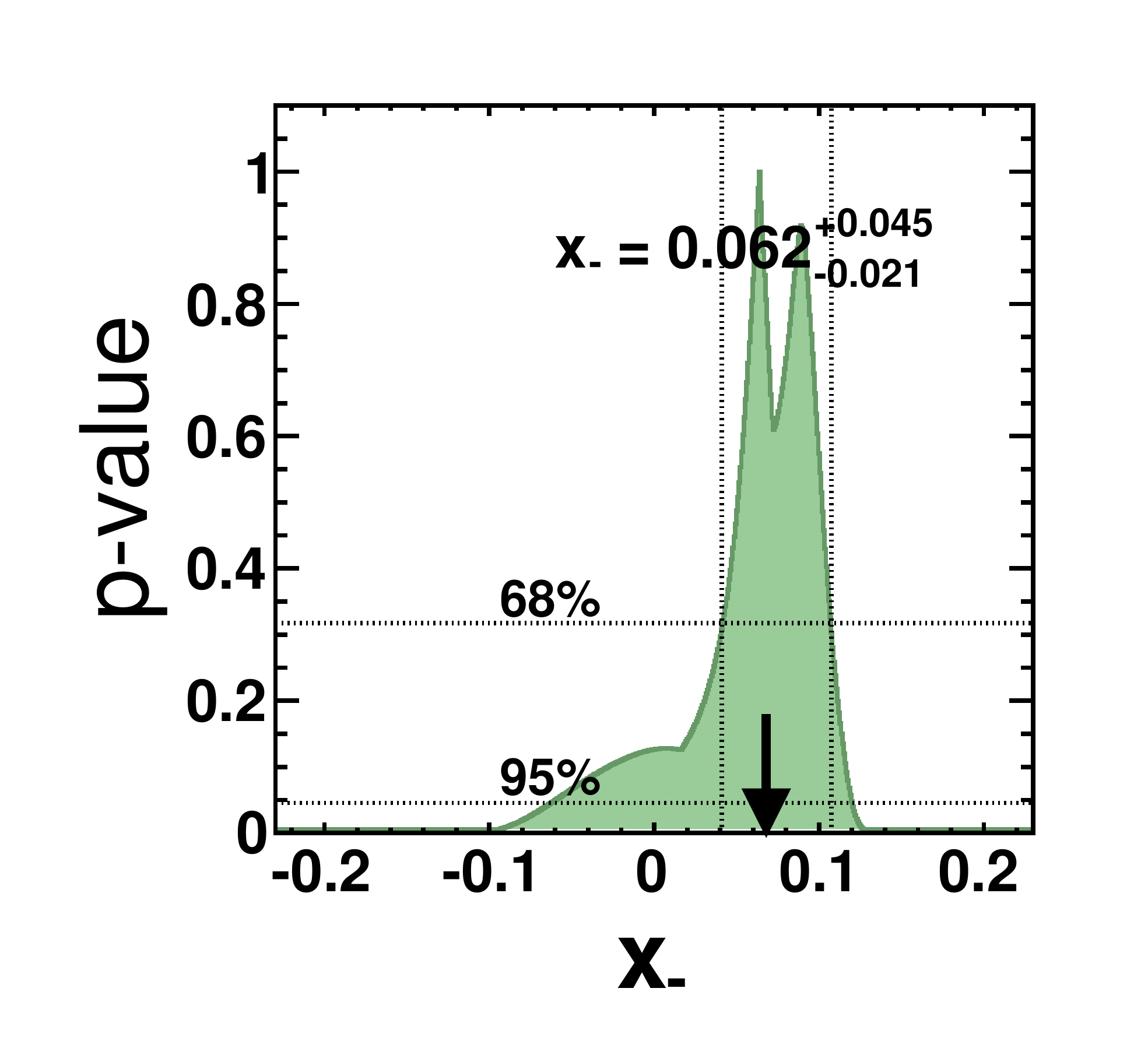} &
\CLScanBox{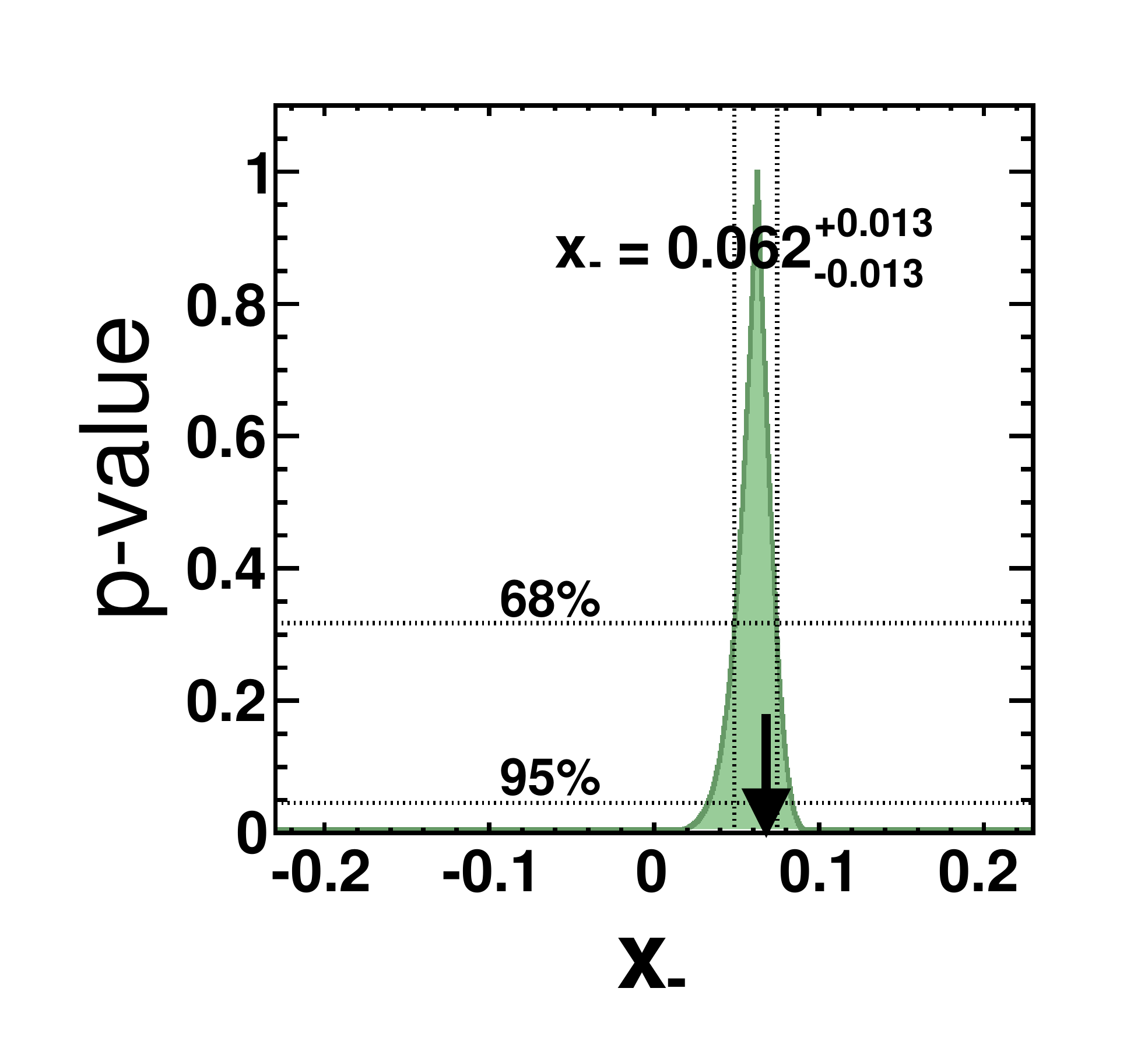}
\end{tabular}
\caption{ The $p$-value (see~\eqnref{eq:PGaussian}) versus \gam, $x_+$,
  and $x_-$ for different charm inputs for estimated LHCb run~II
  statistics. The arrow indicates the input value with which the
  experiment was simulated. The numbers inside the scans represent the
  best fit value $\pm 1\sigma$, as described in the
  text.\label{fig:gammaScans}}
\end{figure}
We perform one-dimensional $p$-value (see~\eqnref{eq:PGaussian}) scans
of the parameters of interest. To translate a scan into a numerical
result for the uncertainty $\sigma$ on a given parameter, we choose
the peak associated to the fit result nearest the input value
with which the data were generated, and take half its width at
$1-p=68\%$. We ignore multiple solutions, unless two solutions merge
at the $68\%$ CL level, in which case we take the width of the
merged double-peak to calculate $\sigma$. This is illustrated for a
few examples in \figref{fig:gammaScans}.

\subsection{Summary of results}
\label{sec:ResultSummary}
\begin{table}
\centering
\input{theTable}
\caption{Uncertainties on key parameters, obtained based on the
  default amplitude model in
  different configurations, averaged over 50 simulated
  experiments. 
  All results are for the binned approach applied to \BtoDK\ and,
  where used, charm mixing data. The first column refers to the
  scenarios defined in \tabref{tab:eventYields}. The second column
  defines whether charm mixing input was used (Y), or not (N). The third column describes additional input from the
  charm threshold. ``CLEO global'' refers to the phase-space integrated input
  from~\cite{Libby:2014rea}. ``BES~III global'' is the same, but
  uses the uncertainties predicted in~\cite{Libby:2014rea} 
  for a data sample $3.5$ times as large as that collected by \cleoc. 
  ``CLEO binned'' and ``BES~III binned''
  extrapolate to a potential binned analysis of the charm
  threshold data described
  in~\secref{sec:binnedInput}.
  \label{tab:allResults}}
\end{table}
\Tabref{tab:allResults} summarises our estimates of the uncertainties
on the parameters describing \CP\ violation in \BtoDK, measured in
\BtoDK, \prt{\Dee \to K\pi\pi\pi} for different charm inputs and data
taking scenarios. These estimates are obtained from $p$-value scans as
described above, averaged over $50$ simulated experiments, generated
using the default amplitude model.

The results indicate that an interesting precision on these parameters
 (especially $x_-$ and $y_-$) can
be achieved solely based on a combined analysis of \BtoDK, \prt{\Dee
  \to K\pi\pi\pi} and charm mixing data in several bins of the \Dee\
decay's phase space. Such a result would not provide a competitive
measurement of $\gam$ by itself, but would be expected to make a
valuable contribution to a combined fit, such as the ones described
in~\cite{Libby:2014rea, LHCb2013GammaCombination, Lowery:2009id}.

However, using both charm input from mixing and from threshold data
transforms this into a precision measurement of \gam. While precise
predictions are impossible until we have a better understanding of the
\DzerotoKpipipiDCS\ amplitude structure, the above results suggests
that, with the approach proposed here applied to LHCb run~1 data, this
channel can reach a similar precision as the combined analysis of
\prt{\BtoDK} with \prt{\Dee \to K_S\pip\pim} and \prt{\Dee \to
  K_S K^+K^-} on LHCb run~1 data~\cite{ModelIndepGammaDalitz_LHCb2014},
currently the most precise individual measurement of \gam\ in
tree-level decays. Conversely, the inclusion of information from charm
mixing leads to a vastly improved precision compared to that
achievable based on charm input from threshold data alone, by about an
order of magnitude for the upgrade scenario, emphasising the crucial
role of the information from charm mixing.

Finally, our results indicate that the input from \besIII\ has the
potential to substantially improve the precision on \gam\ over that
achievable with \cleoc's dataset alone, especially if a binned
analysis were to be performed. Further improvements would be expected
from combining \cleoc\ and \besIII\ input, which, in this study, we
only considered separately.

%% file: theTable.tex
\begin{tabular}{lcc | ccc|cccc}
& & & $\sigma(\gamma)$ & $\sigma(\delB)$ & \multicolumn{1}{c|}{$\sigma(\rB)$} 
& $\sigma(x_+)$ & $\sigma(y_+)$ & $\sigma(x_-)$ & $\sigma(y_-)$
\\ \rotatebox{85}{\parbox{0em}{\parbox{5em}{LHCb\newline scenario}}} 
& \rotatebox{85}{\parbox{0em}{\parbox{5em}{\Do\ mix?}}}
& \multicolumn{1}{c|}{\rotatebox{85}{\parbox{0em}{\parbox{5em}{charm threshold?}}}}
& $[\mbox{}^{\circ}]$ & $[\mbox{}^{\circ}]$ & \multicolumn{1}{c|}{$\times 10^2$} 
& $\times 10^2$ & $\times 10^2$ & $\times 10^2$ & $\times 10^2$
\\\hline
run~I &  &  & 26 & 47 & 1.6 & 8.7 & 9.1 & 8.8 & 8.2 \\ 
run~II & Y &  & 22 & 29 & 1.4 & 7.6 & 6.9 & 4.5 & 4.0 \\ 
upgr &  & \rotatebox{85}{\parbox{0em}{none}} & 15 & 14 & 0.17 & 4.7 & 5.2 & 0.56 & 0.98 \\ 
\hline
run~I &  &  & 20 & 29 & 0.82 & 6.4 & 5.7 & 6.6 & 5.9 \\ 
run~II & Y &  & 15 & 19 & 0.62 & 5.4 & 3.9 & 2.5 & 2.7 \\ 
upgr &  & \rotatebox{85}{\parbox{0em}{CLEO global}} & 11 & 10 & 0.16 & 3.8 & 2.8 & 0.44 & 0.50 \\ 
\hline
run~I &  &  & 19 & 25 & 0.78 & 6.4 & 5.5 & 6.5 & 5.8 \\ 
run~II & Y &  & 14 & 18 & 0.57 & 5.4 & 3.9 & 2.4 & 2.7 \\ 
upgr &  & \rotatebox{85}{\parbox{0em}{BESIII global}} & 9.0 & 8.2 & 0.15 & 3.7 & 2.7 & 0.43 & 0.48 \\ 
\hline
run~I &  &  & 46 & 35 & 3.2 & 6.9 & 6.5 & 8.6 & 10 \\ 
run~II & N &  & 50 & 34 & 3.3 & 6.9 & 6.7 & 8.9 & 11 \\ 
upgr &  & \rotatebox{85}{\parbox{0em}{CLEO binned}} & 52 & 35 & 3.3 & 7.6 & 6.7 & 8.9 & 11 \\ 
\hline
run~I &  &  & 40 & 24 & 2.6 & 4.1 & 5.0 & 5.7 & 6.2 \\ 
run~II & N &  & 34 & 17 & 2.5 & 3.6 & 4.1 & 5.0 & 5.1 \\ 
upgr &  & \rotatebox{85}{\parbox{0em}{BESIII binned}} & 39 & 14 & 2.9 & 3.9 & 4.1 & 4.3 & 5.6 \\ 
\hline
run~I &  &  & 16 & 18 & 0.78 & 2.1 & 3.5 & 2.6 & 3.1 \\ 
run~II & Y &  & 12 & 13 & 0.53 & 1.7 & 3.1 & 1.7 & 2.0 \\ 
upgr &  & \rotatebox{85}{\parbox{0em}{CLEO binned}} & 7.8 & 7.2 & 0.15 & 1.1 & 2.6 & 0.40 & 0.46 \\ 
\hline
run~I &  &  & 12 & 14 & 0.68 & 1.6 & 2.6 & 2.0 & 2.5 \\ 
run~II & Y &  & 8.6 & 9.6 & 0.47 & 0.90 & 2.1 & 1.5 & 1.5 \\ 
upgr &  & \rotatebox{85}{\parbox{0em}{BESIII binned}} & 4.1 & 3.9 & 0.14 & 0.53 & 1.3 & 0.35 & 0.38 \\ 
\hline
\end{tabular}

%% file: 6-conclusions.tex
We have presented a new method for the amplitude model-independent
measurement of the \CP\ violation parameter \gam\ from \BtoDK\ decays,
based on a combined analysis of \BtoDK\ and charm mixing. When
analysed in several bins of the \Dee\ decay's phase space, \gam\ can be
measured without additional input from the charm threshold.
We have evaluated the performance of the method in a simulation
study for the case where the \Dee\ decays to \KPlusMinuspipipi, using
sample sizes representing existing and plausible future datasets. The
precision ultimately achievable depends on the \DzerotoKpipipiDCS\
amplitude structure realised in nature, that we do not know. Our
results suggest that the new method we introduced would, even without
input from the charm threshold, provide valuable input to a
global \gam\ combination, although the precision would be insufficient
to provide a competitive \gam\ measurement in its own right.

We compare the performance of our novel method to that of a binned
analysis with charm input from the threshold, as proposed
in~\cite{Atwood:coherenceFactor}. For the run~I scenario, with
\besIII\ statistics, both methods perform similarly well. Assuming no
additional data from the threshold, the mixing-based method introduced
here performs significantly better for the LHCb-upgrade scenario,
benefiting from the vast number of \Dee\ events expected.

For all data taking scenarios we studied, combining the two methods
results in a far superior performance than either can achieve
individually. This is already the case when threshold data enter in
the form of a phase-space integrated constraint on \Z, but by far the
best results are obtained if \Dee\ mixing, \BtoDK\ and charm threshold
data are analysed in the same phase space bins. Such a combined
approach transforms this into a highly competitive precision
measurement of \gam, on par with the best existing constraints from
individual channels. Its precision keeps improving with charm mixing
and \BtoDK\ event yields projected into the foreseeable future, even
if no new data from the charm threshold become available.


Once a \DzerotoKpipipiDCS\ amplitude model is available to inform the
binning, the techniques we introduced here can be used to significantly
improve the precision on \gam\ and related parameters that can be
obtained from \BtoDK, \DtoKPlusMinuspipipi. Such a measurement would
benefit greatly from an update of the $\ZKpipipi = \RDKpipipi
e^{-\delDKpipipi}$ measurement~\cite{Libby:2014rea,Lowery:2009id} with
\besIII's larger dataset, and, even more so, a binned \ZOmKpipipi\
analysis. With all of the above ingredients in place, the methods
introduced in this letter, applied to \BtoDK, \DtoKPlusMinuspipipi,
could lead to one of the most precise individual \gam\
measurements.

 Its potential for other decay channels is yet to be
evaluated.

%% file: 7-acknowledgements.tex
We thank our colleagues from CLEO-c and LHCb for their helpful input
to this paper, in particular Tim Gershon, Patrick Koppenburg, and Jim
Libby. We acknowledge the support from CERN, the Science and
Technology Facilities Council (United Kingdom) and the European
Research Council under FP7.

%% file: 0-main.bbl
\providecommand{\href}[2]{#2}\begingroup\raggedright\begin{thebibliography}{10}

\bibitem{GLW1}
M.~Gronau and D.~Wyler, {\it {On determining a weak phase from CP asymmetries
  in charged B decays}},  {\em Phys.Lett.} {\bf B265} (1991) 172--176.

\bibitem{GLW2}
M.~Gronau and D.~London, {\it {How to determine all the angles of the unitarity
  triangle from $B_d \to D K_s$ and $B_{s}^{0} \to D \phi$}},  {\em Phys.Lett.}
  {\bf B253} (1991) 483--488.

\bibitem{ADS}
D.~Atwood, I.~Dunietz, and A.~Soni, {\it {Enhanced CP violation with $B
  \rightarrow \mathrm{KD}^{0}(\overline{\mathrm{D}}^{0})$ modes and extraction
  of the Cabibbo-Kobayashi-Maskawa Angle $\gamma$}},  {\em Phys.Rev.Lett.} {\bf
  78} (1997) 3257--3260, [\href{http://arxiv.org/abs/hep-ph/9612433}{{\tt
  hep-ph/9612433}}].

\bibitem{DalitzGamma1}
A.~Giri, Y.~Grossman, A.~Soffer, and J.~Zupan, {\it Determining $\gamma$ using
  ${B}^{\pm} \rightarrow \mathrm{DK}^{\pm}$ with multibody {D} decays},  {\em
  Phys. Rev. D} {\bf 68} (Sep, 2003) 054018.

\bibitem{DalitzGamma2}
{\bf Belle} Collaboration, A.~Poluektov et~al., {\it Measurement of $\phi_{3}$
  with dalitz plot analysis of ${B}^{\pm} \rightarrow {D}^{(*)}{K}^{\pm}$
  decays},  {\em Phys. Rev. D} {\bf 70} (Oct, 2004) 072003.

\bibitem{Rademacker:2006zx}
J.~Rademacker and G.~Wilkinson, {\it {Determining the unitarity triangle gamma
  with a four-body amplitude analysis of $B^{+}\rightarrow (K^{+} K^{-} \pi^{+}
  \pi^{-})_{D}K^{\pm}$ decays}},  {\em Phys.Lett.} {\bf B647} (2007) 400--404,
  [\href{http://arxiv.org/abs/hep-ph/0611272}{{\tt hep-ph/0611272}}].

\bibitem{Atwood:coherenceFactor}
D.~Atwood and A.~Soni, {\it {Role of charm factory in extracting CKM phase
  information via B$\to$DK}},  {\em Phys.Rev.} {\bf D68} (2003) 033003,
  [\href{http://arxiv.org/abs/hep-ph/0304085}{{\tt hep-ph/0304085}}].

\bibitem{ModelIndepGammaTheory}
A.~Giri, Y.~Grossman, A.~Soffer, and J.~Zupan, {\it {Determining gamma using
  $B^{\pm} \rightarrow DK^{\pm}$ with multibody $D$ decays}},  {\em Phys.Rev.}
  {\bf D68} (2003) 054018, [\href{http://arxiv.org/abs/hep-ph/0303187}{{\tt
  hep-ph/0303187}}].

\bibitem{Lowery:2009id}
{\bf CLEO} Collaboration, N.~Lowrey et~al., {\it {Determination of the
  $D^{0}\rightarrow K^- \pi^+ \pi^0$ and $D^0 \rightarrow K^- \pi^+ \pi^+
  \pi^-$ Coherence Factors and Average Strong-Phase Differences Using
  Quantum-Correlated Measurements}},  {\em Phys.Rev.} {\bf D80} (2009) 031105,
  [\href{http://arxiv.org/abs/0903.4853}{{\tt arXiv:0903.4853}}].

\bibitem{Libby:2010nu}
{\bf CLEO} Collaboration, J.~Libby et~al., {\it {Model-independent
  determination of the strong-phase difference between $D^0$ and $\bar{D}^0 \to
  K^0_{S,L} h^+ h^-$ ($h=\pi,K$) and its impact on the measurement of the CKM
  angle $\gamma/\phi_3$}},  {\em Phys.Rev.} {\bf D82} (2010) 112006,
  [\href{http://arxiv.org/abs/1010.2817}{{\tt arXiv:1010.2817}}].

\bibitem{Briere:2009aa}
{\bf CLEO} Collaboration, R.~A. Briere et~al., {\it {First model-independent
  determination of the relative strong phase between $D^{0}$ and
  $\bar{D}^{0}\rightarrow K^{0}_s \pi^+ \pi^-$ and its impact on the CKM Angle
  $\gamma / \phi_3$ measurement}},  {\em Phys.Rev.} {\bf D80} (2009) 032002,
  [\href{http://arxiv.org/abs/0903.1681}{{\tt arXiv:0903.1681}}].

\bibitem{CLEO:DeltaKpi}
{\bf CLEO} Collaboration, D.~M. Asner et~al., {\it {Determination of the $D^0
  \to K^{+} \pi^{-}$ Relative Strong Phase Using Quantum-Correlated
  Measurements in $e^{+} e^{-} \to D^0 \bar{D}^0$ at CLEO}},  {\em Phys.Rev.}
  {\bf D78} (2008) 012001, [\href{http://arxiv.org/abs/0802.2268}{{\tt
  arXiv:0802.2268}}].

\bibitem{Insler:2012pm}
{\bf CLEO} Collaboration, J.~Insler et~al., {\it {Studies of the decays $D^0
  \rightarrow K_S^0K^-\pi^+$ and $D^0 \rightarrow K_S^0K^+\pi^-$}},  {\em
  Phys.Rev.} {\bf D85} (2012) 092016,
  [\href{http://arxiv.org/abs/1203.3804}{{\tt arXiv:1203.3804}}].

\bibitem{Libby:2014rea}
J.~Libby et~al., {\it {New determination of the $D^{0} \to K^{-} \pi^{+}
  \pi^{0}$ and $D^{0} \to K^{-} \pi^{+} \pi^{+} \pi^{-}$ coherence factors and
  average strong-phase differences}},  {\em Phys.Lett.} {\bf B731} (2014)
  197--203, [\href{http://arxiv.org/abs/1401.1904}{{\tt arXiv:1401.1904}}].

\bibitem{selfcite}
S.~Harnew and J.~Rademacker, {\it {Charm mixing as input for model-independent
  determinations of the CKM phase $\gamma$}},  {\em Phys.Lett.} {\bf B728}
  (2014) 296--302, [\href{http://arxiv.org/abs/1309.0134}{{\tt
  arXiv:1309.0134}}].

\bibitem{Bondar:CharmMixingCP}
A.~Bondar, A.~Poluektov, and V.~Vorobiev, {\it {Charm mixing in the
  model-independent analysis of correlated $D^0$ $\bar{D}^0$ decays}},  {\em
  Phys.Rev.} {\bf D82} (2010) 034033,
  [\href{http://arxiv.org/abs/1004.2350}{{\tt arXiv:1004.2350}}].

\bibitem{Belle_uses_CLEO_2011}
{\bf Belle} Collaboration, I.~Adachi et~al., {\it {First measurement of
  $\phi_3$ with a binned model-independent Dalitz plot analysis of
  $\mathit{B^{+-} \to DK^{+-}, D \to K_s^0 \pi^+\pi^-}$ decay}},
  \href{http://arxiv.org/abs/1106.4046}{{\tt arXiv:1106.4046}}.

\bibitem{LHCb2012DalitzGamma}
{\bf LHCb} Collaboration, R.~Aaij et~al., {\it {A model-independent Dalitz plot
  analysis of $B^\pm \to D K^\pm$ with $D \to K^0_{\rm S} h^+h^-$ ($h=\pi, K$)
  decays and constraints on the CKM angle $\gamma$}},  {\em Phys. Lett.} {\bf
  B718} (2012) 43--55, [\href{http://arxiv.org/abs/1209.5869}{{\tt
  arXiv:1209.5869}}].

\bibitem{HFAG2013}
{\bf Heavy Flavor Averaging Group} Collaboration, Y.~Amhis et~al., {\it
  {Averages of B-Hadron, C-Hadron, and tau-lepton properties as of early
  2012}},  \href{http://arxiv.org/abs/1207.1158}{{\tt arXiv:1207.1158}}.

\bibitem{LHCbMixingAndCPV}
{\bf LHCb} Collaboration, R.~Aaij et~al., {\it {Measurement of
  $D^0$--$\bar{D}^0$ mixing parameters and search for CP violation using
  $D^0\to K^+\pi^-$ decays}},  {\em Phys.Rev.Lett.} {\bf 111} (2013) 251801,
  [\href{http://arxiv.org/abs/1309.6534}{{\tt arXiv:1309.6534}}].

\bibitem{CDF:Mixing2008}
{\bf CDF} Collaboration, T.~Aaltonen et~al., {\it {Evidence for $D^0 -
  \bar{D}^0$ mixing using the CDF II Detector}},  {\em Phys.Rev.Lett.} {\bf
  100} (2008) 121802, [\href{http://arxiv.org/abs/0712.1567}{{\tt
  arXiv:0712.1567}}].

\bibitem{Belle:Mixing2007}
{\bf Belle} Collaboration, M.~Staric et~al., {\it {Evidence for $D^0$ -
  $\bar{D}^0$ Mixing}},  {\em Phys.Rev.Lett.} {\bf 98} (2007) 211803,
  [\href{http://arxiv.org/abs/hep-ex/0703036}{{\tt hep-ex/0703036}}].

\bibitem{BaBar:Mixing2007}
{\bf BaBar} Collaboration, B.~Aubert et~al., {\it {Evidence for $D^0$ -
  $\bar{D}^0$ Mixing}},  {\em Phys.Rev.Lett.} {\bf 98} (2007) 211802,
  [\href{http://arxiv.org/abs/hep-ex/0703020}{{\tt hep-ex/0703020}}].

\bibitem{BaBar:Mixing2008}
{\bf BaBar} Collaboration, B.~Aubert et~al., {\it {Measurement of $D^0 -
  \bar{D}^0$ mixing from a time-dependent amplitude analysis of $D^0 \to K^{+}
  \pi^{-} \pi^0$ decays}},  {\em Phys.Rev.Lett.} {\bf 103} (2009) 211801,
  [\href{http://arxiv.org/abs/0807.4544}{{\tt arXiv:0807.4544}}].

\bibitem{BaBar:Mixing2009}
{\bf BaBar} Collaboration, B.~Aubert et~al., {\it {Measurement of $D^0 -
  \bar{D}^{0}$ Mixing using the Ratio of Lifetimes for the Decays $D^{0}
  \rightarrow K^- \pi^+$ and $K^+ K^-$}},  {\em Phys.Rev.} {\bf D80} (2009)
  071103, [\href{http://arxiv.org/abs/0908.0761}{{\tt arXiv:0908.0761}}].

\bibitem{LHCb:Mixing}
{\bf LHCb} Collaboration, R.~Aaij et~al., {\it {Observation of
  $D^0$--$\overline{D}^0$ oscillations}},  {\em Phys. Rev. Lett.} {\bf 110}
  (2013) 101802, [\href{http://arxiv.org/abs/1211.1230}{{\tt
  arXiv:1211.1230}}].

\bibitem{LHCb2013GammaCombination}
{\bf LHCb} Collaboration, R.~Aaij et~al., {\it {Measurement of the CKM angle
  $\gamma$ from a combination of $B^{\pm} \to Dh^{\pm}$ analyses}},  {\em
  Phys.Lett.} {\bf B726} (2013) 151--163,
  [\href{http://arxiv.org/abs/1305.2050}{{\tt arXiv:1305.2050}}].

\bibitem{LHCb-CONF-2013-006}
{\bf LHCb} Collaboration, {\it {Improved constraints on $\gamma$ from $B^\pm\to
  DK^\pm$ decays including first results on 2012 data}},  {\em
  {LHCb-CONF-2013-006}} (2013).

\bibitem{Rama:2013voa}
M.~Rama, {\it {Effect of D-Dbar mixing in the extraction of gamma with $B^- \to
  D^0 K^-$ and $B^- \to D^0 \pi^-$ decays}},  {\em Phys.Rev.} {\bf D89} (2014)
  014021, [\href{http://arxiv.org/abs/1307.4384}{{\tt arXiv:1307.4384}}].

\bibitem{MarkIII_K3piModel}
{\bf Mark III} Collaboration, D.~Coffman et~al., {\it Resonant substructure in
  $\overline{K}\pi\pi\pi$ decays of $d$ mesons},  {\em Phys. Rev. D} {\bf 45}
  (Apr, 1992) 2196--2211.

\bibitem{LHCb2013:Miranda}
{\bf LHCb} Collaboration, R.~Aaij et~al., {\it {Model-independent search for CP
  violation in $D^{0} \to K^- K^+ \pi^- \pi^+$ and $D^{0} \to \pi^- \pi^+ \pi^-
  \pi^+$ decays}},  \href{http://arxiv.org/abs/1308.3189}{{\tt
  arXiv:1308.3189}}.

\bibitem{Bondar:2007ir}
A.~Bondar and A.~Poluektov, {\it {On model-independent measurement of the angle
  phi(3) using Dalitz plot analysis}},
  \href{http://arxiv.org/abs/hep-ph/0703267}{{\tt hep-ph/0703267}}.

\bibitem{LHCb2013ADSObservation}
{\bf LHCb} Collaboration, R.~Aaij et~al., {\it {Observation of the suppressed
  ADS modes $B^\pm \to [\pi^\pm K^\mp\pi^+\pi^-]_D K^\pm$ and $B^\pm \to
  [\pi^\pm K^\mp \pi^+\pi^-]_D \pi^\pm$}},  {\em Phys. Lett.} {\bf B723} (2013)
  44, [\href{http://arxiv.org/abs/1303.4646}{{\tt arXiv:1303.4646}}].

\bibitem{LHCbUpgradeCDR}
{\bf LHCb} Collaboration, I.~Bediaga et~al., {\it {Framework TDR for the LHCb
  Upgrade: Technical Design Report}},  Tech. Rep. CERN-LHCC-2012-007, CERN,
  Apr, 2012.

\bibitem{HFAG_CHARM2013}
{\bf Heavy Flavor Averaging Group} Collaboration, Y.~Amhis et~al., {\it
  {Averages of B-Hadron, C-Hadron, and tau-lepton properties as of early 2012.
  }},  \href{http://arxiv.org/abs/1207.1158}{{\tt arXiv:1207.1158}}. Regular
  online updates, we use CHARM 2013 averages.

\bibitem{PDG2014}
{\bf {P}article {D}ata {G}roup} Collaboration, J.~A. Olive et~al., {\it
  {R}eview of {P}article {P}hysics},  {\em Chin.Phys.} {\bf C38} (Aug, 2014)
  090001.

\bibitem{geneva}
D.~M. Kunze, D.~S. Gabriel, and D.~A. Garcia, {\it Distributed parametric
  optimization with the {G}eneva library},  in {\em Data Driven e-Science}
  (S.~Lin and E.~Yen, eds.), p.~303, Springer, 2011.

\bibitem{MINUIT}
F.~James, {\it {MINUIT Function Minimization and Error Analysis: Reference
  Manual Version 94.1}},  {\em {CERN-D-506}} (1994).

\bibitem{ModelIndepGammaDalitz_LHCb2014}
{\bf LHCb} Collaboration, R.~Aaij et~al., {\it {Measurement of the CKM angle
  $\gamma$ using $B^\pm \to D K^\pm$ with $D \to K^0_{\rm S} \pi^+\pi^-,
  K^0_{\rm S} K^+ K^-$ decays}},  {\em JHEP} {\bf 1410} (2014) 97,
  [\href{http://arxiv.org/abs/1408.2748}{{\tt arXiv:1408.2748}}].

\end{thebibliography}\endgroup
